\documentclass{aa}  

\usepackage{graphicx}
\usepackage{txfonts}

\usepackage{natbib} 
\usepackage{rotating} 
\graphicspath{{figures/}}
\usepackage{gensymb} 
\usepackage{pdflscape}
\usepackage[titletoc,title]{appendix}

\begin{document} 

   \title{LOFAR MSSS: The Scaling Relation between AGN Cavity Power and Radio Luminosity at Low Radio Frequencies}   
   	\titlerunning{The Scaling Relation between AGN Cavity Power and Radio Luminosity at Low Radio Frequencies}

   \author{G. Kokotanekov \inst{1}\thanks{E-mail: g.d.kokotanekov@uva.nl}
          \and
          M. Wise \inst{1,2} 
          \and
          G.~H. Heald \inst{3,2,4}
          \and
          J.~P. McKean \inst{2,4}
          \and
          L. B\^irzan \inst{5}
          \and 
          D.~A. Rafferty \inst{5}
          \and
          L.~E.~H. Godfrey \inst{2}
          \and
          M.~ de Vries \inst{1}
          \and
          H.~T. Intema \inst{6}
          \and
          J. W. Broderick \inst{2}
          \and
          M.J. Hardcastle  \inst{7}
          \and
          A. Bonafede \inst{8,5}
          \and
          A. O. Clarke \inst{9}
          \and
          R. J. van Weeren \inst{10}
          \and
          H. J. A. R\"ottgering \inst{6}
          \and
          R. Pizzo \inst{2}
          \and
          M. Iacobelli \inst{2}
          \and
          E. Orr\'u \inst{2,11}
          \and
          A. Shulevski \inst{2}
          \and
          C. J. Riseley \inst{3}
          \and
          R. P. Breton \inst{9}
          \and
          B. Nikiel-Wroczy\'nski \inst{12}
          \and
          S. S. Sridhar \inst{2,4}
          \and
          A. J. Stewart \inst{13}
          \and
          A. Rowlinson \inst{2,1}
          \and
          A. J. van der Horst \inst{14}
          \and
          J. J. Harwood \inst{2}
          \and
          G. G\"urkan \inst{3}
          \and
          D. Carbone \inst{1}
          \and
          M. Pandey-Pommier \inst{15}
          \and
          C. Tasse \inst{16,17}
          \and
          A. M. M. Scaife \inst{9}
          \and
          L. Pratley \inst{18,19}
          \and
          C. Ferrari \inst{20}
          \and
          J. H. Croston \inst{21}
          \and
          V. N. Pandey \inst{2,4}
          \and
          W. Jurusik \inst{12}
          \and
          D. D. Mulcahy \inst{9}
           \vspace{-2ex}
          }

   \institute{Anton Pannekoek Institute for Astronomy, University of Amsterdam, Postbus 94249, 1090 GE Amsterdam, The Netherlands
         \and
          Netherlands Institute for Radio Astronomy (ASTRON), Postbus 2, 7990 AA Dwingeloo, The Netherlands
          \and
          CSIRO Astronomy and Space Science, 26 Dick Perry Ave., Kensington WA 6151, Australia
          \and
          Kapteyn Astronomical Institute, Rijksuniversiteit Groningen, Landleven 12, 9747 AD Groningen, The Netherlands
          \and
          Universit\"at Hamburg, Hamburger Sternwarte, Gojenbergsweg 112, 21029, Hamburg, Germany
          \and
          Leiden Observatory, Leiden University, Niels Bohrweg 2, NL-2333CA, Leiden, The Netherlands
         \and
         School of Physics, Astronomy and Mathematics, University of Hertfordshire, College Lane, Hatfield AL10 9AB, UK
         \and
         Radio astronomy institute (IRA) - INAF - via Gobetti 101 40100 Bologna, Italy   
         \and 
         Jodrell Bank Centre for Astrophysics, University of Manchester, Manchester, UK, M139PL      
         \and
         Harvard-Smithsonian Center for Astrophysics, 60 Garden Street, Cambridge, MA 02138, USA
         \and
         Department of Astrophysics, Institute for Mathematics, Astrophysics and Particle Physics (IMAPP), Radboud University Nijmegen, P.O. Box 9010, 6500 GL Nijmegen, The Netherlands
         \and
         Astronomical Observatory, Jagiellonian University, ul. Orla 171, 30-244 Krak\'ow, Poland
         \and
         Astrophysics, Department of Physics, University of Oxford, Keble Road, Oxford OX1 3RH, UK
         \and
         Department of Physics, The George Washington University, 725 21st Street NW, Washington, DC 20052, USA
         \and
         Centre de Recherche Astrophysique de Lyon, Observatoire de Lyon, 9 Av. Charles Andr\'e, F-69230, Saint-Genis-Laval, France
         \and
         GEPI, Observatoire de Paris, CNRS, Universit\'e Paris Diderot, 5 place Jules Janssen, 92190 Meudon, France
         \and
         Department of Physics \& Electronics, Rhodes University, PO Box 94, Grahamstown, 6140, South Africa
         \and
         Mullard Space Science Laboratory (MSSL), University College London (UCL), Holmbury St Mary, Surrey RH5 6NT, UK
         \and
         School of Chemical and Physical Sciences, Victoria University of Wellington, PO Box 600, Wellington 6140, New Zealand
         \and
         Laboratoire Lagrange, Universit\'e C\^ote d’Azur, Observatoire de la C\^ote d’Azur, CNRS, Bvd de l’Observatoire, CS 34229, 06304 Nice Cedex 4, France
         \and
         School of Physics and Astronomy, University of Southampton, Southampton, SO17 1BJ, UK
             }

   \date{Received April 06, 2017; accepted May 30, 2017}

 
\abstract{

We present a new analysis of the widely used relation between cavity power and radio luminosity in clusters of galaxies with evidence for strong AGN feedback. We study the correlation at low radio frequencies using two new surveys -- the First Alternative Data Release of the TIFR GMRT Sky Survey (TGSS ADR1) at 148~MHz and LOFAR's first all-sky survey, the Multifrequency Snapshot Sky Survey (MSSS) at 140~MHz.
We find a scaling relation $P_{\rm cav} \propto L_{148}^{\beta}$, with a logarithmic slope of $\beta = 0.51 \pm 0.14$, which is in good agreement with previous results based on data at 327~MHz. The large scatter present in this correlation confirms the conclusion reached at higher frequencies that the total radio luminosity at a single frequency is a poor predictor of the total jet power. Previous studies have shown that the magnitude of this scatter can be reduced when bolometric radio luminosity corrected for spectral aging is used. We show that including additional measurements at 148\,MHz alone is insufficient to improve this correction and further reduce the scatter in the correlation.
For a subset of four well-resolved sources, we examine the detected extended structures at low frequencies and compare with the morphology known from higher frequency images and \textit{Chandra} X-ray maps. In Perseus we discuss details in the structures  of the radio mini-halo, while in the 2A 0335+096 cluster we observe new diffuse emission associated with multiple X-ray cavities and likely originating from past activity. For A2199 and MS 0735.6+7421, we confirm that the observed low-frequency radio lobes are confined to the extents known from higher frequencies.
This new low-frequency analysis highlights the fact that existing cavity power to radio luminosity relations are based on a relatively narrow range of AGN outburst ages. We discuss how the correlation could be extended using low frequency data from the LOFAR Two-metre Sky Survey (LoTSS) in combination with future, complementary deeper X-ray observations.

}
   \maketitle


\section{Introduction}

High-resolution X-ray images have revealed many large-scale interactions between the intracluster medium (ICM) and the central AGN in galaxy cluster cores (e.g. Perseus: \citealp{Boehringer1993, Fabian2006, Fabian2011, Zhuravleva2015} and Hydra A: \citealp{McNamara2000, Nulsen2002, Wise2007}).  
In these systems, the radio jets of the AGN have pushed out cavities in the cluster's atmosphere, creating surface-brightness depressions. 
The energy released by the AGN required to create these cavities appears to be sufficient to balance the cooling observed in the X-rays \citep{Birzan2004, Rafferty2006, McNamaraNulsen2007}.
Therefore, the X-ray cavities provide a unique way of measuring the amount of energy dissipated into the ICM from AGN activity.
This feedback process is believed to moderate the availability of fuel for the accretion process in a homeostatic way that regulates both the growth of the black hole and the formation of stars in the surrounding galaxy \citep{Galaxy_Formation_Silk, MBH_Msigma_Gebhardt, SMBH_Galaxy_Ferrarese}. 

In many cavity systems, the depressions in X-ray surface brightness are found to be filled with radio emitting plasma.
This spatial anti-correlation between the X-rays and radio provides strong circumstantial evidence that the AGN activity is responsible for the observed X-ray cavities.
Given this common origin, X-rays directly probe the mechanical effects of the feedback process, while radio observations directly reveal the radiative output of the lobes.
Combined X-ray and radio observations can provide constraints on the radio radiative efficiencies, radio lobe and ICM properties.

Evidence for this common origin is found in the observed correlation between the power required to create the X-ray cavities  and the luminosity of the radio plasma associated with them. Using a sample of 24 systems with pronounced cavities, \citet{Birzan2008} find that the scaling relation is well described by a power law of the form $P_{\rm cav} \propto L_{\rm{rad}}^{\beta}$  with a logarithmic slope of  $0.35 \leq \beta \leq 0.70$. They further find that the correlation is steeper at 327~MHz than at 1.4~GHz ($\beta_{327}=0.51 \pm 0.07$ vs. $\beta_{1400}=0.35\pm0.07$), albeit with similarly large scatters of 0.80~dex and 0.85~dex, respectively. 
Subsequent investigations of \cite{Cavagnolo2010} and \cite{O'Sullivan2011} expand the sample size and essentially confirm the $P_{\rm cav} - L_{\rm{rad}}$ scaling relation found by \cite{Birzan2008}.
\cite{HardcastleKrause2013, HardcastleKrause2014} show that a significant scatter is physically expected in this correlation.

Although now well established, this correlation suffers from several limitations related to both the radio and the X-ray data.
In radio, all of the analysis to date has been based on data from higher frequencies, above 300~MHz.
Yet, in objects where low-frequency data has previously been available, the observed emission tends to be more diffuse and extended \citep[e.g.][]{Lane2004}.
At the same time, the original analysis in X-rays was based on a sample of bright nearby objects that show a clear single pair of cavities.
In objects with deeper X-ray data, however, we often see evidence of multiple surface brightness depressions at larger radii \citep[Table 3 in][]{Vantyghem2014}.
Further, these more extended structures  are often poorly described by simple spherical geometry and are usually not as well correlated spatially with high frequency radio emission as the inner cavity structures.

Obtaining sufficiently deep X-ray data for a large sample of these systems is problematic. 
Extending these studies to higher redshift is also difficult, as it becomes increasingly difficult to both detect and resolve the cavity structures.
With the advent of new low-frequency all-sky surveys, however, we can obtain maps of the extended diffuse emission for a large sample of sources.
If properly calibrated, the $P_{\rm cav} - L_{\rm{rad}}$ scaling relation can be a powerful tool in statistical studies of the AGN activity and its impact on the surrounding medium over time.

In this work, we employ low-frequency observations at 140 -- 150~MHz in order to pursue a more complete picture of AGN feedback signatures.
Our goal is two-fold: to extend the $P_{\rm cav} - L_{\rm{rad}}$ scaling relation to low radio frequencies, and to understand and reduce the observed scatter in this correlation.
For the statistical study we derive fluxes from the publicly available First Alternative Data Release of the TIFR GMRT Sky Survey \citep[TGSS ADR1;][hereafter TGSS]{Intema2017} at 148~MHz. 
In order to resolve individual clusters and examine the structure of their extended radio emission in the context of the X-ray cavities, we reprocess data from LOFAR’s first all-sky imaging survey, the Multifrequency Snapshot Sky Survey \citep[MSSS;][]{Heald2015}.
We focus our analysis on the \cite{Birzan2008} sample since it consists of very well known nearby sources which already have a deep ensemble of multi-wavelength data
and it includes primarily very bright sources, easily detectable in the shallow low-frequency surveys available so far.

In Section 2 we describe the characteristics of the cluster sample, the radio observations, and the X-ray data used.
Section 3 presents a statistical analysis of the cavity power to radio luminosity relation including the new low frequency data. A detailed discussion comparing these results to previous analyses is also presented. 
In Section 4, we present images for a subset of objects well-resolved in MSSS and discuss their detailed morphology in comparison with existing X-ray data.
We conclude in Section \ref{sec:conclusions} with a summary of our analysis and a discussion of the implications of these results.

We adopt $H_{0}=70$ km s$^{-1}$ Mpc$^{-1}$, $\Omega_{\rm{M}}=0.3$, and $\Omega_{\Lambda}=0.7$ for all calculations throughout this paper. \\


\section{Data Sources and Sample Selection}

\begin{table*}[ht]
	\begin{center}
	\caption{Characteristics of the sample sources and the field images from MSSS at 140~MHz and TGSS at 148~MHz.}
	\resizebox{1.0\textwidth}{!}{%
	\begin{tabular}{l|c| c| c c c| c c c}
		\hline
		\hline
									& 				& 								&					& TF-23 Sample		& 							&					& MF-14 Sample & \\
		\hline
		Source		& $z$			& Coordinates			&	Ext$\rm{^a}$& TGSS Res.$\rm{^b}$ & Noise$\rm{^c}$	&Ext$\rm{^d}$ & MSSS Res.$\rm{^e}$ & Noise$\rm{^f}$\\
									& 				& 	 (RA Dec)	 			&					& (arcsec) 		    & (mJy/beam)   &    &   (arcsec) 	 & (mJy/beam)   \\
		\hline		
		A2199					&	0.030	& 16 28 38.0 +39 32 55	&	Y		&	25.00	&	10		&	Y		&	21.63	&	40		\\
		MS 0735.6+7421	& 0.216		& 07 41 44.8 +74 14 52	&	Y		&	25.00	&	3		&	Y		&	27.77	&	30		\\
		2A 0335+096			&  0.035	& 03 38 35.3 +09 57 55	&	Y		&	25.31	&	5		&	Y		&	23.58	&	11         \\
		Perseus					& 0.018		& 03 19 47.2 +41 30 47	&	Y		&	25.00	&	10		&	Y		&	20.81	&	20		\\
		A262						&  0.016	& 01 52 46.8 +36 09 05	&	N		&	25.00	&	5		&	N		&	20.75	&	15		\\
		MKW3s					&  0.045	& 15 21 51.9 +07 42 31	&	Y		&	25.42	&	7		&	N		&	223.05	&	100			\\
		A2052					&  0.035	& 15 16 44.0 +07 01 07	&	N		&	25.42	&	7		&	N		&	251.05	&	150			\\		
		A478						& 0.081		& 04 13 25.6 +10 28 01	&	N		&	25.31	&	5		&	N		&	178.05	&	30		\\ 				
		Zw 3146				& 0.291		& 10 23 39.6 +04 11 10	&	N		&	25.85	&	3.5	&	N		&	236.05	&	30		\\		
		Zw 2701				&	0.214	& 09 52 49.2 +51 53 05	&	N		&	25.00	&	2.5	&	N		&	172.05	&	10		\\  	
		A1795					&  0.063	& 13 48 53.0 +26 35 44	&	N		&	25.00	&	7		&	N		&	269.05	&	40		\\
		RBS 797				&	0.350	& 09 47 12.9 +76 23 13	&	N		&	25.00	&	6		&	N		&	185.05	&	15		\\ 
		MACS J1423.8		& 0.545		& 14 23 47.6 +24 04 40	&	N		&	25.00	&	7		&	N		&	270.05	&	35		\\
		A1835					& 0.253		& 14 01 02.3 +02 52 48	&	N		&	26.03	&	4		&	N		& 	246.05	&	40		\\
		M84						& 0.0035	& 12 25 03.7 +12 53 13	&	Y		&	25.14	&	20		&				&					\\
		M87						& 0.0042	& 12 30 49.4 +12 23 28	&	Y		&	25.31	&	70		&				&					\\
		A133						& 0.060		& 01 02 42.1 $-$21 52 25	&	Y		&	32.53	&	6		&				&					\\
		Hydra A					& 0.055		& 09 18 05.7 $-$12 05 44	&	Y		&	29.09	&	30		&				&					\\
		Centaurus				& 0.011		& 12 48 47.9 $-$41 18 28	&	Y		&	49.90	&	6		&				&					\\
		HCG 62					& 0.014		& 12 53 05.5 $-$09 12 01	&	N		&	28.27	&	3		&				&					\\
		Sersic 159/03			& 0.058		& 23 13 58.6 $-$42 44 02	&	N		&	52.98	&	3		&				&					\\
		A2597					& 0.085		& 23 25 20.0 $-$12 07 38	&	Y		&	29.55	&	7		&				&					\\
		A4059					& 0.048		& 23 57 02.3 $-$34 45 38	&	Y		&	43.02	&	6		&				&					\\
		\hline
		\multicolumn{9}{l}{$\rm{^a}$ Indicates if the source is extended with respect to a point source in the TGSS map.} \\
		\multicolumn{9}{l}{$\rm{^b}$ Resolution of TGSS maps. This column shows one axis of the synthesized beam. The other axis of the TGSS beam is $25.00^{\prime\prime}$.} \\
		\multicolumn{9}{l}{$\rm{^c}$ Local rms noise in TGSS maps, measured within 1 deg from the center of the source.} \\
		\multicolumn{9}{l}{$\rm{^d}$ Indicates if the source is extended with respect to a point source in either the default or reprocessed MSSS map.} \\
		\multicolumn{9}{l}{$\rm{^e}$ Resolution of MSSS maps. MSSS maps have a circular synthesized beam with the stated diameter.} \\
		\multicolumn{9}{l}{$\rm{^c}$ Local rms noise in MSSS maps, measured within 1 deg from the center of the source.} \\
	\end{tabular}
	
	}
	
	\label{tab:source_properties}
	\end{center}
\end{table*}

We base our study on the \cite{Birzan2008} sample of 24 feedback systems (hereafter B-24). 
The sample consists of relaxed cool-core clusters showing evidence of AGN activity. 
This is an X-ray selected sample for which the available X-ray observations have shown clear signatures of cavities and at the same time radio data has demonstrated strong lobes.
However, in the radio, these clusters have been primarily studied at higher frequencies which tend to reveal emission associated with the most recent epoch of AGN activity. 

Throughout this work we use MSSS (Section \ref{subsec:MSSS} and \ref{label:reprocessing}) and TGSS (Section \ref{lab:TGSS}) data to study the sample of feedback systems. 
Based on the data from those two surveys we select two subsamples of the B-24 sample that are described in Section \ref{lab:subsamples}.
The $P_{\rm cav}$ literature values we use for the correlation studies are summarized in Section \ref{lab:Xray_Pcav_data}.
We do not include the VLA Low-Frequency Sky Survey Redux at 74 MHz \citep[VLSSr;][]{Cohen2007, Lane2012, Lane2014} in our analysis due to its low sensitivity combined with low resolution and insufficient sky coverage (see Section \ref{subsec:VLSSr}).
The Galactic and Extragalactic All-sky Murchison Widefield Array survey \citep[GLEAM;][]{Wayth2015, Hurley-Walker2017} was released shortly before the submission of this work and we do not include it in our study.

\subsection{MSSS}
\label{subsec:MSSS}

MSSS is the first major imaging campaign with the Low Frequency Array \citep[LOFAR;][]{vanhaarlem2013}.
The main goal of MSSS is to produce a broadband catalog of the brightest sources in the low-frequency northern sky, creating a calibration sky model for future observations with LOFAR.
It covers two frequency windows: one within the low-band antenna range (LBA; 30 – 75 MHz) and the other in the high-band antenna range (HBA; 119 – 158 MHz).
The LBA survey is a work in progress and will be examined in a separate publication.
In this paper we focus exclusively on the HBA part of the survey, 
where each one of the 3616 fields required to survey the entire northern sky is observed in two 7-minute scans separated by 4 hours to improve the $uv$-coverage.

In this work we use a set of preliminary images (hereafter default images) used by the MSSS team to produce the first internal version of the MSSS catalog.
The preliminary MSSS processing strategy includes primary flux calibration based on a bright, compact calibrator observed before the target snapshot. One round of phase-only, direction-independent calibration is performed using a VLSSr-based sky model \citep{Heald2015} and then imaging is performed with the AWImager \citep{Tasse2013} with a simple, shallow deconvolution strategy using 2500 CLEAN iterations. 
The imaging run per field incorporates projected baselines shorter than 2 k$\lambda$. 
Baselines shorter than 100 $\lambda$ were excluded from the imaging for fields at declination $\delta \leq 35$ degrees in order to exclude contamination from incompletely sampled large-scale galactic plane structures and thus provide a smoother background \citep[see][]{Heald2015}.
A correction based on VLSSr and the NRAO VLA Sky Survey \citep[NVSS;][]{Condon1998} was applied to the MSSS images to compensate for errors in the default flux density scale dependent on the position of the source on the sky \citep{Hardcastle2016}.

\subsection{Reprocessing of MSSS data}
\label{label:reprocessing}

Although the characteristic resolution of the default MSSS images is $\sim2^{\prime}$, the either high or low declination of the majority of the B-24 systems visible in MSSS results in an average resolution of $\sim3.5^{\prime}$ due to the limited subset of the data imaged as described in Section \ref{subsec:MSSS}.
Thus, the default MSSS images do not allow us to resolve the sources and study the radio features corresponding to the observed X-ray structures.
For this reason we developed a strategy to reprocess the data and produce custom images with  20 -- $30^{\prime\prime}$ resolution 
that allow us to study the morphology of the most extended systems in the sample.
Furthermore, the resolution of the reprocessed MSSS data matches the resolution of the TGSS image products (discussed in more detail below), which allows for easy and reliable comparison between the two surveys.

In order to reach angular resolution of 20 -- $30^{\prime\prime}$, we reprocessed the long baselines which were not used so far in the default MSSS imaging strategy but were included as part of the MSSS observation. 
The primary step of the reprocessing is an additional round of phase-only, direction-independent self-calibration in order to fine-tune the calibration for higher-resolution imaging capability.
We accessed the MSSS archive to obtain the flagged, demixed, and flux-calibrated target snapshot observations/data sets.
Each snapshot consists of 8 measurement sets containing the individual 2~MHz-wide bands at 120, 125, 129, 135, 143, 147, 151, and 157~MHz. These bands are treated separately throughout the phase-only self-calibration process accomplished with Black Board Selfcal \citep[BBS;][]{Pandey2009}.

The first step of our reprocessing procedure is to image the default pre-calibrated MSSS data to a higher resolution. 
We use AWImager including projected baselines up to 10~k$\lambda$ employing Briggs weighting \citep{Briggs1995} with a robust parameter of $-1$. 
From the resulting map, we extract a high-resolution skymodel, which we use to perform phase-only self-calibration. 
After the self-calibration loop, the two snapshots are combined (for better $uv$-coverage) and each of the 8 bands is imaged separately using the same imaging parameters as in the first imaging round.
Individual band images are later smoothed to match resolution and weighted by rms noise before they are combined to create the final full-band images presented in Section \ref{sec:resolved}.

To make sure that there are no significant astrometric shifts between the separately-calibrated bands, we crossmatched catalogs derived from the individual-band images against TGSS, and assessed the typical difference in positions for the sources common to both. 
Each field typically contains several tens of sources detected both in MSSS and TGSS. 
Although the overall astrometry of each individual field can differ from TGSS by up to $1-2^{\prime\prime}$, the relative astrometric difference between the MSSS bands was seen to be negligible within the uncertainties, which are $<1^{\prime\prime}$ for all fields. 
We do find a small systematic astrometric shift with frequency only in the field of A262, but even in this case the difference between the two most distant bands is at the $1^{\prime\prime}$ level, far smaller than the synthesized beam ($20.75^{\prime\prime}$).

We focused our reprocessing efforts on six fields (A2199, MS 0735.6+7421, 2A 0335+096, A262, Perseus, and MKW3s/A2052) since, based on the 330 MHz and 1.4~GHz VLA maps, only they had sizes that could be potentially resolved at a resolution of 20 -- 30$^{\prime\prime}$. 
Unfortunately, we could not obtain a reliable image at this resolution for the field of MKW3s and A2052. Due to the presence of three very strong sources in that field (MKW3s, A2052 and 3C313), the 10~k$\lambda$ image contained many strong artifacts, which distort faint features and bias the flux measurements. Thus we use the default MSSS images to measure the flux of those two systems.
Furthermore, A262 was not well resolved at 20$^{\prime\prime}$ resolution, so we used the reprocessed image only to measure the flux of the source.
The successfully reprocessed images of A2199, MS 0735.6+7421, 2A 0335+096, and Perseus are shown and discussed in detail in Section \ref{sec:resolved}.

For several objects in the studied sample deep LOFAR observations are already available. 
As a sanity check we imaged a small frequency chunk ($\nu=142$~MHz, $\Delta\nu=4$~MHz) of a full 8-hour LOFAR observation on A2199.
This image revealed the same morphology as observed in the reprocessed MSSS image (shown in Section \ref{sec:A2199}). 
Furthermore, the total flux of A2199 measured in the test LOFAR image coincides with the flux derived from the MSSS map within 2\%.
This gives us confidence that the quoted flux densities and the features outlined below are not a product of processing artefacts or poor $uv$-coverage but real physically existing characteristics of the sources.

\subsection{TGSS}
\label{lab:TGSS}

TGSS was carried out at 148~MHz with the Giant Metrewave Radio Telescope (GMRT). 
Each pointing was observed for $\sim$15 minutes, split over 3 or more scans spaced out in time to improve $uv$-coverage. 
The TGSS data products have gone through a fully automated pipeline \citep{Intema2009, Intema2014}, which includes direction-dependent calibration, modeling and imaging to suppress mainly ionospheric phase errors.
As a result the flux density accuracy is estimated to be $\approx$10\% and the noise level is below 5 mJy/beam for the majority of the pointings.
TGSS and MSSS have a comparable bandwidth (16.7 and 16 MHz) and integration time per field (15 and 14 min, respectively).
The current data release of TGSS covers the sky between $-53$ and $+90$ deg declination. 
By including the TGSS data in our analysis, we can look at the full B-24 sample, for which 30\% of the sources are in the southern hemisphere.

Compared to MSSS, TGSS has a number of advantages and disadvantages, which follow from the different $uv$-coverage between the two surveys and the different processing schemes used to produce the images.
While MSSS is more sensitive to extended diffuse emission, TGSS has a higher resolution than the default MSSS ($\sim25^{\prime\prime}$ and $\sim3.5^{\prime}$, respectively).
This makes TGSS better at resolving the morphology of the more distant sources and correctly isolating their emission from the contaminating emission of neighboring sources (e.g. Zw3146 and A478).
On the other hand, being much more sensitive to extended diffuse emission, MSSS allows us to get a more complete picture of the integrated AGN activity over time.

\subsection{Sample Selection}
\label{lab:subsamples}

Since radio galaxies of Fanaroff-Riley type I and II \citep[FRI and FRII;][]{Fanaroff1974} are likely to have different particle content, we do not expect them to follow the same relationship \citep{GodfreyShabala2013}. 
Thus, we exclude Cygnus A \citep[e.g.][]{McKean2016} as being the only FRII.
In total the sample comprises 23 systems: 21 galaxy clusters, one galaxy group (HCG 62), and one elliptical galaxy (M84). They range in redshift from 0.0035 (M84) to 0.545 (MACS J1423.8+2404). 
Since we will study this sample with TGSS, we will refer to this sample as the TGSS Feedback sample, shortly TF-23.
In this work we use the TF-23 sample to study the $P_{\rm cav} - L_\nu$ relation.

We further define a subsample of B-24 including only the systems observed by MSSS.
This subsample again excludes Cygnus A for being the only FRII as well as M87 and M84 for not having MSSS observations of reasonable quality. 
In total the sample comprises 14 galaxy clusters and ranges in redshift from 0.016 (A262) to 0.545 (MACS J1423.8+2404). We will call this sample the MSSS Feedback sample, shortly MF-14.
Table \ref{tab:source_properties} lists the redshift and coordinates of the sources as well as the properties of the corresponding maps in MSSS and TGSS.

\subsection{X-rays}
\label{lab:Xray_Pcav_data}

The X-ray data utilized in this work are taken from the literature or based on archival \textit{Chandra} observations. For the correlation analysis discussed in Section~\ref{analysis_and_correlations}, we have adopted the values calculated for $P_{\rm cav}$ by \cite{Rafferty2006}. These estimates were determined from the existing \textit{Chandra} exposures for the sample at that time and assume a simple geometrical model to calculate the mean cavity power based on the cavity's size, pressure, and position relative to the cluster center. This technique for estimating cavity powers has been employed routinely in other studies of feedback systems and can yield variations in the derived values for $P_{\rm cav}$ of $\sim2$--4 due to uncertainties in the cavity geometry as well as the methods used to estimate the cavity ages. In this work, we take the literature values from \cite{Rafferty2006} as reported and discuss some of the caveats associated with these estimates below in Section~\ref{analysis_and_correlations}.

For the four resolved sources in our MSSS sample, Perseus, 2A0335+096, MS0735.6+7421, and A2199, we have created X-ray surface brightness images and residual maps using the current \textit{Chandra} archival data. In all four cases, the objects were imaged multiple times with the ACIS detector and we have extracted all of the existing exposures from the Chandra Data Archive. The data were reprocessed using CIAO 4.7 and CALDB 4.6.7 to apply the latest gain and other calibration corrections as well as filtered to remove any contamination due to background flares. Instrument and exposure maps were created for each exposure individually with the spectral weighting determined by the fitting a single temperature, MEKAL thermal model \citep{Mewe85,Liedahl95} plus foreground Galactic absorption to the total integrated spectrum from the central region of each cluster. Individual background event files were created for each dataset from the standard ACIS blank-sky event files following the procedure described in \cite{Vikhlinin05}.

Finally, combined X-ray surface brightness maps were constructed by reprojecting the data for each exposure to a common target point on the sky for each object and then combining each exposure into a final mosaic. Corresponding mosaics of the exposure maps and background counts were also constructed and used to form the final background-subtracted, exposure-corrected X-ray surface brightness images. The X-ray images shown in Section~\ref{sec:resolved} have been filtered in energy to the range 0.5-7.0\,keV. For each object, unsharp-masked images have been constructed by subtracting images which have been Gaussian smoothed on two length-scales. The resulting residual images are used in Section~\ref{sec:resolved} in conjunction with our reprocessed MSSS radio maps at 140\,MHz to study the correspondence, or lack thereof, between the observed, diffuse low-frequency emission and the presence of feedback signatures in the X-ray.

\subsection{VLSSr}
\label{subsec:VLSSr}

In principle, data at low frequencies such as VLSSr at 74 MHz \citep{Cohen2007, Lane2012, Lane2014} could be used to provide extra information which helps constrain the turnover frequencies in the radio spectra (as discussed in Section \ref{lab:Spectrum}).
With this in mind we have obtained the VLSSr maps, with a resolution of $75^{\prime\prime}$ and an average rms noise of $\sim$0.1 Jy/beam, and attempted to compile VLSSr flux density measurements for our sample.
Unfortunately, we could not obtain reliable measurements for more than a third of our sample.
Two sources are outside of the VLSSr coverage (Centaurus and Sersic 159/03), three systems are below the detection limit (A478, MACS J1423.8, and HCG 62), and two sources are only marginally detected (RBS 797 and A1835). Due to the low resolution of VLSSr, Zw3146 is blended with the emission from close-by sources. 
For the rest of the sources we analyzed the spectra from which was clear that the available VLSSr points do not help constrain the turnover frequency.
Given the incomplete coverage of the sample and the lack of improvement in the fitting results we do not include these points in the subsequent analysis.
The radio spectra of the sample systems are discussed in detail in Section \ref{lab:Spectrum}.

\section{Sample Analysis and Correlations}
\label{analysis_and_correlations}

In this section we use the TF-23 sample of bright feedback systems in order to derive the  $P_{\rm cav} - L_\nu$ scaling relation at 148~MHz.
We also study the radio spectra of the sample sources and comment on the correlation between cavity power and bolometric radio luminosity corrected for spectral aging.

\subsection{Radio Flux Measurements}

Since more than 40\% of the TF-23 sample sources are not resolved at 20 -- $30^{\prime\prime}$ scale, and in many of the resolved sources it is difficult to separate the lobes and the core, we only measure the total flux densities from the sources.
In principle, this could introduce a systematic offset to higher radio luminosities in the correlation.
However, based on the analysis of \cite{Birzan2008}, the core emission comprises $\lesssim20\%$ of the total flux density of these sources for frequencies $\geq 300$~MHz.
Due to the steep spectrum of the lobe emission, we expect the core contribution to be even smaller at lower frequencies (140--150~MHz).

To obtain the flux densities from TGSS, we used the publicly available mosaics at 148~MHz.
We used the reprocessed MSSS fields to measure the flux density of A2199, MS 0735.6+7421, 2A 0335+096, Perseus, and A262. 
These reprocessed maps provide higher resolution which allows us to more easily isolate the region of interest and disentangle contaminating emission from nearby sources. 
For the remaining nine sources in the MF-14 sample, the fluxes were extracted from the default MSSS mosaics.

Our general recipe for obtaining the flux densities consists of two steps.
First, we visually inspect the maps and select a nearby region devoid of sources which we use to define the local rms noise (Table \ref{tab:source_properties}).
Then we measure the flux density of the source in a region confined by the 3$\sigma$ contour. 
We used this strategy for well-resolved and marginally resolved sources. 
For point sources, we measured the flux in a region with the size of the FWHM of the synthesized beam.
This was done for HCG62 in TGSS and MACS~J1423.8 in MSSS.

In several fields we noted the presence of strong artifacts in the images. For these objects we use a higher cut-off for the background in order to exclude surrounding artifacts from the measurements.
We used a 5$\sigma$ threshold for A4059 and Hydra A in TGSS; and for MS 0735.6+7421, Zw 3146, A2199, and A1795 in MSSS.

In some cases we clearly detect more extended diffuse emission at low frequencies that was not observed in the high frequency maps of \cite{Birzan2008}. 
In order to be consistent with the analysis at high frequencies, we restrict the flux density measurements to the morphological features considered at 325~MHz. 
In TGSS we applied this strategy for Perseus and M87; in MSSS, for Perseus and 2A0335+096.

We report the measured fluxes in Table \ref{tab:all_freq_flux}.
In accordance with \cite{Birzan2008}, the listed uncertainties include the statistical error (thermal noise) plus the uncertainty of the absolute flux scale. 
The flux scale uncertainty is 10\% in MSSS, as well as in TGSS \citep{Intema2017}.

\begin{center}
\begin{table*}
\caption{Flux density measurements from MSSS, TGSS, and higher frequencies from literature used for the spectral fitting discussed in Section \ref{lab:Spectrum}. The values for MSSS, and TGSS derive from our analysis. The unreferenced values at higher frequencies come from \cite{Birzan2008}. The numbers in superscripts reference the high frequency flux density measurements coming from the literature.}

\resizebox{1.0\textwidth}{!}{
\centering
\begin{tabular}{l|l|l|ll|ll|ll|ll|ll}
\hline
\hline
Source & MSSS 140~MHz & TGSS 148~MHz &       & P-band &       & 750~MHz &       & L-band &       & S-band &       & C-band \\
		& $S_{\nu}$ & $S_{\nu}$ & $\nu$ & $S_{\nu}$ & $\nu$& $S_{\nu}$ & $\nu$ & $S_{\nu}$ & $\nu$ & $S_{\nu}$ & $\nu$ & $S_{\nu}$ \\
		& (Jy) & (Jy) & (MHz) & (Jy) & (MHz) & (Jy) & (MHz) & (Jy) & (MHz) & (Jy) & (MHz) & (Jy) \\
\hline

A2199 & 54.3   $\pm$       6.1 & 52.9   $\pm$       5.6 & 324   & 24.00 $\pm$ 0.96 & $750^{1}$ & 9.1900 $\pm$ 0.0049 & $1400^2$ & 3.68 $\pm$ 0.12 & $2700^3$ & 1.21 $\pm$ 0.06 & 4675  & 0.313 $\pm$ 0.012 \\
MS 0735.6+7421 & 5.30    $\pm$       0.83 & 3.95   $\pm$       0.46 & 330   & 0.80 $\pm$ 0.03 &       &       & 1425  & 0.021 $\pm$ 0.001 &       &       &       &  \\
2A 0335+096      & 1.10     $\pm$       0.21 & 1.05   $\pm$       0.19 & 324   & 0.21 $\pm$ 0.01 &       &       & $1490^2$ & 0.0367 $\pm$ 0.0018 &       &       & 4860  & 0.01 $\pm$ 0.001 \\
Perseus & 58.8    $\pm$       6.3 & 42.4   $\pm$       4.4 & $408^4$ & 29.00 $\pm$ 0.04 & $750^{1}$ & 21.82 $\pm$ 0.28 &       &       &       &       &       &  \\
A262     & 0.57    $\pm$      0.15 & 0.540   $\pm$       0.091 & 324   & 0.299 $\pm$ 0.012 &       &       & 1365  & 0.0734 $\pm$ 0.0030 &       &       &       &  \\
MKW3s & 17.3    $\pm$       2.2 & 15.6   $\pm$       1.8 & 327   & 4.73 $\pm$ 0.18 &       &       & 1440  & 0.0897 $\pm$ 0.0040 &       &       & 4860  & 0.0025 $\pm$ 0.0001 \\
A2052 & 58.5     $\pm$      6.7 & 59.4   $\pm$       6.3 & 330   & 30.9 $\pm$ 1.2 & $750^{1}$ & 11.76 $\pm$ 0.041 & 1490  & 5.7 $\pm$ 0.2 & $2700^{5}$ & 2.02 $\pm$ 0.04 & 4860  & 0.72 $\pm$ 0.03 \\
A478     & 0.53   $\pm$       0.14* & 0.305   $\pm$       0.050 & 327   & 0.11 $\pm$ 0.01 &       &       & 1440  & 0.027 $\pm$ 0.001 &       &       &       &  \\
Zw 3146 & 0.259   $\pm$      0.056** & 0.0174   $\pm$       0.0056 & 324   & 0.016 $\pm$ 0.004*** &       &       &       &       &       &       & 4860  & 0.00139 $\pm$ 0.00007 \\
Zw 2701 & 1.36    $\pm$       0.18 & 1.35   $\pm$       0.15 & 324   & 0.21 $\pm$ 0.01 &       &       &       &       &       &       & 4860  & 0.0043 $\pm$ 0.0002 \\
A1795 & 7.47    $\pm$       0.90 & 6.08   $\pm$       0.67 & 327   & 3.36 $\pm$ 0.14 &       &       & 1465  & 0.88 $\pm$ 0.04 & $2700^3$ & 0.48 $\pm$ 0.03 & $4850^6$ & 0.261 $\pm$ 0.034 \\
RBS 797 & 0.133   $\pm$       0.035 & 0.174   $\pm$       0.033 & 324   & 0.104 $\pm$ 0.006 &       &       & 1475  & 0.021 $\pm$ 0.001 &       &       & 4860  & 0.0042 $\pm$ 0.0003 \\
MACS J1423.8 & 0.103   $\pm$       0.044 & 0.054   $\pm$       0.013 & 327   & 0.0269 $\pm$ 0.002 &       &       & 1425  & 0.0044 $\pm$ 0.0002 &       &       &       &  \\
A1835 & 0.150 $\pm$    0.055 & 0.107   $\pm$       0.021 & 327   & 0.095 $\pm$ 0.007 &       &       & 1400  & 0.031 $\pm$ 0.001 &       &       & 4760  & 0.0099 $\pm$ 0.0004 \\
M84   &       & 14.9   $\pm$       1.9 & 324   & 11.1 $\pm$ 0.5 &       &       & 1425  & 05.6 $\pm$ 0.2 &       &       & 4860  & 2.28 $\pm$ 0.09 \\
M87   &       & 436   $\pm$       45 & 324   & 124 $\pm$ 5 &       &       & 1400  & 138 $\pm$ 6 &       &       & 4860  & 59 $\pm$ 2 \\
A133  &       & 9.6   $\pm$       1.0 & 330   & 3.60 $\pm$ 0.10 &       &       & 1425  & 0.132 $\pm$ 0.005 &       &       &       &  \\
Hydra A &       & 317   $\pm$       36 & 333   & 152 $\pm$ 6 & $750^7$ & 79.9 $\pm$ 3.2 & 1423  & 39.2 $\pm$ 1.6 & $2700^{8}$ & 23.50 $\pm$ 0.93 & 4760  & 15.0 $\pm$ 0.6 \\
Centaurus &       & 19.5   $\pm$       2.0 & 327   & 12.3 $\pm$ 0.5 & $625^7$ & 7.160 $\pm$ 0.040 & 1565  & 3.4 $\pm$ 0.1 & $2700^{9}$ & 2.458 $\pm$ 0.048 & 4760  & 1.37 $\pm$ 0.06 \\
HCG 62 &       & 0.0090   $\pm$       0.0039 & 324   & 0.008 $\pm$ 0.002 &       &       & 1440  & 0.0050 $\pm$ 0.0004 &       &       &       &  \\
Sersic 159/03 &       & 4.23   $\pm$       0.45 & 324   & 1.53 $\pm$ 0.06 &       &       & 1425  & 0.22 $\pm$ 0.01 &       &       & 4860  & 0.056 $\pm$ 0.002 \\
A2597 &       & 14.7   $\pm$       1.5 & 328   & 8.3 $\pm$ 0.3 &       &       & 1400  & 1.86 $\pm$ 0.07 &       &       & 4985  & 0.37 $\pm$ 0.02 \\
A4059  &       & 22.8   $\pm$       2.4 & 328   & 9.93 $\pm$ 0.40 &       &       & $1440^2$ & 1.285 $\pm$ 0.043 & $2700^{10}$ & 0.35 $\pm$ 0.07 & $4860^{11}$ & 0.12 $\pm$ 0.04 \\

\hline

\multicolumn{13}{l}{* A478 is not resolved in the MSSS map and is blended with neighboring cluster members. Thus, the flux measurement at 140~MHz includes serious contributions from several surrounding sources.} \\	
\multicolumn{13}{l}{** Zw3146 is not resolved in the MSSS map and is blended with neighboring sources. Thus, the flux measurement at 140~MHz includes serious contributions from several surrounding sources.} \\	
\multicolumn{13}{l}{*** Value not identical with the one cited in \cite{Birzan2008}. We used the P-band map from \cite{Birzan2008} to measure the flux ourselves.} \\
\multicolumn{13}{l}{References: (1) \cite{Pauliny1966}, (2) \cite{Condon1998}, (3) \cite{Andernacht1981}, (4) \cite{Burbidge1979} } \\
\multicolumn{13}{l}{ (5) \cite{Wall1985}, (6) \cite{Gregory1991}, (7) \cite{Haynes1975}, (8) \cite{Wright1990},  } \\	
\multicolumn{13}{l}{ (9) \cite{Sadler1984}, (10) \cite{Vollmer2005}, (11) \cite{Wright1994}  } \\
\end{tabular}
}
\label{tab:all_freq_flux}
\end{table*}
\end{center}

\subsection{Cavity Power Measurements}

\cite{Rafferty2006} have utilized \textit{Chandra} X-ray data to calculate the amount of work required to create the observed cavities, which gives a measure of the total mechanical energy produced by the AGN. 
The total energy, combined with a buoyancy timescale estimate of the age of the bubble, provides an estimate of the time averaged cavity/jet power.
We adopt the cavity power estimates of \cite{Rafferty2006} for our correlation analysis.

Those cavity power estimates have two main limitations.
For all systems except two (Perseus and Hydra A) the cavity power is estimated for a single or a pair of cavities; thus the measurements are potentially limited to the energy output of a single outburst.
Another mechanism that carries energy into the ICM is the mild shocks, with Mach number between 1.2 and 1.7, which are observed in many systems \citep{Fabian2006, McNamara2005, Nulsen2005, Nulsen2005b, Forman2005, Forman2007, SandersFabian2006, Wise2007, McNamaraNulsen2007}. Therefore, the cavity estimates are only lower limits on the energy output of the AGN, and as a consequence, the radiative efficiencies are overestimated.

\subsection{Luminosity and Cavity Power Uncertainties}

The monochromatic radio luminosity was calculated as 
\begin{equation} \label{equ:L_nu_formula}
	L_{\nu}= 4 \pi D_{L}^2 S_{\nu} (1+z)^{-(\alpha+1)},
\end{equation}
where $D_{L}$ is luminosity distance, $z$ is redshift, and $S_{\nu}$ is the radio continuum flux density,
for which we have assumed a power-law spectrum of the form $S_{\nu} \propto \nu^{\alpha}$.  
The values of the spectral index $\alpha$ are presented in Table \ref{tab:fitting_L_bol} and are derived from our spectral analysis described in Section \ref{lab:Spectrum}. 

To estimate the errors in the luminosity values we propagate the uncertainties of the flux density, luminosity distance, and spectral index using
\begin{equation}
\Delta L_{\rm{\nu}} = L_{\rm{\nu}} ~ \sqrt{\left(  \frac{2 \Delta D_{\rm{L}}}{D_{\rm{L}}} \right)^2 + \left( \frac{\Delta S_{\rm{\nu}}}{S_{\rm{\nu}}} \right)^2  + \left( \ln (1+z) \Delta\alpha \right)^2.} 
\end{equation}
 
To compute the luminosity distance uncertainties we follow the recipe used by \cite{Leith2016}.
The sources lying at $D_L > 70$~Mpc have redshift-derived distance estimates and for them we assume $\Delta D_L = 7$~Mpc corresponding to peculiar velocities of $\sigma_{\rm v} \approx 500$~km/s. 
For sources with $D_L < 70$~Mpc we assume $\Delta D_L = 0.1 D_L$, corresponding to the estimated uncertainty in redshift independent distance measurements  \citep{Cappellari2011}. 
The distances of M84 and M87 are accurate within $\sim 3 \%$ since they have been measured with the \textit{Hubble Space Telescope} using the surface brightness fluctuation method \citep{Blakeslee2009}. 
Thus, the uncertainties in luminosity are greater than the ones of \cite{Birzan2008}.

The cavity power measurements that we take from \cite{Rafferty2006} have asymmetric errors.
In order to simplify the fitting, we assume Gaussian uncertainties in the cavity power measurements and calculate their standard deviation as the average of the positive and negative uncertainties given by \cite{Rafferty2006}.
The distance uncertainties are not propagated with cavity power, since the cavity power uncertainties are strongly dominated by other sources of error such as volume estimates \citep{O'Sullivan2011}.

\subsection{$P_{\rm{cav}}$ vs. $L_{148}$}
\label{sec:PvsL148}

For our correlation analysis we use the flux density values derived from TGSS at 148~MHz for the TF-23 sample. 
For comparison we performed the same analysis at 327~MHz using the flux density values published by \cite{Birzan2008}.

\begin{figure*}[t]
\begin{center}
\includegraphics[width=3.60in]{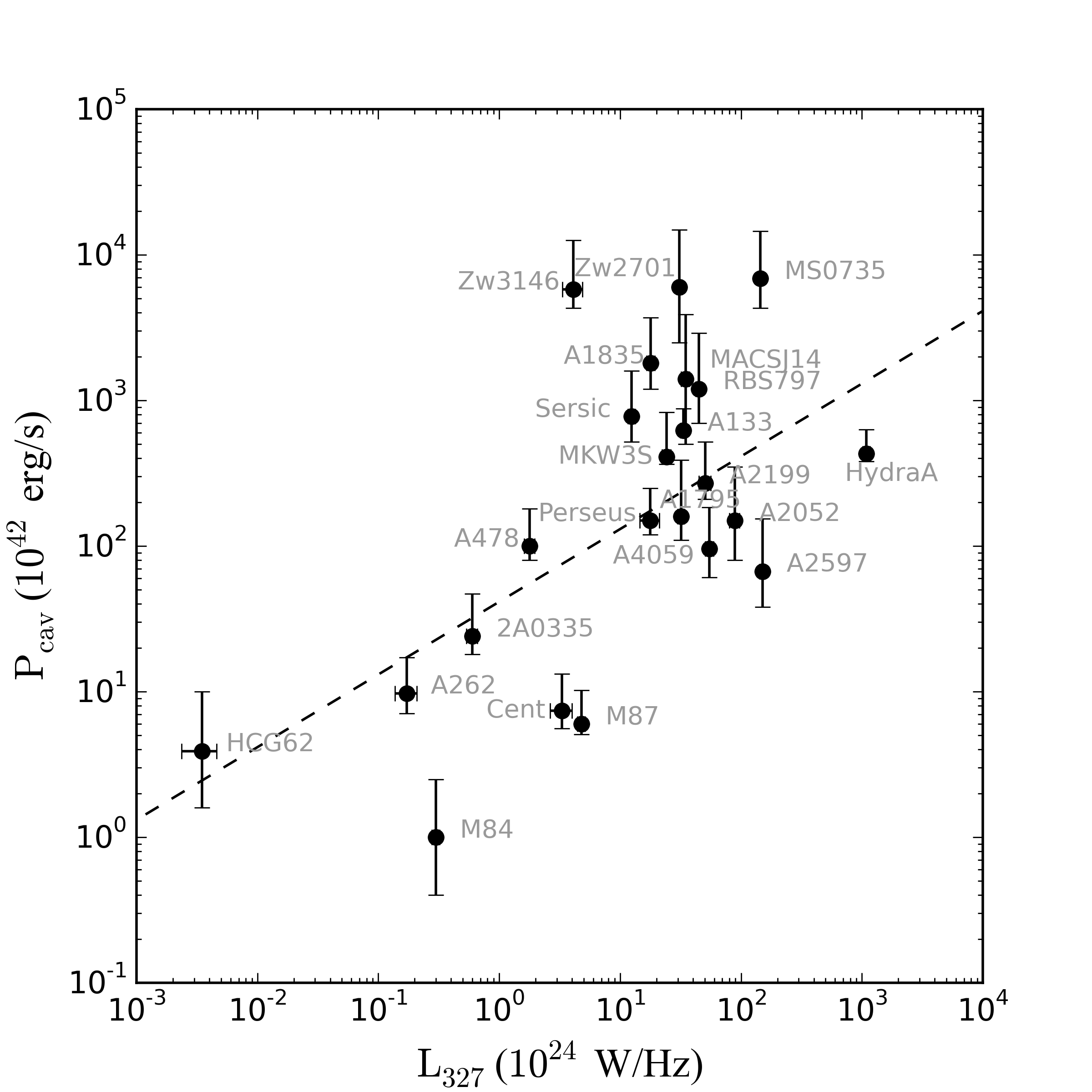}
\includegraphics[width=3.60in]{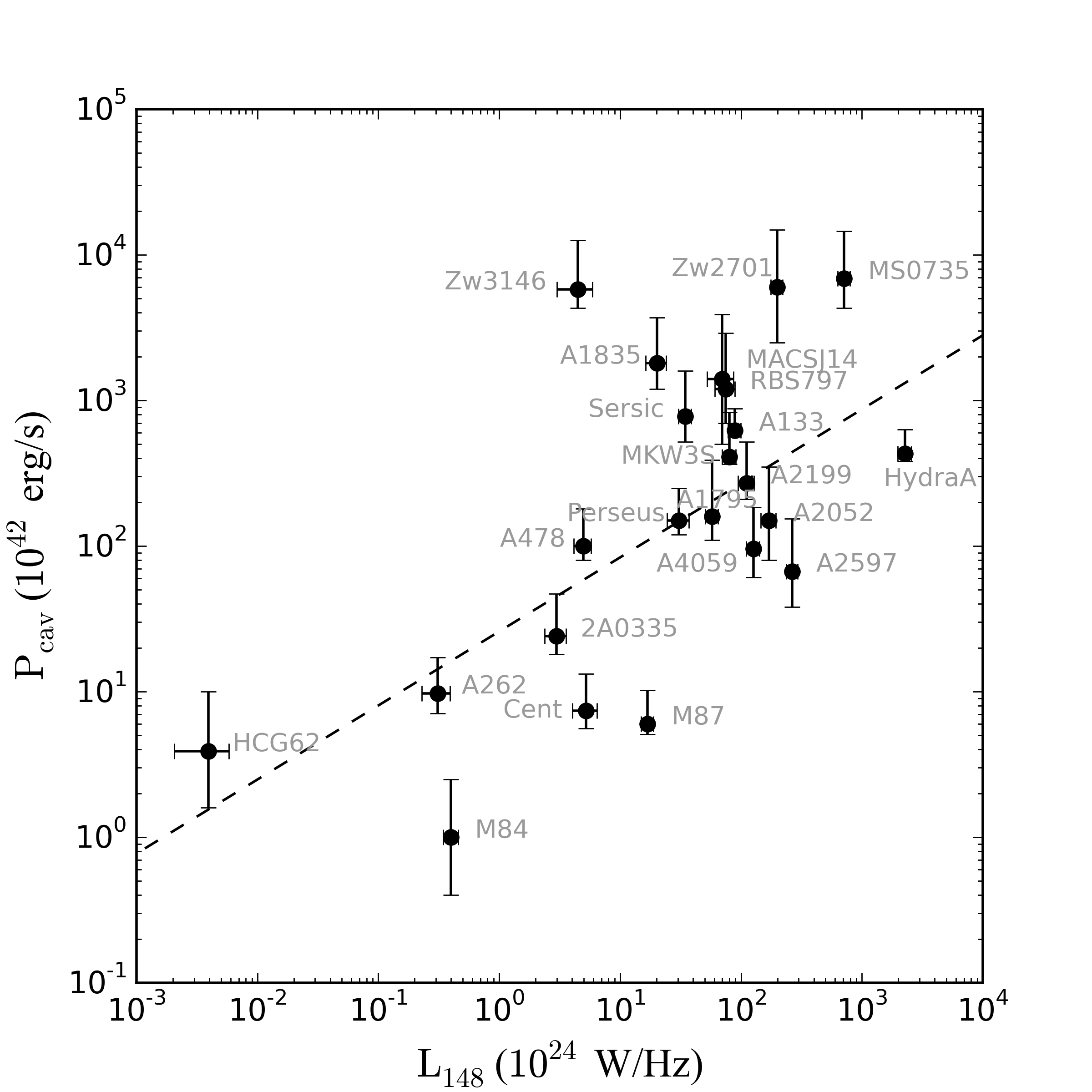}
\end{center}
\caption{\small Cavity (jet) power vs. the total radio luminosity for the TF-23 sample at 327~MHz (left) and 148~MHz (right). The dashed line shows the best-fit power law. It is $\log{P_{\rm{cav}}} = (1.6 \pm 0.2) + (0.50 \pm 0.14)\log{L_{327}}$ with intrinsic scatter of 0.86 dex at 327~MHz and $\log{P_{\rm{cav}}} = (1.4 \pm 0.3) + (0.51 \pm 0.14)\log{L_{148}}$ with intrinsic scatter of 0.85 dex at 148~MHz. 
\label{fig:P_vs_L148_L327}}
\vspace{0.15in}
\end{figure*}

We utilized orthogonal distance regression to fit a model of the form
\begin{equation}
\label{equ:Pcav_LD_fit}
\log{P_{\rm{cav}}} = \log{P_{0}} + \beta\log{L_{\rm{\nu}}},
\end{equation}
where $L_{\rm{\nu}}$ is in units of $10^{42}$ erg s$^{-1}$ and $P$ is in units of $10^{24}$ W Hz$^{-1}$.
The uncertainties of both radio luminosity and cavity power were taken into account as weights in the fit.
The result of the regression analysis is presented in Table \ref{tab:regression}. The $P_{cav}$ vs. $L_{\rm{\nu}}$ plots are shown in Figure \ref{fig:P_vs_L148_L327}.

\begin{table}[ht]
	\begin{center}
	\caption{Results of fitting Equation \ref{equ:Pcav_LD_fit} for the TF-23 sample.}
	\begin{tabular}{l c c c }
		\hline
		\hline
		$\nu^{a}$		&	$P_{0}^{b}$		&	$\beta^{c}$		&	$\sigma^{d}$\\
		\hline
		327			&	$10^{1.6 \pm 0.2}$		& 	0.50 $\pm$ 0.14		&	0.86	 \\
		148			&	$10^{1.4 \pm 0.3}$ 	&  0.51 $\pm$ 0.14 	&	0.85	 \\			
		\hline	
		\multicolumn{4}{l}{$\rm{^a}$ Frequency in MHz.} \\
		\multicolumn{4}{l}{$\rm{^b}$ Normalization.} \\
		\multicolumn{4}{l}{$\rm{^c}$ Luminosity slope.} \\
		\multicolumn{4}{l}{$\rm{^d}$ Intrinsic scatter in dex.} \\
	\end{tabular}
	\label{tab:regression}
	\end{center}
\end{table}

First, we must note that thare is no significant difference between the fit of \cite{Birzan2008} on the sample of 24 sources and our results for the TF-23 sample.
This suggests that removing Cygnus A from the sample does not significantly bias the correlation.
The $\beta$ reported by \cite{Birzan2008} at 327~MHz is 0.51 $\pm$ 0.07 with a scatter (standard deviation) of 0.81 dex, while for the TF-23 sample we get 0.50 $\pm$ 0.14 with a scatter of 0.86 dex.
We get the same slope but higher uncertainty and slightly higher scatter. This small difference is a cumulative effect from several factors; most importantly, omitting Cygnus A and using different spectral indices when computing the $L_{\nu}$.
Other contributors are the different luminosity errors we compute, which change the weighting of points during the fit, and the lower flux of Zw3146 we measure from the map at 327~MHz of \cite{Birzan2008} (see Table \ref{tab:all_freq_flux}).   

We show that the $P_{cav}$ -- $L_{\nu}$ scaling relation holds at low radio frequencies.
As can be seen from Table \ref{tab:regression}, there is no difference between the correlation at 327~MHz and 148~MHz. They have virtually identical slope and scatter. 
The comparison between the results at 148~MHz and 327~MHz shows that merely moving to $\sim$150~MHz does not provide us with a better understanding of the correlation between cavity power and monochromatic radio luminosity. 
This is not surprising since the shift by a factor of $\approx2$ in frequency corresponds to only a factor of $\approx1.4$ in terms of electron energy.

In only 2 of the 12 resolved sources in TGSS (Perseus and M87) we detect significantly more extended low-frequency emission than observed at 327~MHz.
For the remainder of the sample, the observed low-frequency emission is well correlated spatially with higher frequency emission seen at 327~MHz.
Therefore, we conclude that for $\sim$90\% of the systems in TF-23 sample, TGSS retrieves the flux density at 148~MHz associated with the same episode(s) of AGN activity seen at higher frequency by \cite{Birzan2008} and in the X-rays by \cite{Rafferty2006}.
Only Perseus and Hydra A include cavity power measurements for more than one pair of cavities, and thus more than one episode of activity. 
Therefore, our analysis at 148~MHz, as well as the analysis of \cite{Birzan2008} at higher frequencies, is effectively restricted to a fairly limited range of outburst ages. 

In the reprocessed MSSS images of Perseus and 2A0335+096 (presented in Section \ref{sec:resolved}) we clearly detect diffuse emission extending well beyond the regions associated with the X-ray cavities studied by \cite{Rafferty2006}. 
The X-ray morphology of these objects on this large scale is complex and not well correlated spatially with the observed low-frequency emission.
Associating these more extended structures in the radio and X-rays with single, well-defined episodes of AGN activity is not trivial.
This complication is most evident in the case of Perseus where even in the radio map, it is difficult to separate emission associated with relic AGN outbursts from the surrounding mini-halo.
In the absence of well-defined spatial correlation, including the flux from this more extended radio emission without measurements of corresponding X-ray structures would artificially result in a correlation with flatter slope and large scatter.
This situation indicates that in order to study the correlation over multiple episodes of activity, we would need a sample with deeper X-ray data with multiple cavities and the corresponding low-frequency radio observations.

Studying the previously published correlations between $P_{\rm{cav}}$ and $L_{\nu}$, \cite{Leith2016} have raised and explored the question if the observed correlation is caused by the underlying physical mechanisms or is a result of distance effect related to a selection bias. Isolating the luminosity distance effect, they find a much weaker correlation between $P_{\rm{cav}}$ and $L_{\nu}$.
The selection effects discussed by \cite{Leith2016} apply equally to the sample discussed in the present paper, since
they considered a larger sample, which includes the objects studied here.
We refer the reader to the work of  \cite{Leith2016} for a deeper analysis of the distance dependence of the two studied parameters.
Although, it seems that at least some of the correlation we measure here is due to this effect, regression taking distance into account does not necessarily correctly account for potential measurement bias.
A more complete way to investigate a likely distance dependence would be to perform a population synthesis simulation in which to create a theoretical sample and study its properties over distance.
Such analysis is however beyond the scope of this paper and we leave it for further work.

\subsection{Spectrum and $P_{\rm{cav}}$ vs. $L_{\rm{bol}}$}
\label{lab:Spectrum}

\begin{table*}[ht]
	\begin{center}
	\caption{Spectral information. Points at 148~MHz derive from TGSS. If not designated differently, the higher frequency points come from \cite{Birzan2008}. 
	}
	\begin{tabular}{l l l l l l}
		\hline
		\hline
		Source		 		& Points$\rm{^a}$	& $\alpha\rm{^b}$ 			& $\sigma$ $\rm{^c}$ 	& $\chi^{2}_{\rm{r}}$ $\rm{^d}$  & Radio data used in the fit$\rm{^e}$\\
								&							&				 					&			(dex)				&	   		 								  &			(MHz)					\\
		\hline		
		A2199					& 6			&	$-$1.65 $\pm$ 0.02	&	0.24	&	73		&				 	148, 324, 750 (1), 1400 (2), 2700 (3), 4675\\
		MS 0735.6+7421	& 3			&	$-$2.43 $\pm$ 0.04	&	0.14	&	9.9	&					148, 330, 1425\\
		2A 0335+096			& 4			&	$-$1.17 $\pm$ 0.03	&	0.19	&	6.8	&					148, 324, 1490 (2), 4860\\
		A262						& 3			&	$-$0.96 $\pm$ 0.04	&	0.07	&	1.0	&					148, 324, 1365\\
		Perseus					& 3			& 	$-$0.46 $\pm$ 0.02	&	0.04	&	0.9	&					148, 408 (4), 750 (1)\\
		MKW3s					& 4			&	$-$2.72 $\pm$ 0.02	&	0.28	&	43		&					148, 327,1440, 4860\\
		A478						& 3			&	$-$1.00 $\pm$ 0.05	&	0.07	&	1.5	&				 	148, 327,1440\\ 
		A2052					& 6			&	$-$1.33 $\pm$ 0.01	&	0.14	&	31		&					148, 330, 750 (1), 1490, 2700 (5), 4860\\
		Zw 3146				& 3			&	$-$0.81 $\pm$ 0.07	&	0.17	&	1.9	&					148, 324, 4860\\		
		Zw 2701				& 3			&	$-$1.49 $\pm$ 0.02	&	0.24	&	30		&					148, 324, 4860\\  	
		A1795					& 5			&	$-$0.90 $\pm$ 0.03	&	0.04	&	0.8	&					148, 327, 1465, 2700 (3), 4850 (6)\\
		RBS 797				& 4			&	$-$1.14 $\pm$ 0.03	&	0.14	&	6.8	&					148, 324, 1475, 4860\\ 
		MACS J1423.8		& 3			&	$-$1.20 $\pm$ 0.05	&	0.10	&	0.9	&					148, 327, 1425\\		
		A1835					& 4			&	$-$0.84 $\pm$ 0.03	&	0.18	&	7.5	&					148, 327, 1400, 4760\\
		M84						& 4			&	$-$0.57 $\pm$ 0.02	&	0.07	&	9.4	&					148, 324, 1425, 4860 \\	
		M87						& 3			&  $-$0.60 $\pm$ 0.03    &	0.08	&  5.3  &                    148, 1400, 4860 \\

		A133						& 3			&	$-$2.18 $\pm$ 0.03	&	0.31	&	52		&					148, 330, 1425 \\
		Hydra A					& 6			&	$-$0.89 $\pm$ 0.02	&	0.02	&	1.9	&					148, 333, 750 (7), 1423, 2700 (8), 4760\\
		Centaurus				& 6			&	$-$0.76 $\pm$ 0.01	&	0.04	&	4.0	&					148, 327, 625 (7), 1565, 2700 (9), 4760\\
		HCG 62					& 3			&	$-$0.29 $\pm$ 0.13	&	0.04	&	0.1	&					148, 325, 1440\\
		Sersic 159/03			& 4			&	$-$1.09 $\pm$ 0.02	&	0.04	&	10		&					148, 325, 1425, 4860\\
		A2597					& 4			&	$-$1.09 $\pm$ 0.02	&	0.11	&	10		&					148, 328, 1400, 4985\\
		A4059					& 5			&	$-$1.37 $\pm$ 0.03	&	0.22	&	5.0	&					148, 328, 1440 (2), 2700 (10), 4860 (11)\\
		\hline
		\multicolumn{6}{l}{$\rm{^a}$ Number of frequencies used for the power-law fit.} \\	
		\multicolumn{6}{l}{$\rm{^b}$ Spectral index derived from the power-law fit.} \\	
		\multicolumn{6}{l}{$\rm{^c}$ Intrinsic scatter of the points towards the power-law fit.} \\
		\multicolumn{6}{l}{$\rm{^d}$ Reduced $\chi^2$ of the power-law fit.} \\	
		\multicolumn{6}{l}{$\rm{^e}$ The frequencies in MHz used for power-law fitting of the radio spectra. The numbers in parentheses } \\	
		\multicolumn{6}{l}{\,\,\,  are the references for the literature values and have the same meaning as in Table \ref{tab:all_freq_flux}. } \\	
	\end{tabular}
	\label{tab:fitting_L_bol}
	\end{center}
\end{table*}

As the bolometric radio luminosity ($L_{\rm{bol}}$) is expected to be a better gauge of the total radiative power, \cite{Birzan2008} also study the relation between $L_{\rm{bol}}$ and cavity power.
They conclude that the total bolometric radio power is not a better proxy for cavity power than the monochromatic luminosity at 327~MHz due to the similarly large scatter.
They further show that the scatter is reduced from 0.83 to 0.64~dex if only the lobe emission is considered.  
This selection, however, reduces the number of studied systems to 13 \citep[Table 3][]{Birzan2008}.
To further investigate whether the scatter is due to radio aging, they include the spectrum break frequency ($\nu_{\rm{C}}$) in the regression analysis and derive a tight correlation \cite[Equation 17;][]{Birzan2008}.
They show that the knowledge of the lobe break frequency improves the scatter by $\approx$50\% (to 0.33~dex), which significantly increases the accuracy by which one can estimate the cavity power of the AGN when cavity data are unavailable.

Although now widely used, the $P_{\rm{cav}}$ -- $L_{\rm{bol}}$,$\nu_{\rm{C}}$ scaling relation is based on a number of simplifying assumptions.
\cite{Birzan2008} report that the continuous injection model \citep[CI;][]{Kardashev1962} does not provide acceptable fits to many of the spectra. As an alternative they adopt the single injection KP model \citep[][]{Kardashev1962, Pacholczyk1970}.
It is, however, by now well established that the KP model makes unrealistic physical assumptions especially for studies of extended FRI type sources \citep[e.g.][]{Tribble1993, Harwood2013, Hardcastle2013}.
From the 13 points used to derive the $P_{\rm{cav}}$ -- $L_{\rm{bol}}$,$\nu_{\rm{C}}$ relation, six systems have only lower or upper limits on the break frequency.
However, these limits are treated as detections in the regression analysis of \cite{Birzan2008}. The actual break frequencies could be far from the limit values, and the resulting impact of these limits on the reduction in the correlation scatter has not been assessed.

From the remaining seven systems used by \cite{Birzan2008} to derive the $P_{\rm{cav}}$ -- $L_{\rm{bol}}$,$\nu_{\rm{C}}$ relation, three include break frequencies obtained by fitting a KP model to the data. 
For one of these sources (MS 0735.6+7421) the KP fit results in a very high injection index of $\alpha_i =1.3$, while the typical value for $\alpha_i$ in the literature is in the range 0.5 -- 0.8 \citep{KomissarovGubanov1994}.
The other four systems have their break frequency computed using the recipe of \citet{Myers1985} assuming a fixed injection index of $\alpha_i =0.5$. 
This approximation is based on the KP model and allows estimation of the break frequency by assuming an injection index and providing a measurement of the spectral index between 1.4 and 4.9~GHz. While this strategy conveniently provides break frequencies based on two data points, the presented values for those four systems do not include uncertainty estimates.
Although correcting for the effects of spectral aging can in principle reduce the scatter in the observed correlation, large uncertainties in the derived break frequencies may distort the reduction in the scatter this correction can yield.

In an attempt to derive reliable break frequencies and bolometric luminosities, we revisit the radio spectra adding our new low frequency measurements.
Plotting the total flux values at 330~MHz, 1.4~GHz, 4.5~GHz, and 8.5~GHz from \cite{Birzan2008} we find that in several cases the fluxes at 8.5~GHz appeared significantly higher than expected by extrapolating the lower frequency points. 
This difference is especially evident in the SED's of MKW3s, MS0735, and Zw2701.  
Inspecting the VLA contours provided in Fig. 10 of \cite{Birzan2008} we determined that no extended structures are observed on the 8.5~GHz maps of ten systems -- 2A0335, A262, A478, A1795, A2052, MKW3, MS0735, Perseus, RBS797, A133. The emission at this frequency clearly derives from the compact core. 
The lobes are partially exposed on the 8.5~GHz map of only eight systems -- A2199, Hydra A, Centaurus, Sersic 159/03, A2597, A4059, and M84. Four systems (Zw 2701 and Zw 3146, A1835, and HCG 62) are basically unresolved at all frequencies and there is no observation of MACS J1423.8 at 8.5~GHz.
We argue that the total luminosity at 8.5~GHz primarily derives from the core, as opposed to the aged radio lobes which clearly dominate below 1.4~GHz.
For this reason, we have neglected the 8 GHz data points in our analysis and considered only data below 5 GHz in our SED fits.

Given the agreement in flux scales between MSSS and TGSS, and the more complete sky coverage of the TGSS sample, we have restricted ourselves to the TGSS fluxes in our spectral fitting analysis. We note that in four systems, MSSS shows significantly higher flux densities relative to TGSS. These systems include Perseus, A478, Zw3146, and MACS J1423.8 with flux densities that are factors of 1.4, 1.7, 10.0, and 1.8 higher, respectively. These discrepancies may represent new emission components, issues in the flux density calibration, or, in the case of A478 and Zw3146, blending with neighboring objects due to insufficient angular resolution. Although excluded from the fitting process, we have included the MSSS points in the SED plots in Figure \ref{fig:sed_msss_tgss_vlss} and \ref{fig:nonMSSS_sed_tgss_vlss} in Appendix \ref{appendix:sed} for the purposes of comparison. 

As an alternative to the KP model used by \cite{Birzan2008}, we could consider fitting a continuous injection (CI) model to the SEDs to correct for the effects of spectral curvature. However, we note recent work has shown that fitting integrated spectra with the CI model should be treated with caution \citep{Harwood2017}. In addition, the number of frequency points for objects in our sample range from three to six, with 40\% of the sample having just three data points for fitting. With these limitations, we find we cannot fit a single, well-constrained and physically justified electron aging model consistently for all sources. As a result, we are unable to derive a reliable and uniform set of break frequencies and bolometric luminosities for the sample. Given the available data, we therefore find that the addition of a single low-frequency constraint at 148 MHz is insufficient to improve upon the spectral aging correction of \cite{Birzan2008}.

For this reason, and in order to treat all sources consistently, we have fit a simple power law model to all spectral points. Reconstructing the SEDs, we include the TGSS 148 MHz flux densities, the 330 MHz, 1.4 GHz, and 4.5 GHz VLA measurements, as well as the high frequency literature values used by \cite{Birzan2008}. We perform the spectral analysis using total flux measurements since the majority of the TF-23 sources have only total flux density measurements at both low and high frequencies.
Table \ref{tab:fitting_L_bol} presents the derived spectral index, the intrinsic scatter, the reduced chi-squared, the number of fitted points, and their origin.
Plots of the fits themselves can be seen in Figure \ref{fig:sed_msss_tgss_vlss} and \ref{fig:nonMSSS_sed_tgss_vlss} in Appendix \ref{appendix:sed}. We note that, based on the reduced chi-squared values, the SED's for many of the sources in the sample are not well fit by a simple power-law. However, as already discussed above, the available data is currently insufficient to consistently fit a more complicated spectral model. The derived spectral indices are used for computing the monochromatic luminosities employing Equation \ref{equ:L_nu_formula}.

\section{Resolved Systems}
\label{sec:resolved}

This section describes the morphology of the resolved sources from our reprocessed MSSS fields. 
We review 2A0335+096, MS 0735.6+7421, A2199, and the Perseus cluster and discuss the new features revealed at 140~MHz. 
We present both the MSSS and TGSS images and discuss the differences between the two surveys.
We compare the low frequency radio images with \textit{Chandra} X-ray maps and study the correspondence between the observed radio and X-ray structures.

\subsection{Perseus}

\begin{figure*}[t]
\begin{center}
\includegraphics[height=3.2in]{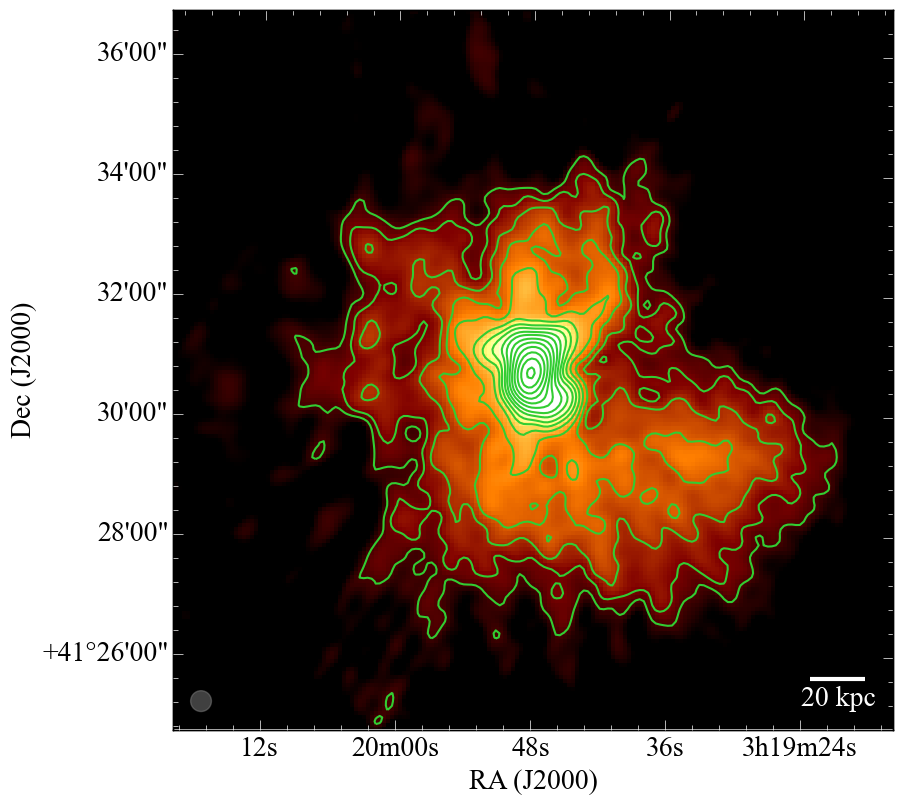}
\includegraphics[height=3.2in]{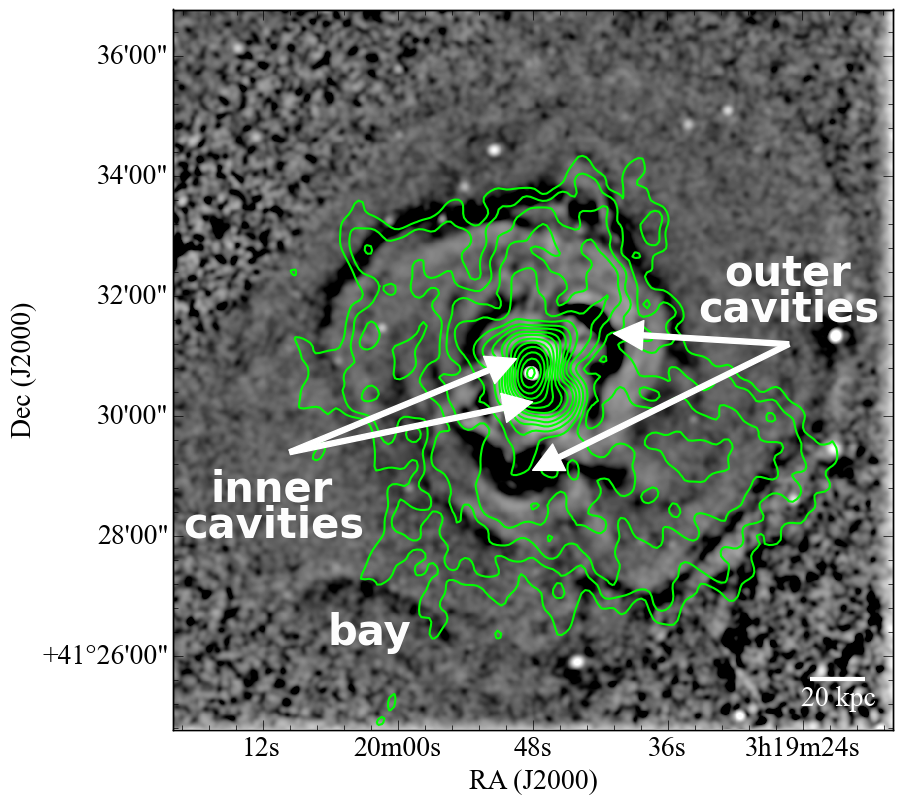} \\
\includegraphics[height=3.20in]{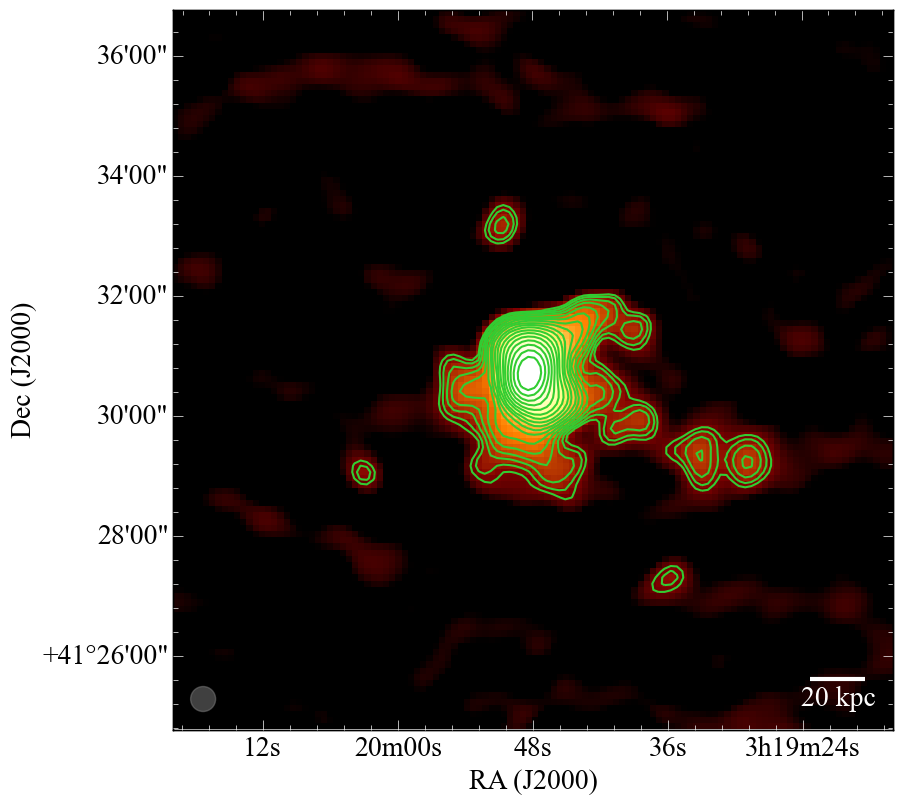} 
\end{center}
\caption{\small  Perseus cluster. Top left: Reprocessed MSSS map with resolution $20.8^{\prime\prime}\times20.8^{\prime\prime}$ and rms noise 20~mJy/beam. The contours start at 5$\sigma$ level and are drawn at 100~mJy/beam $\times$ [1, 1.4, 2, 2.8, 4, 5.7, 8, 11, 16, 22, 32, 45, 64, 91, 128]. Top right:  \textit{Chandra} X-ray surface brightness residual map. The image is produced by unsharp masking using archival data in the 0.5 -- 7 keV band with total exposure of 1.4~Msec after standard filtering. Contours correspond to the MSSS image. Bottom:  TGSS map with resolution $25.0^{\prime\prime}\times25.0^{\prime\prime}$ and rms noise 10~mJy/beam. The contours start at 5$\sigma$ level and are drawn at 50~mJy/beam $\times$ [1, 1.4, 2, 2.8, 4, 5.7, 8, 11, 16, 22, 32, 45, 64, 91, 128, 182, 256].
\label{fig:perseus_all}}
\vspace{0.15in}
\end{figure*}

The Perseus cluster, A426, is the brightest cluster in the X-ray sky and has therefore been well studied by all X-ray telescopes. 
The X-ray emission is sharply peaked on the cluster core, centered on the cD galaxy NGC~1275 (Perseus A). 
A pair of X-ray cavities in north (N) and south (S) from the center are coincident with the  FRI radio source, 3C84 \citep{Pedlar1990, Boehringer1993, Fabian2000, Churazov2000}. 
More distant bubbles, presumably a product of past activity, are seen to the NW and S.
Further to the north, X-rays have revealed a region of flux drop, named the northern trough by \cite{Fabian2011}.
At 50 -- 80~kpc from the center, \cite{Fabian2006, Fabian2011} identify a south 'bay' of hot gas which is in approximate pressure equilibrium.
They argue that rising bubbles from energetic past outburst have accumulated in the northern trough and the south bay.

A semi-circular cold front \citep{Markevitch2007}, accompanied by a sharp drop of metallicity, is distinguished at $\sim$100~kpc to the west and south-west of the nucleus of NGC 1275 \citep{Fabian2011}. 
The northern trough and the southern bay lie along a continuation of the west cold front. 
While the morphology inside the cold front is dominated by the AGN activity and turbulence in the gas \citep{Hitomi2016}, the structures at larger radii are most probably associated with a subcluster merger \citep{Churazov2003}, which also accounts for the east-west asymmetry in the X-ray surface brightness.

It is known from higher frequency observations that the center of Perseus hosts a rare \citep{Feretti2012} radio mini-halo, which may be generated by turbulence \citep{Gitti2004} or gas sloshing \citep{ZuHone2011, Mazzotta2008}, presumably induced by an off-axis merger.
In our MSSS map at 148~MHz we successfully recover the known general morphology of the inner part of the mini-halo and distinguish new structures in this region.

While the TGSS image only shows the radio emission around the inner cavities (Figure \ref{fig:perseus_all}), the reprocessed MSSS map reveals the inner part of the mini-halo (Figure \ref{fig:perseus_all}).
Being a short integration survey, MSSS is not deep enough to detect the full scale of the mini-halo, as observed at 330 MHz \citep{Burns1992, DeBruyn2005} and 610 MHz \citep{Sijbring1993}, but it does reveal interesting new features within 100~kpc around NGC 1275.

The inner part of the radio emission follows the general NS orientation of the jets (Figure \ref{fig:perseus_all}).
The outer part of the diffuse emission continues up to 60 kpc north from center of Perseus A, reaching up to an arch-like drop in X-ray brightness (Figure \ref{fig:perseus_all}).
South from the core the mini-halo bends towards the west and stretches up to 100~kpc  towards SW, reaching as far as the cold front.

The northern part of the diffuse emission shows two pronounced features elongated towards N and NW.
The NW feature corresponds to the outer NW X-ray cavity. 
The structure stretching north lies along the main jet axis and is cospatial with a known optical filament \citep{Conselice2001, Fabian2011}.
It extends along the X-ray spur, which continues further than the optical filament.
What we see in our MSSS map is presumably the base of the radio structure observed at higher frequencies \citep{Burns1992, Sijbring1993, DeBruyn2005}, which reaches the X-ray northern trough.

We further identify two radio `streamers' within the mini-halo.
We distinguish an elongated structure which starts at the eastern edge of the inner north cavity and stretches NE reaching the arch-like drop in X-ray brightness $\sim$80~kpc from the center.
Our MSSS map also shows an enhancement of the radio flux to the south of the core which corresponds to the southern old cavity.

The outlined radio features once again exemplify the complicated morphology at the center of the Perseus cluster.
Our shallow MSSS map demonstrates the potential of low-frequency observations on Perseus and promises that the deep LOFAR data will reveal more details of the complex morphology of the mini-halo.
We will present a full track LOFAR observation in a followup paper.
The deep LOFAR map is expected to help us constrain the integrated energy output of the AGN over the last few Myr and verify if the low-frequency radio mini-halo is energetically consistent with the turbulence recently measured with \textit{Hitomi} \citep{Hitomi2016}.

\subsection{2A 0335+096}
\label{lab:2A0335_resolved}

\begin{figure*}[t]
\begin{center}
\includegraphics[height=3.10in]{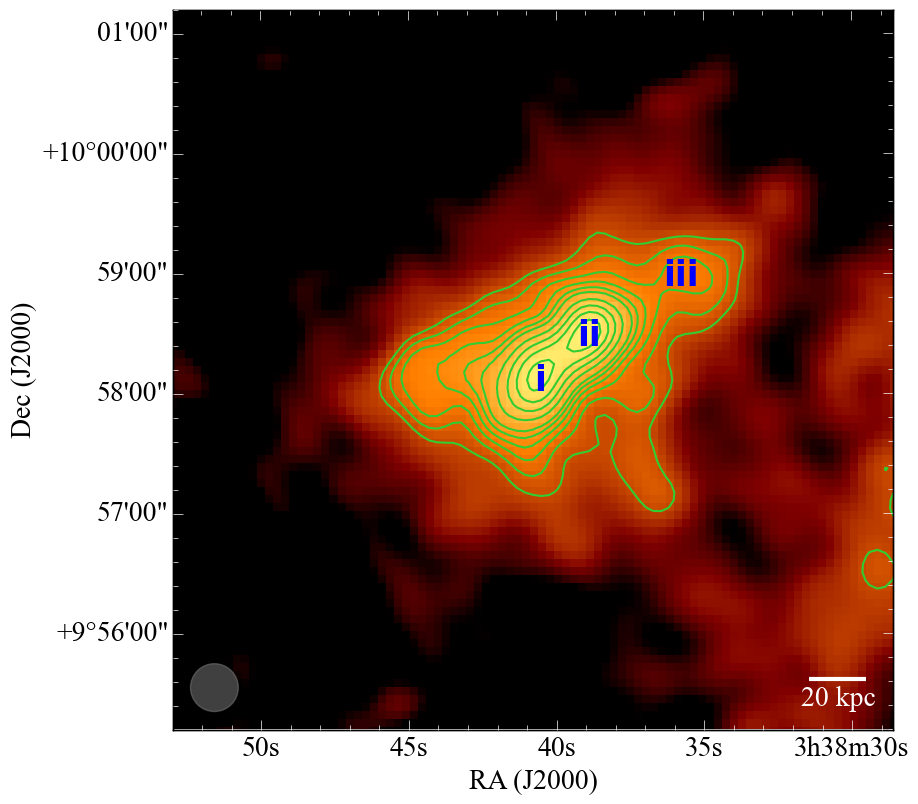} 
\includegraphics[height=3.10in]{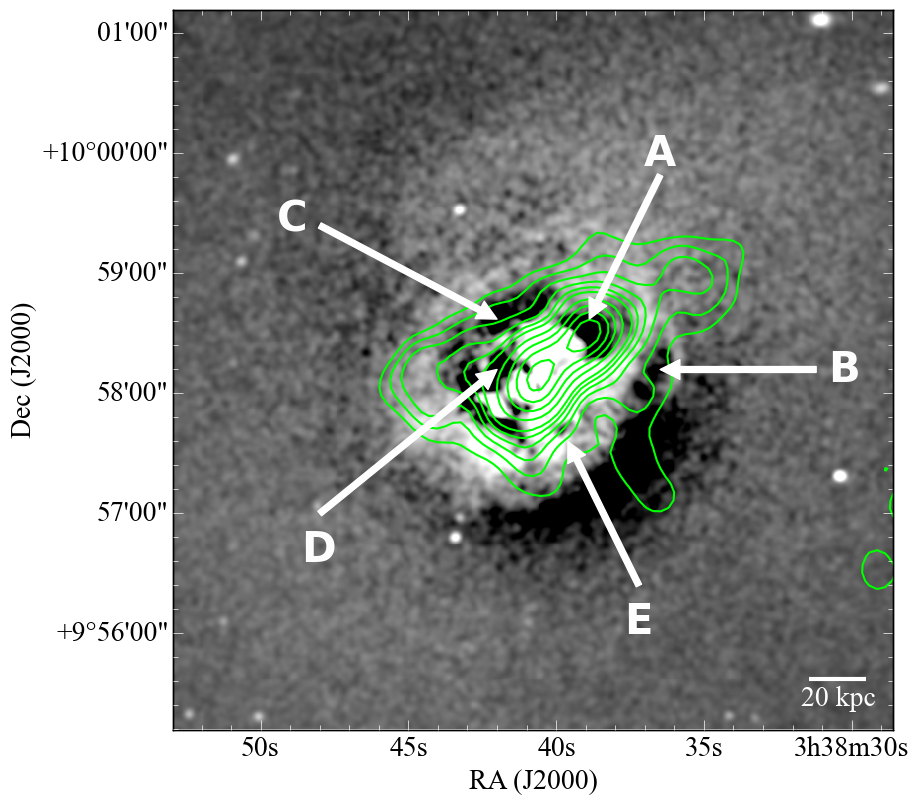} \\
\includegraphics[height=3.10in]{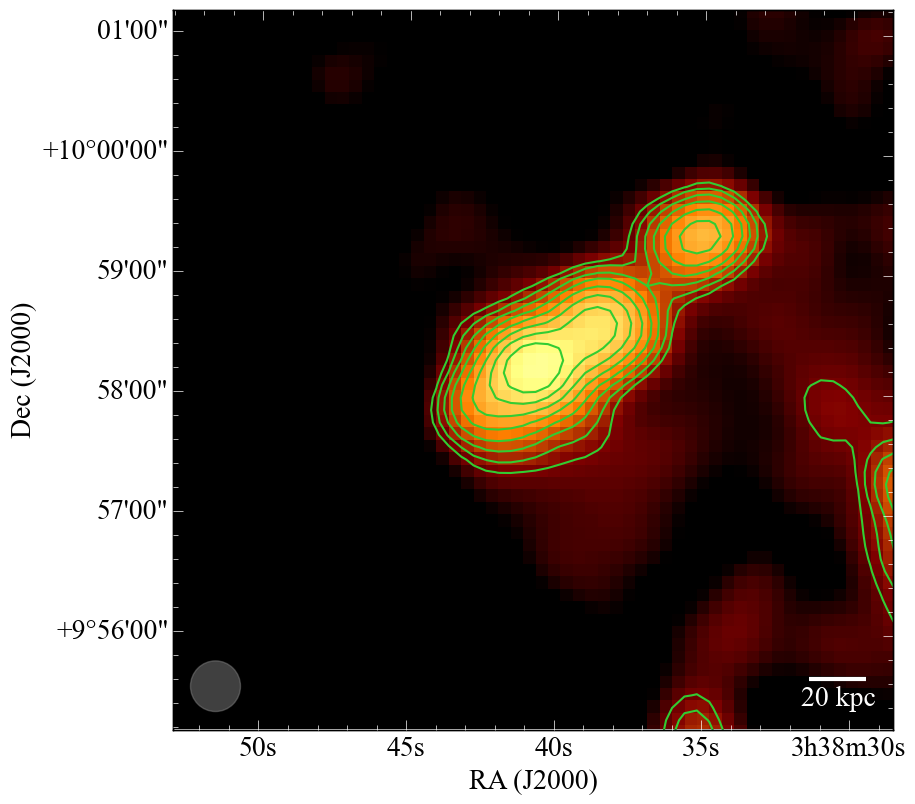} 
\includegraphics[height=3.10in]{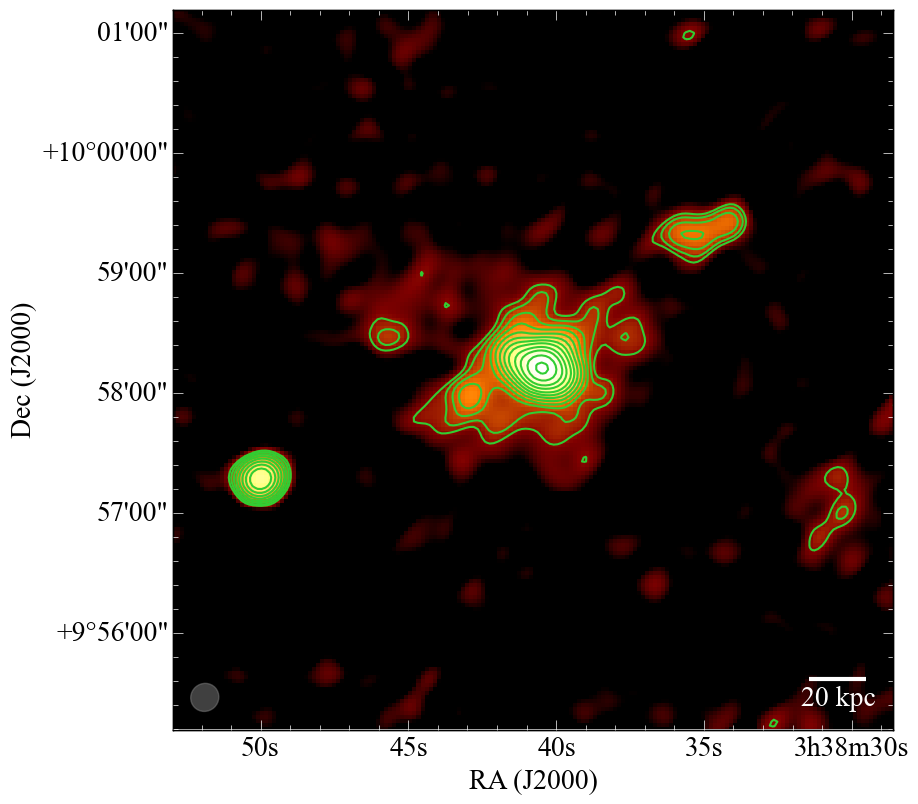} 
\end{center}
\caption{\small  2A0335+096. Top left: Reprocessed MSSS map with resolution $23.6^{\prime\prime}\times23.6^{\prime\prime}$ and rms noise 11~mJy/beam. Contours start at 5$\sigma$ level and correspond to 55~mJy/beam $\times$ [1, 1.2, 1.4, 1.6, 1.8, 2, 2.4, 2.8, 3.2, 3.6]. Individual features are labeled i, ii, and iii. These are discussed in the text. Top right:  \textit{Chandra} X-ray surface brightness residual map. The image is produced by unsharp masking using archival data in the 0.5 -- 7 keV band with total exposure of 101~ksec after standard filtering. Green contours correspond to the MSSS image. The five cavities identified by \cite{Sanders2009} are marked. The correspondence between these features and our radio map is discussed in the text. Bottom left: TGSS map with resolution $25.3^{\prime\prime}\times25.0^{\prime\prime}$ and rms noise 5~mJy/beam. Contours  start at 3$\sigma$ level and correspond to 15~mJy/beam $\times$ [1, 1.4, 2, 2.8, 4, 5.7, 8, 11]. Bottom right: Radio map at 1.5~GHz from \cite{Birzan2008}. Contours correspond to 0.25~mJy/beam $\times$ [1, 1.4, 2, 2.8, 4, 5.7, 8, 11, 16, 22, 32, 45].
\label{fig:2a0335_all}}
\vspace{0.15in}
\end{figure*}

2A 0335+096 is a compact, radio-bright, nearby ($z = 0.035$) galaxy cluster with a luminous X-ray core \citep{Sanders2009}.
The central galaxy has a cD morphology that hosts a weak radio source \citep{Farage2012}.
It has a nearby companion elliptical galaxy situated  $\sim7^{\prime\prime}$ ($\sim$5~kpc) to the north-west \citep{Sanders2009}.
The velocities of the central galaxy and the companion derived from optical spectroscopy studies suggest they are merging \citep{Gelderman1996, Donahue2007, Hatch2007}. 

2A0335+096 shows a very complicated morphology in both radio and X-rays.
\textit{XMM-Newton} and \textit{Chandra} have revealed evidence for an active history of interaction between the AGN and ICM in the core of 2A 0335+096.
X-ray observations show a complex system of X-ray-emitting structures in the core, including filaments, cool clumps, a metal-rich spiral, and at least five distinct cavities at varying distances and position angles relative to the central galaxy \citep{Mazzotta2003, Birzan2004, Kaastra2004, SandersFabian2006, Sanders2009}. 
The total $4pV$ enthalpy associated with the five cavities is estimated to be $5\times10^{59}$~erg \citep{Sanders2009}.
There is a sharp drop in X-ray surface brightness at $1^{\prime}$ radius to the south-east of the core, identified as a cold front by \cite{Mazzotta2003}.
Since the feature shows an abrupt density decline but no significant temperature change over the edge, \cite{SandersFabian2006} classify it as an isothermal shock, as found in the Perseus cluster.

The radio source shows a very different morphology at different frequencies. At $\sim5$~GHz twin lobes associated with oppositely directed radio jets are observed in NE -- SW direction extending $\sim12^{\prime\prime}$ in length \citep{Sanders2009}.
Using the \cite{Myers1985} relation, \cite{Donahue2007} estimate an age for the radiating electrons of 25~Myr assuming equipartition or 50~Myr if the magnetic field is a factor of 4 less than the equipartition value.
They propose that an interaction between the central and the nearby companion elliptical galaxy may have triggered the current episode of radio activity.

At 1.5~GHz (Figure \ref{fig:2a0335_all}) the main observable feature is a clear single peak at the center of the galaxy surrounded by a steep-spectrum emission roughly symmetric around the center \citep{Sarazin1995, Birzan2008, Sanders2009}. 
The observed shape and spectral index has led the observed emission to be classified as a mini-halo.
At 330~MHz the mini-halo is not well observed, but a second peak of emission is distinguished $30^{\prime\prime}$ ($\sim$20~kpc) NW from the center \citep{Birzan2008} and the morphology appears clearly elongated in NW-SE direction.

The cluster hosts an unusual nearby Narrow Angle Tail (NAT) radio source which has been observed at 327~MHz \citep{Patnaik1988} and 1.5~GHz \citep{Odea1985, Sarazin1995}. Its center is situated at $\sim$380~kpc in the N-W direction from the center of the cluster, but its long tail propagates south getting as close as $\sim$150 kpc ($\sim3.5^{\prime}$) from the central galaxy.
No interaction between the NAT source and the core of 2A0335+096 has been observed on previously published images.
At low significance level the reprocessed MSSS map (Figure \ref{fig:2a0335_all}) shows some evidence that the diffuse emission surrounding the core of 2A0335+096 is connected to the emission from the NAT source. This suggest that the two sources might be interacting, however, the quality of the data does not allow us to be conclusive about this scenario.

The diffuse radio emission at 140~MHz (Figure \ref{fig:2a0335_all}) extends in all directions further than observed at 1.5~GHz (Figure \ref{fig:2a0335_all}).
Similar to the Perseus cluster, the reprocessed MSSS map (Figure \ref{fig:2a0335_all}) reveals much more diffuse structure than the TGSS image (Figure \ref{fig:2a0335_all}).
Although the reprocessed MSSS map includes significant artifacts around point sources, our visual inspection confirmed that the observed extended emission of 2A0335 is authentic since no other source in the maps shows emission with similar extent or morphology.
The 140~MHz map (Figure \ref{fig:2a0335_all}) shows pronounced elongated shape along the NW - SE direction.
The diffuse emission at 140~MHz spans $\sim$130~kpc in NW--SE direction and $\sim$60~kpc in NE--SW direction. 
The boundaries of the observed emission at 140~MHz towards south reach as far as the isothermal shock at $\sim1^{\prime}$ from the center.

Figure \ref{fig:2a0335_all} shows that, besides the central peak of emission, there is clearly a second peak of emission situated $30^{\prime\prime}$ ($\sim$20~kpc) in the NW direction from the center.
In the MSSS map these two peaks appear equally bright.
The central maximum (i) corresponds to the very center of the cluster.
The second peak (ii) is clearly associated with the most pronounced cavity in the system (cavity A, Figure \ref{fig:2a0335_all}). 

We observe a significant extension of the emission towards north-west which includes the distant radio peak (iii).
The observations at 140~MHz (Figure \ref{fig:2a0335_all}) and 148~MHz  for the first time reveal the bridge between the radio emission associated with the central mini-halo and the radio maximum iii, which appeared disconnected from the core at higher frequencies.
Despite the fact that emission at peak iii was observed in deep high-frequency maps, it is nonetheless found to be a steep spectrum source.
We measured the spectral index in a circular aperture centered at peak iii and found a value of $\sim -1.8$ between 147~MHz and 1.5~GHz. 
This spectral index seems consistent with the idea that this plasma originated in an older outburst of AGN activity.
The distant peak iii is situated $\sim$110~kpc from the central AGN and for it we calculate an age of 150~Myr based on the sound speed estimate for the system by \cite{Birzan2004}.

The residual map we calculate based on archival X-ray data (Figure \ref{fig:2a0335_all}) agrees reasonably well with the image published by \cite{Sanders2009}. 
The only major difference is that cavity E appears much less pronounced in our residual map.
Based on our MSSS image, all of the five cavity structures identified by \cite{Sanders2009} contain low frequency emission.
There is a clear extension of the radio emission along cavity D, which continues beyond the cavity.
A spur of emission starts at cavity B and continues south through the isothermal shock.
Although cavities C and E are also fully covered by the radio emission, they are not clearly associated with a particular features of the radio morphology. 
The association between the radio emission and the X-ray cavities implies that the former is mostly formed from multiple previous generations of AGN radio outbursts.
However, the morphology of the cavities is complex and is unclear what the order of generation of the bubbles was. 

In general, what we observe with LOFAR appears much more extended and complicated than the picture at higher frequencies. 
The overall shape of the source does not resemble the traditional symmetric round morphology associated with a mini-halo.
The structure is elongated to the NW, following the direction defined by the two main radio enhancements (i and ii), i.e the direction between the center and cavity A.
This direction is consistent with the elongation observed at 330~MHz 
and matches the axis between the central galaxy and the nearby companion, as well as the direction of the filament observed in H$\rm{\alpha}$ \citep{Sanders2009}.

If the source is not a mini-halo the other most likely candidate is structure formed by activity over time from the AGN.
We see evidence for this in the correlation between this extended low-frequency radio structure and the X-rays.
There is a clear correspondence between peak i and ii and the X-ray cavities
The correspondence between radio peak iii and the X-ray morphology is of lower significance, which might be due to the quality of the X-ray data.
The observed structures seem consistent with relics of past AGN outbursts.
Although we cannot definitively say with the currently available data, the evidence seems to favor the AGN interpretation over a simple mini-halo.

The observed X-ray cavities do not follow the usual simple linear progression, but resemble a network of structures that are presumably related to multiple outbursts \citep{Sanders2009}.
It might be that density inhomogeneity of the surrounding environment has deflected the straight buoyancy rise path of the bubbles. 
Based on the fact that the orientation of the low frequency radio peaks i and ii is perpendicular to the small scale radio lobes observed by \cite{Donahue2007}, the idea that the radio source has changed orientation over time also seems quite plausible.
This interpretation would imply that the source has been deflected in the recent past.
Based on its circular shape and proximity to the center, X-ray cavity A (coincident with low frequency radio peak ii) seems to be the newest one. 
Considering the buoyancy age of cavity A from \cite{Rafferty2006}, the change in direction of the radio source must have happened in the last $\sim60$ Myr. 
This value is consistent with the age estimate for the NE-SW radio lobes of \cite{Donahue2007} presented earlier.  

Unfortunately, with the currently available data we cannot be conclusive about the origin of the newly observed diffuse radio emission.
The proximity of the NAT source additionally complicates the interpretation of the faintest and most extended structure to the west and NW.
We expect that the future map from the LOFAR Two-metre Sky Survey \citep[LoTSS;][]{Shimwell2017} will be able to shed light onto the outburst history of this AGN by showing in greater detail the radio structures corresponding to the numerous X-ray cavities and by revealing the region of interaction between the NAT source and the AGN.

 \subsection{MS 0735.6+7421}

\begin{figure*}[t]
\begin{center}
\includegraphics[height=3.20in]{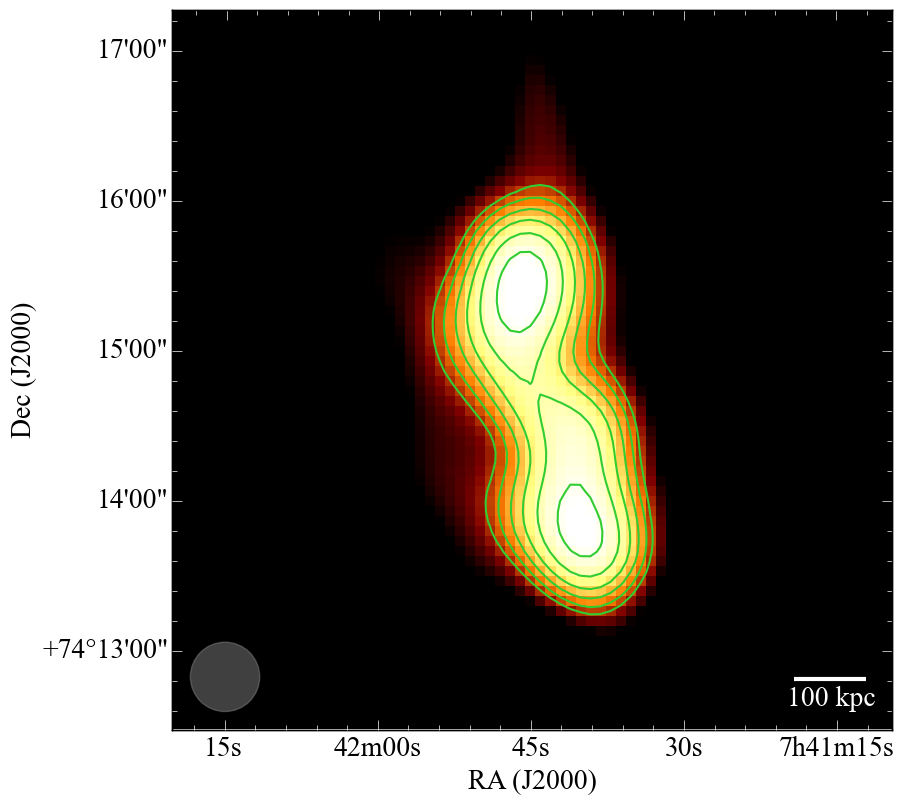}
\includegraphics[height=3.20in]{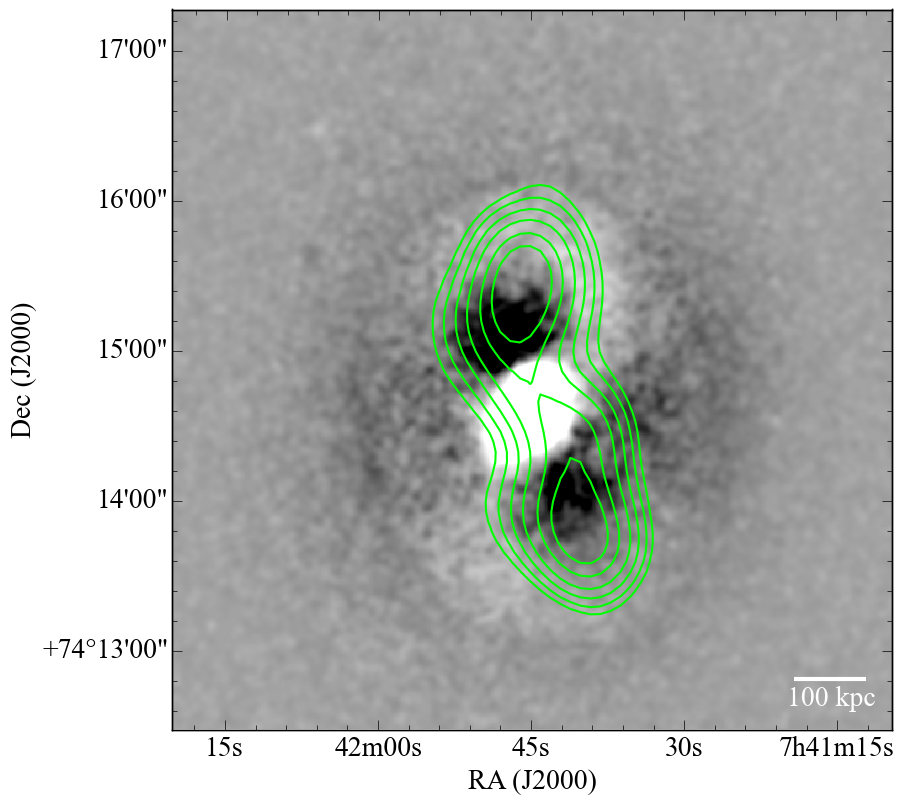} \\
\includegraphics[height=3.20in]{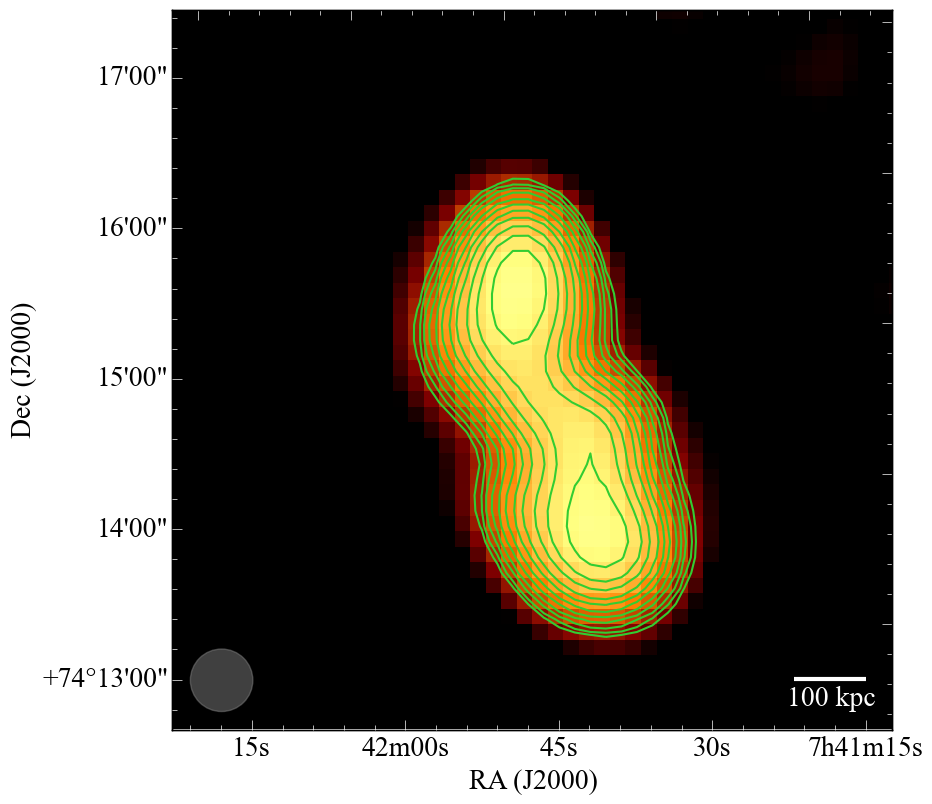} 
\end{center}
\caption{\small MS0735+7421. Top left: Reprocessed MSSS map with resolution $27.8^{\prime\prime}\times27.8^{\prime\prime}$ and rms noise 30~mJy/beam. The contours start at 5$\sigma$ level and are drawn at 0.150~mJy/beam $\times$ [1, 1.4, 2, 2.8, 4, 5.7]. Top right: \textit{Chandra} X-ray surface brightness residual map. The image is produced by unsharp masking using archival data in the 0.5 -- 7 keV band with total exposure of 520~ksec after standard filtering. Green contours correspond to the MSSS image. Bottom: TGSS map with resolution $25.0^{\prime\prime}\times25.0^{\prime\prime}$ and rms noise 3~mJy/beam. The contours start at 5$\sigma$ level and are drawn at 15~mJy/beam $\times$ [1, 1.4, 2, 2.8, 4, 5.7, 8, 11, 16, 22, 32]. 
\label{fig:ms0735_all}}
\vspace{0.15in}
\end{figure*}

The cool-core cluster MS 0735.6+7421 (hereafter MS0735) hosts one of the most energetic radio AGN known. 
It is the most distant ($z = 0.22$) among the sub-sample of clusters that we resolve with MSSS. 
MS0735 hosts large X-ray cavities in an otherwise relaxed system \citep{Gitti2007}. 
Each cavity has a diameter of $\sim$200~kpc and is filled with synchrotron emission from the radio lobes. 
A weak but powerful shock front encloses the cavities and the radio lobes in a cocoon \citep{McNamara2005}. 
The total energy required to inflate the cavities and drive the shock front is above $10^{62}$~erg \citep{McNamara2009, Vantyghem2014}.

Both the MSSS map and the TGSS image (Figure \ref{fig:ms0735_all}) retrieve the main morphology of the source known from higher frequencies.
In this system, two symmetric lobes are propagating in north-south direction.
We observe a change of direction of the lobes at 140~MHz consistent with the bending visible at 324~MHz.
As opposed to higher frequencies, at 140 MHz the compact core is not distinguished between the bright steep-spectrum lobes, which clearly dominate radio structure.
The 140~MHz map does not show more extended diffuse radio emission associated with this object.

The data from TGSS and MSSS is consistent with the picture from higher frequencies showing the radio emission still trapped in the surrounding cocoon (Figure \ref{fig:ms0735_all}). 
This phenomenon may be related to the fact that MS0735.6+7421 is the most energetic outburst observed so far.
The energy of the outburst has clearly displaced and compressed material in the cluster atmosphere which may have piled up faster and created a more effective confining shell to stop the advance of any radio plasma.
We find no evidence in this data for more extended emission corresponding to older outbursts.
If present, that emission will presumably peak at even lower frequencies detectable with LOFAR LBA.

\subsection{A2199}
\label{sec:A2199}

\begin{figure*}[t]
\begin{center}
\includegraphics[width=3.50in]{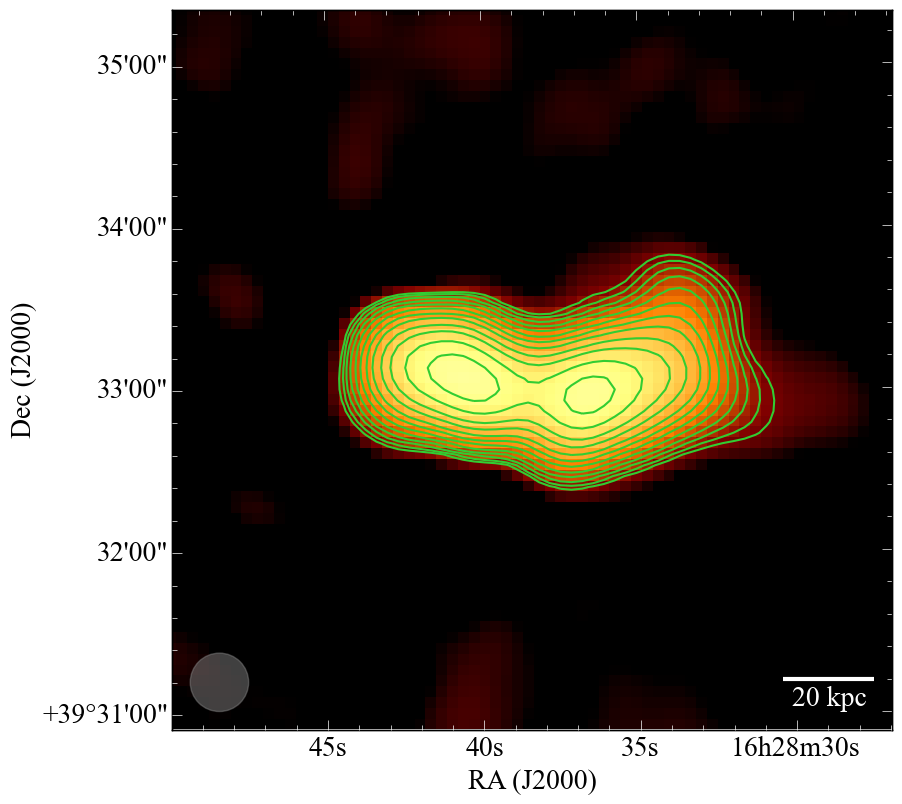}
\includegraphics[width=3.5in]{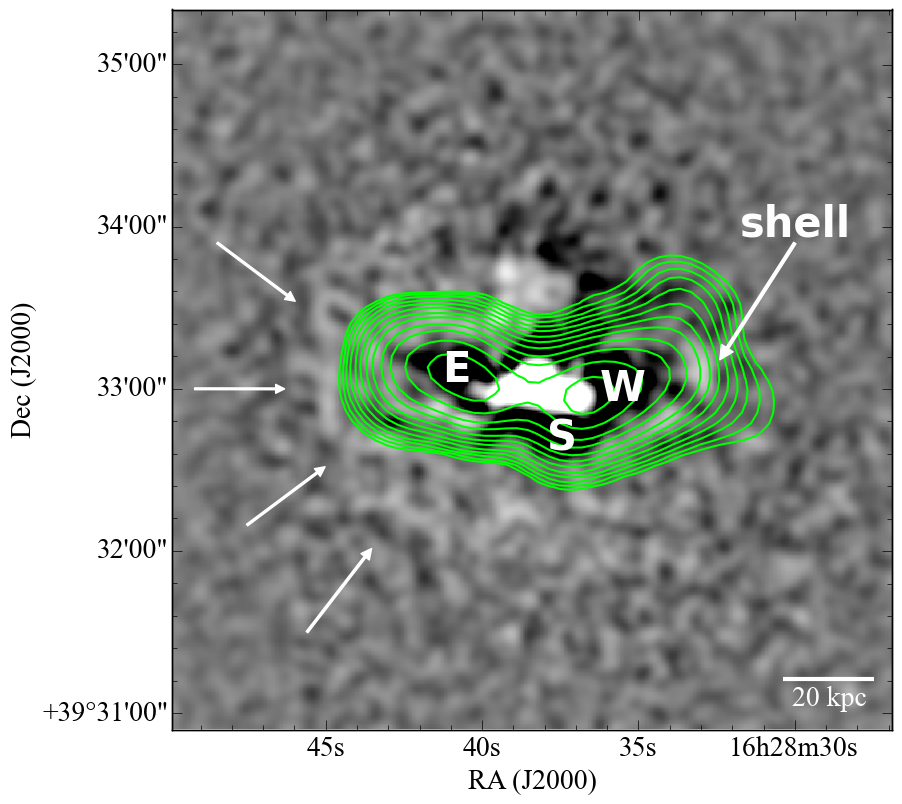} \\
\includegraphics[width=3.5in]{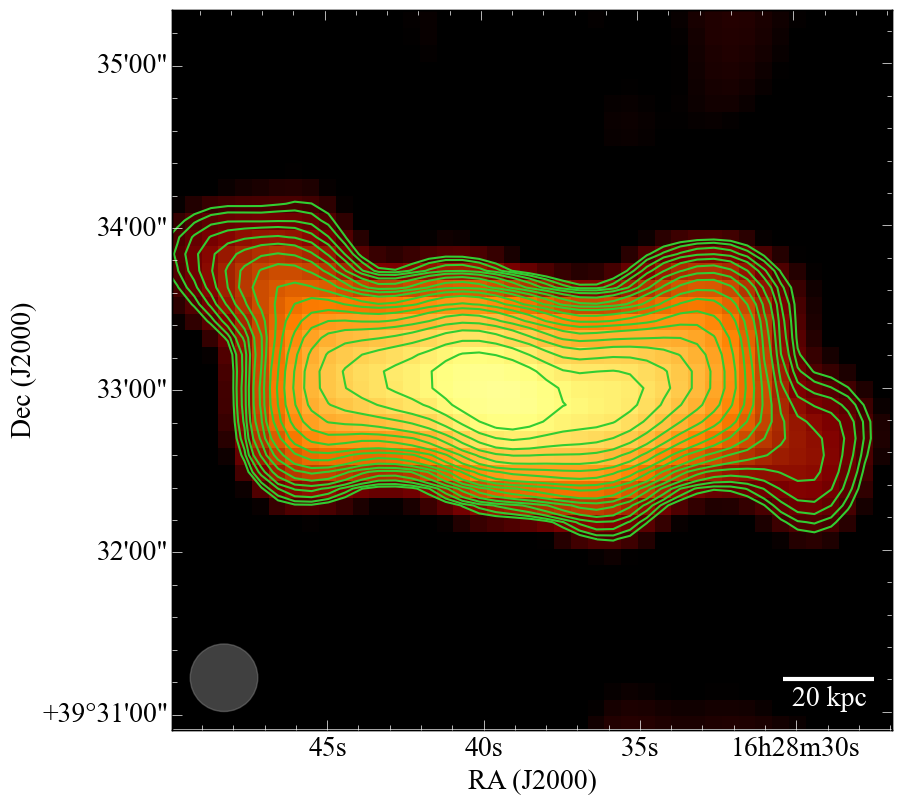}
\end{center}
\caption{\small A2199. Top left: Reprocessed MSSS map with resolution $21.6^{\prime\prime}\times21.6^{\prime\prime}$ and rms noise 40~mJy/beam. The contours start at 5$\sigma$ level and are drawn at 200~mJy/beam $\times$ [1, 1.4, 2, 2.8, 4, 5.7, 8, 11, 16, 22, 32, 45]. Top right: \textit{Chandra} X-ray surface brightness residual map. The image is produced by unsharp masking using archival data in the 0.5 -- 7 keV band with total exposure of 154~ksec after standard filtering. The east, west, and south cavities are denoted by 'E', 'W', and 'S', respectively. The SE surface brightness edge is denoted by arrows. Contours correspond to the MSSS image. Bottom: TGSS map with resolution $25.0^{\prime\prime}\times25.0^{\prime\prime}$ and rms noise 10~mJy/beam. The contours start at 3$\sigma$ level and are drawn at 30~mJy/beam $\times$ [1, 1.4, 2, 2.8, 4, 5.7, 8, 11, 16, 22, 32, 45, 64, 91, 128, 182]. 
\label{fig:a2199_all}}
\vspace{0.15in}
\end{figure*}

A2199 is a nearby cool-core cluster at $z=0.030$.   
The central dominant galaxy NGC 6166 hosts the unusual restarted radio source 3C338. 
The large-scale structure of 3C338 can be separated into two regions. The active region includes the core and two symmetric jets terminating in two faint hot spots. 
The older region is displaced to the south and consists of two extended steep-spectrum radio lobes connected by a bright filamentary structure \citep{Burns1983, Giovannini1998}. 
\cite{Burns1983} propose that the shift between the large-scale structure and the restarted jets could indicate a motion of the central AGN inside the galaxy. \cite{Vacca2012}, on the other hand, suggest that the displacement could also be due to an interaction between the old radio lobes with bulk motions in the surrounding medium caused by the sloshing of the cluster core \citep{Markevitch2007}.

The diffuse radio lobes clearly coincide with two large X-ray cavities 25 kpc either side of the nucleus \citep[e.g.][]{Johnstone2002, Gentile2007, Nulsen2013}. 
The low-frequency data at 330~MHz shows the presence of an extension to the south that corresponds to a third X-ray cavity \citep{Gentile2007}. 
Similar to 2A0335+096 and the Perseus cluster, A2199 hosts an isothermal shock - a sharp drop in X-ray brightness with a pressure jump but no temperature change across it \citep{SandersFabian2006}. It is seen in X-rays as a surface brightness edge $100^{\prime\prime}$ SE from the cluster center and \cite{Nulsen2013} argue that it is most probably a result of a shock produced by an older, significantly more powerful AGN outburst than the one that produced the current outer radio lobes and cavities.

Our 140~MHz MSSS map is shown in Figure~\ref{fig:a2199_all}. 
The new low-frequency data are consistent with previously published results.
In general, the emission at this frequency appears slightly more extended than it shows on the maps at 327~MHz presented by \cite{Gentile2007} and \cite{Birzan2008}, even though the comparison is difficult due to the lower angular resolution of our map.	
Interestingly, at a comparable resolution, the emission on the MSSS map also extends further from the center than on the 74 MHz image of \cite{Gentile2007}. 
This is most probably a result of the low sensitivity of VLA at 74~MHz at the time of their observation.

The radio emission at 140~MHz fully covers the southern X-ray cavity (Figure \ref{fig:a2199_all}).
Similar to the observed morphology at higher frequency maps \citep{Birzan2008, Gentile2007}, the eastern lobe appears well confined. 
We do not detect radio signatures of the powerful outburst which presumably created the isothermal shock.
On the other hand, since the SE X-ray edge and the outer radio lobes have comparable scales, it seems consistent that the shock front restricts the expansion of the eastern lobe further east.

The plume extending north-ward from the northern tip of the western lobe, seen at the 330 MHz by \cite{Gentile2007} and \cite{Birzan2008}, is also present at 140~MHz. 
It appears diluted in our map due to the lower resolution, which is also the case at 74 MHz \citep{Gentile2007}.  
\cite{Nulsen2013} argue that the indistinct appearance of the $100^{\prime\prime}$ front to the northwest would then imply that the turbulence is greater there than on the southeastern side of the cluster center.
A more disturbed environment could well explain why the western lobe has a more untypical shape than the eastern one.
\cite{Nulsen2013} show that a shell of denser gas is situated to the west of the western lobe at 4.9~GHz.
This denser structure seems to inhibit the growth of the lobe towards the west. However, since the shell is discontinued to the north, it leaves room for the lobe to expand in this direction.

In Figure \ref{fig:a2199_all} we show the TGSS image only for completeness.
On this image, the shape of the system looks very different.
The lobes do not have their expected morphology and orientation, and only one central peak of emission is observed.
Based on our experience calibrating the MSSS data of A2199, where we were getting similar deformations, we believe that the TGSS image suffers from significant distortions due to phase calibration problems.
As noted earlier in Section \ref{label:reprocessing}, a deeper full-track LOFAR pointing for A2199 is already available. 
This observation is currently being analyzed and will be published separately.

\section{Discussion and Conclusions}
\label{sec:conclusions}

With facilities such as LOFAR, GMRT, and the Murchison Widefield Array (MWA) online, uniform surveys at low radio frequencies are now becoming available for a large fraction of the sky. The angular resolution and sensitivity of these first all-sky, low-frequency surveys are well-matched to studies of extended, steep spectrum diffuse emission in cluster feedback systems and will ultimately provide the larger samples necessary for statistical population studies. In this pilot work, we have employed data from the first all-sky imaging surveys with LOFAR (MSSS at 140\,MHz) and the GMRT (TGSS ADR1 at 150\,MHz) to study a sample of known, strong AGN feedback sources drawn from \cite{Birzan2008}. Combining data from both surveys, we have searched for the presence of extended, diffuse emission not seen previously at higher radio frequencies and possibly associated with relic emission from previous AGN activity. We have also computed total low-frequency fluxes for the sources in the sample in order to test the well-known correlation between high frequency radio flux and the power required to create the cavity structures seen in the X-ray \citep{Birzan2008, Rafferty2006}. 

For both MSSS and TGSS surveys, images were created and examined for each of the objects in the sample. In the case of the TGSS data, images were extracted from the default mosaics provided as part of the standard TGSS ADR1 data products \citep{Intema2017} and no additional processing was required. The LOFAR images used in this work were based on a preliminary data release of the MSSS all-sky survey \citep{Heald2015}. Although the MSSS survey data included baseline lengths out to 100\,km, the initial imaging products produced during the development of the survey processing utilized a cutoff in baseline length to obtain a nominal angular resolution of $\sim2^{\prime}$. In this work, we have described a simple reprocessing strategy that takes advantage of data from these longer baselines to both improve the noise and angular resolution of the resulting maps. We have applied this procedure to all objects in the sample which showed evidence of extended emission in the default, low-resolution maps resulting in improved images with angular resolutions of $\sim25^{\prime\prime}$. For the subset of resolved objects, these improved MSSS maps were used to characterize the presence of any diffuse, extended emission and in all subsequent comparisons to TGSS, high-frequency radio, and X-ray images.

Based on the resulting TGSS and MSSS maps, we have measured the total radio fluxes for all sources; however, due to differing sky coverages, not all sources were visible in both surveys. For the subset of overlapping sources, we have compared the derived fluxes between the two surveys and find good agreement. The exceptions are Perseus, where the radio mini-halo shows complicated spatial structure over a large area, and A478 and ZW3146, which are not well resolved in the MSSS maps and blended with nearby sources. Given the agreement in flux scales and the more complete sky coverage of the TGSS sample, we have utilized the TGSS fluxes in all subsequent analysis of the $P_{\rm cav} - L_\nu$ correlation known from higher frequency data. Estimates for the values of $P_{\rm cav}$ based on \textit{Chandra} X-ray data were taken from the literature \citep{Rafferty2006} and compared to the radio luminosity at 148\,MHz as inferred from the measured TGSS flux values. At this frequency, we find a correlation of the form $P_{\rm cav} \propto L_{148}^{0.51 \pm 0.14}$, which is in good agreement with the result found by \cite{Birzan2008} at 327\,MHz. 

We find a relatively large scatter in the 148\,MHz correlation of $\sim$0.85~dex which again is consistent with the value obtained at 327\,MHz by \cite{Birzan2008}. This large scatter makes the derived correlation difficult to use as a reliable proxy for jet power in sources where only radio data is available. In the previous high frequency study, \cite{Birzan2008} found that the scatter in the correlation could be significantly reduced by including the effects of spectral aging, although in that analysis the lack of low-frequency data required a number of simplifying assumptions in order to constrain the break frequency. Including our new data points at 148\,MHz with the data at higher frequencies taken from \cite{Birzan2008}, we have examined the spectral energy distributions for our sample in an attempt to provide better constraints on the break frequencies and by extension an improved correction for the scatter in the observed correlation due to spectral aging. Unfortunately, our results show that including additional measurements at 148\,MHz alone is insufficient to fit a consistent, physically justified spectral aging model. We therefore conclude that to improve on the aging correction of \cite{Birzan2008} will likely require additional data below 148\,MHz, such as with the LOFAR LBA at 30 -- 90\,MHz, and data with sufficient angular resolution to clearly resolve emission from the core and lobe regions. 

In four of the sources, Perseus, 2A0335+096, MS0735.6+7421, and A2199, the reprocessed MSSS images show clear evidence for extended, diffuse emission at 140\,MHz. We have compared the observed morphology of the extended, diffuse emission in the MSSS maps with corresponding radio images from TGSS and the VLA, as well as X-ray surface brightness and residual maps based on archival \textit{Chandra} data. In Perseus, we easily resolve the well-known, radio lobes observed at higher frequencies \citep{Pedlar1990} and associated with the inner cavities seen in X-ray images \citep{Fabian2000,Fabian2006}. The MSSS image also clearly recovers the more extended halo structure seen in previous deep radio maps of Perseus out to radii of $\sim100$\,kpc \citep{Burns1992, Sijbring1993, DeBruyn2005}. The overall morphology of the low-frequency radio halo is well-correlated with a number of arc-like edges and ripples visible in the unsharp-masked X-ray image \citep{Fabian2006} reinforcing the current picture whereby the X-ray and radio structures share a common origin in the recurrent AGN activity of Perseus A.

We find a similar situation in the well-known feedback system 2A0335+096. Although the overall morphology of the low-frequency emission present in the reprocessed MSSS image confirms the structure visible in the TGSS maps and hinted at in higher frequency VLA maps \citep{Birzan2008}, the diffuse emission detected in the MSSS map is more extended and exhibits a rough correspondence to the large-scale depressions seen in the X-ray map outside the inner $\sim20$\,kpc \citep{Sanders2009}. A bridge of low-frequency emission is also observed connecting the core of 2A0335+096 with a bright peak of low-frequency emission $\sim110$\,kpc to the NW, parallel to the axis of the inner cavity system. This peak does not appear to be associated with a known radio source and exhibits a steep spectral index of $\alpha \sim -1.8$ consistent with remnant plasma from previous AGN activity. Assuming this emission originated in the core of 2A0335+096, we estimate an age of $\sim150$\,Myr for the original outburst based on estimates of the sound speed in the cluster core \citep{Birzan2004}. This emission peak, however, does not seem to correlate with an obvious depression in the X-ray map.

For the remaining two resolved sources in our sample, MS0735.6+7421 and A2199, we find the low-frequency emission observed in the MSSS maps corresponds closely with what has been seen at 327~MHz. 
In both the TGSS and MSSS images for MS0735.6+7421, the low-frequency emission clearly traces the large-scale, X-ray cavities seen in the X-ray and at higher radio frequencies \citep{McNamara2005}. The low-frequency emission shows no evidence for a strong central core and is completely dominated by emission in the lobes corresponding to the X-ray cavities. We find no evidence for more extended, diffuse low-frequency emission outside the well-known cocoon shock in this system \citep{McNamara2009, Vantyghem2014}, implying that the radio plasma is fully enclosed by the shock. In the case of A2199, the overall morphology of the diffuse emission in both TGSS and MSSS maps agree and is consistent with the structures seen in the 330\,MHz map of \cite{Birzan2008}. Compared to the TGSS image, the final, reprocessed MSSS map has higher quality and angular resolution, which allows us to clearly resolve two peaks of low-frequency emission coincident with the two main X-ray cavities observed in the core along the jet axis \citep{Nulsen2013}, as well as extensions to the S and NW corresponding to an additional cavity and surface brightness jump seen in the X-rays, respectively \citep{Gentile2007, Nulsen2013}. Taken all together, the morphology of the low-frequency emission in A2199 is consistent with the picture of a system having undergone at least two episodes of AGN activity.

While the images presented in this work demonstrate the potential of LOFAR to recover extended, diffuse emission, our analysis highlights some important caveats for future low-frequency radio studies of feedback.
First, the original sample of \cite{Birzan2008} were selected on the basis of containing well-defined cavity systems in the X-ray. With a few exceptions, these cavities and the higher frequency radio emission they enclose can be linked to a single epoch of AGN activity occurring over a fairly narrow range of outburst ages. The same can be said for the "ghost" cavities seen in systems such as Perseus \citep{Fabian2006} and A2597 \citep{Clarke2005} and found to contain lower frequency radio emission at 330\,MHz. At the lower frequency and moderate angular resolutions provided by TGSS and MSSS, however, the distribution and morphology of the observed low-frequency radio emission is quite complicated and not easily resolved into individual components that may be associated with well-defined episodes of AGN activity. As a result, radio flux measurements at these frequencies are difficult to place in the context of the $P_{\rm cav} - L_\nu$ correlation found for single episodes of AGN activity at higher frequency.

The situation is similarly complicated in the X-ray. As even the small sample of objects in this work illustrates, the well-defined surface brightness depressions or "cavities" associated with the radio emission at higher frequencies is harder to identify for older outbursts ($\ge100$\,Myr). Even for objects with deeper X-ray exposures such as Perseus and 2A0035+096, the morphologies of these outbursts are more complicated, difficult to disentangle from other features possibly related to shocks or core sloshing, and not always well-correlated with the low-frequency radio emission. This complexity is further exacerbated by the lack of sufficiently deep exposures for many of the nearby, strong feedback systems where we might expect to resolve the X-ray signatures of older AGN outbursts at larger radii (e.g., Perseus \cite{Fabian2006}). Taken altogether, these factors introduce considerable uncertainty in estimates for the value of $P_{\rm cav}$ associated with older or multiple AGN outbursts.

The LoTSS currently underway will produce images of the full accessible northern sky at angular resolutions of $5^{\prime\prime}$ and sensitivities of $\sim100\mu$Jy. When combined with new high frequency data from surveys such as the VLA Sky Survey \citep[VLASS;][]{VLASS2013}, we will have the radio data necessary to both spatially resolve and spectrally discriminate between different episodes of AGN activity for virtually all nearby feedback systems. In the near-term, these new radio data can be matched to deeper X-ray exposures with \textit{Chandra} and \textit{XMM} to better attempt to separate outburst related signatures from other physical processes operating in the cluster cores. On the longer term, upcoming missions such as \textit{eRosita} \citep{eRosita2012} and \textit{Athena} \citep{Athena2013} will yield larger samples of potential feedback systems as well as information about velocity motions in the gas from high spectral resolution emission line studies. The combination of these new radio surveys, in particular the low-frequency data from LoTSS, and improved X-ray data will allow us to build up a picture of the integrated effects of AGN output on the surrounding environment for a large sample of systems over timescales of several 100\,Myr.

\begin{acknowledgements}

GDK acknowledges support from NOVA (Nederlandse Onderzoekschool voor Astronomie).
BNW acknowledges support from the NCN OPUS UMO-2012/07/B/ST9/04404 funding grant.
DDM acknowledges support from ERCStG 307215 (LODESTONE).
LOFAR, the Low Frequency Array designed and constructed by ASTRON, has facilities in several countries, that are owned by various parties (each with their own funding sources), and that are collectively operated by the International LOFAR Telescope (ILT) foundation under a joint scientific policy. 
GMRT is run by the National Centre for Radio Astrophysics of the Tata Institute of Fundamental Research.
This research has made use of the NASA/IPAC Extragalactic Database (NED) which is operated by the Jet Propulsion Laboratory, California Institute of Technology, under contract with the National Aeronautics and Space Administration. 
We have also used SAOImage DS9, developed by Smithsonian Astrophysical Observatory. 
        
\end{acknowledgements}

\bibliographystyle{aa} 
\bibliography{MSSS_Feedback_Clusters_GDK} 

\begin{appendices}

\section{}
 \label{appendix:sed}

\begin{figure*}[t]
\begin{center}
\includegraphics[width=2.3in]{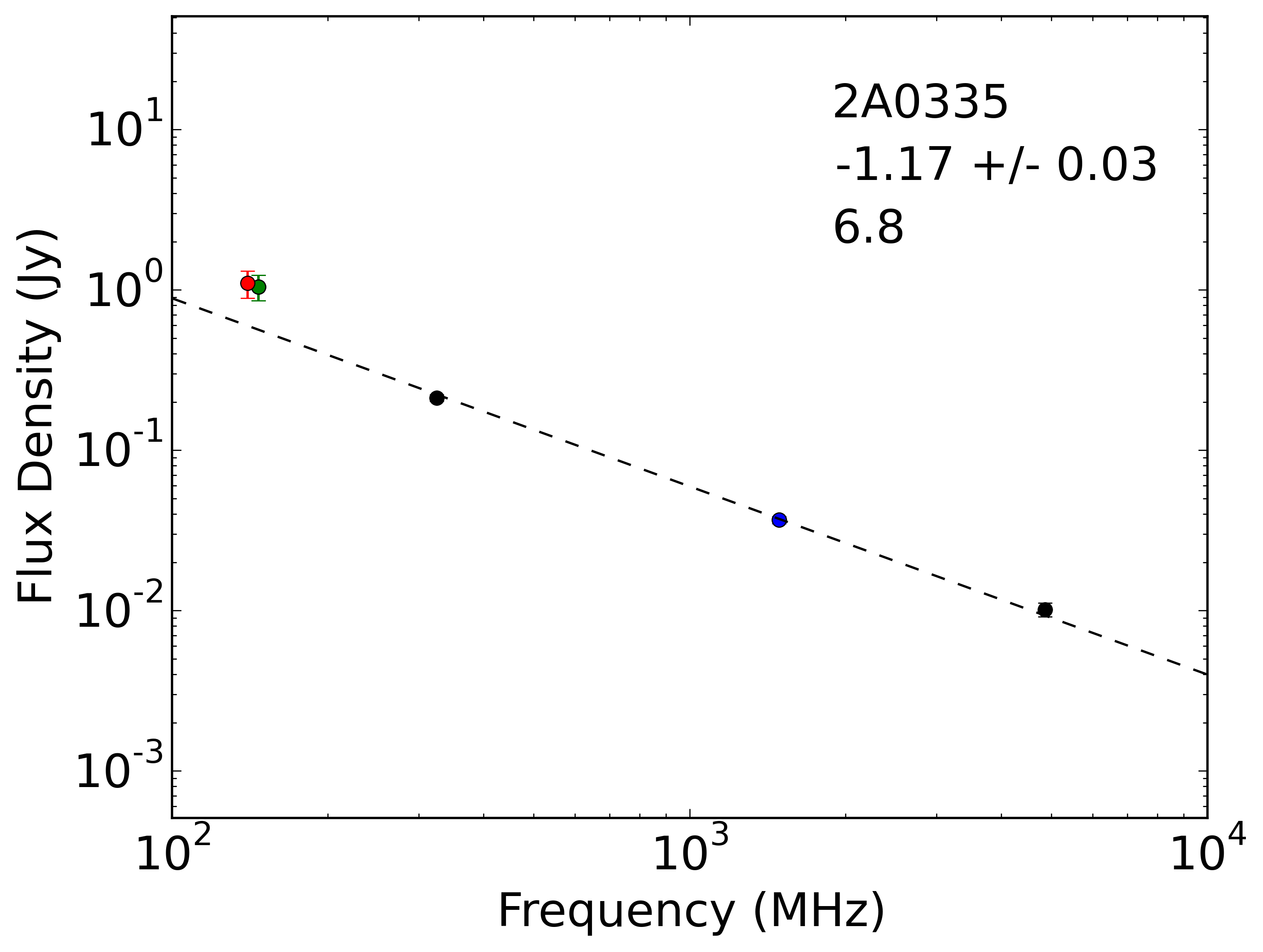}
\includegraphics[width=2.3in]{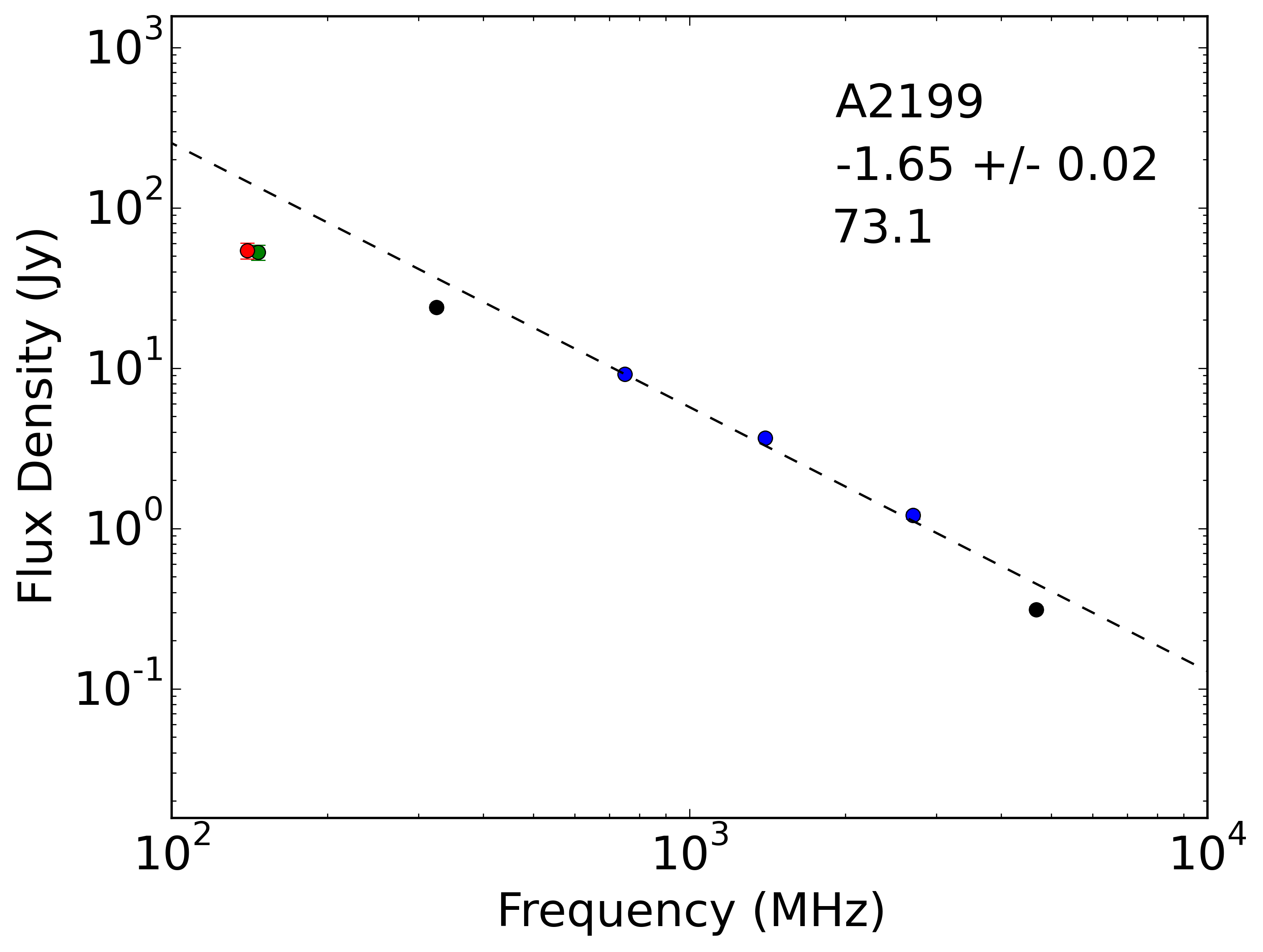}
\includegraphics[width=2.3in]{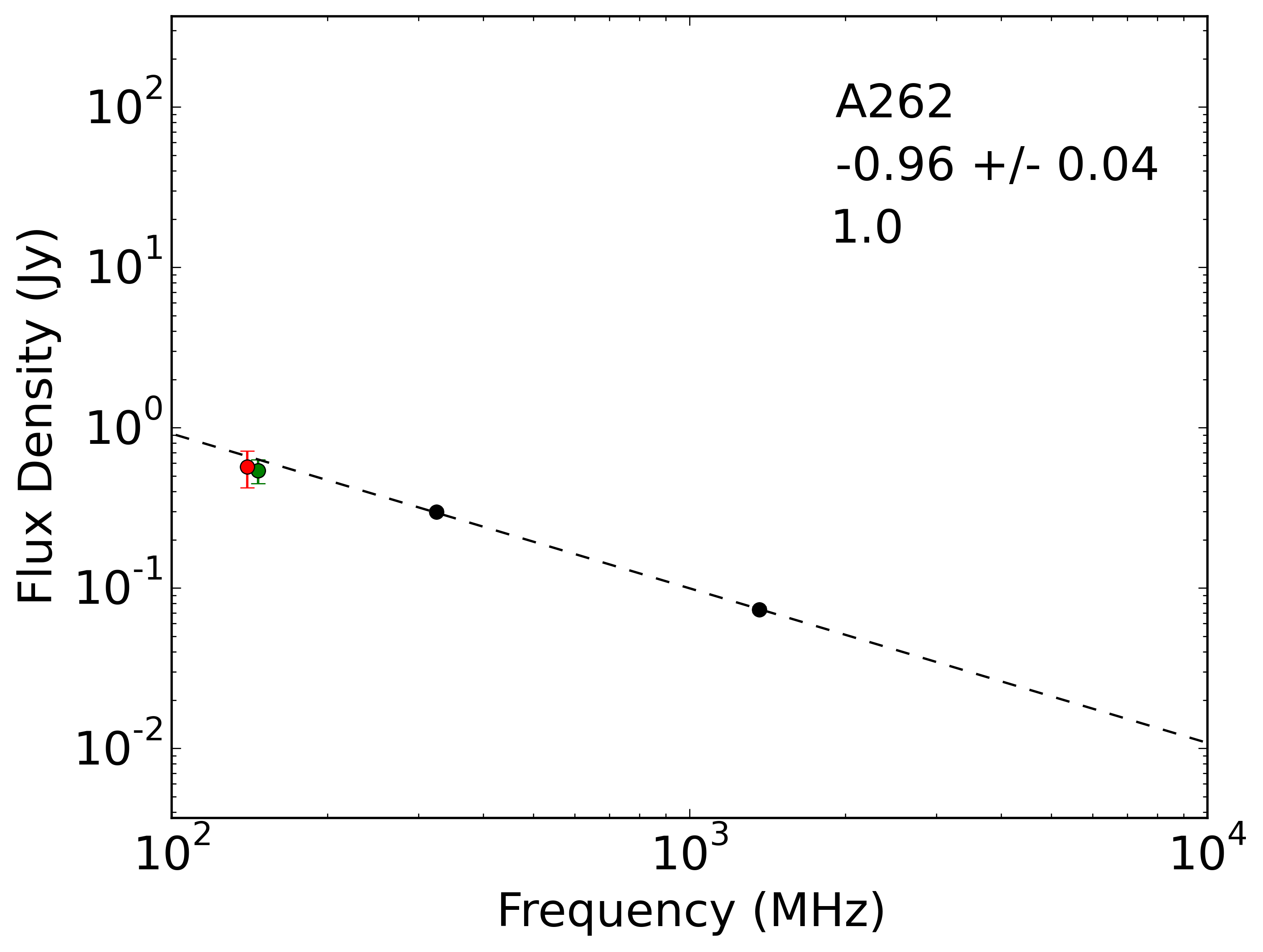}

\includegraphics[width=2.3in]{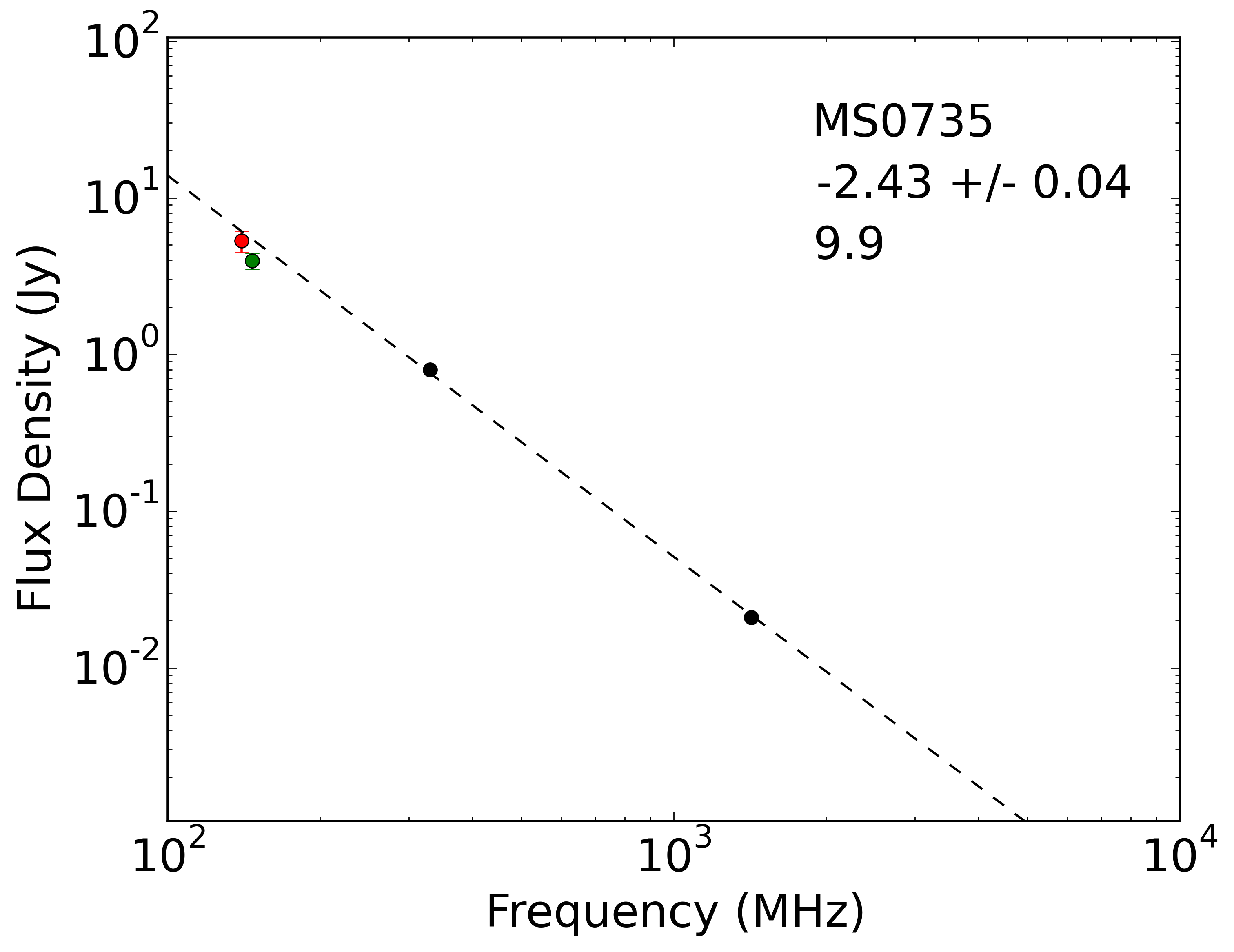}
\includegraphics[width=2.30in]{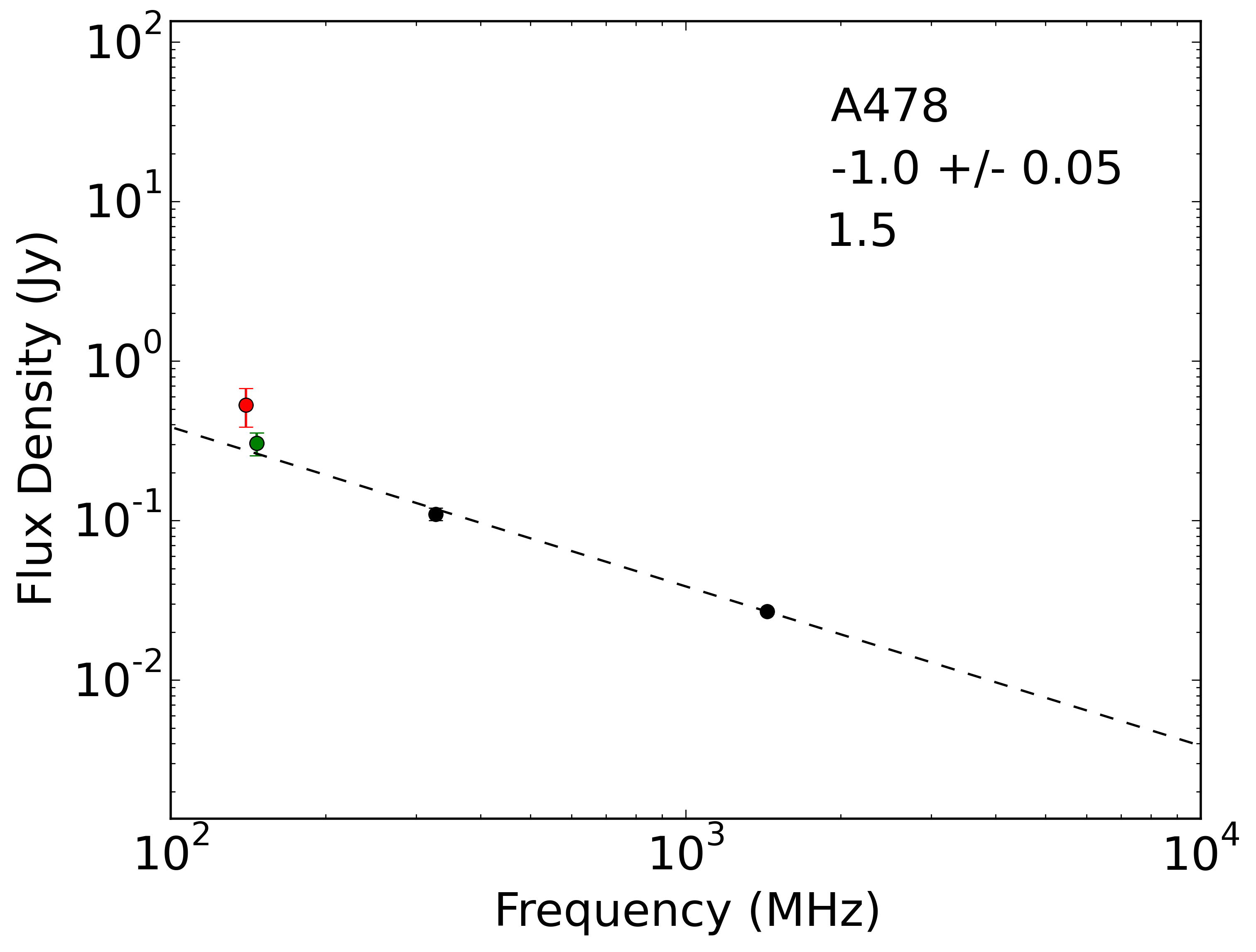}
\includegraphics[width=2.30in]{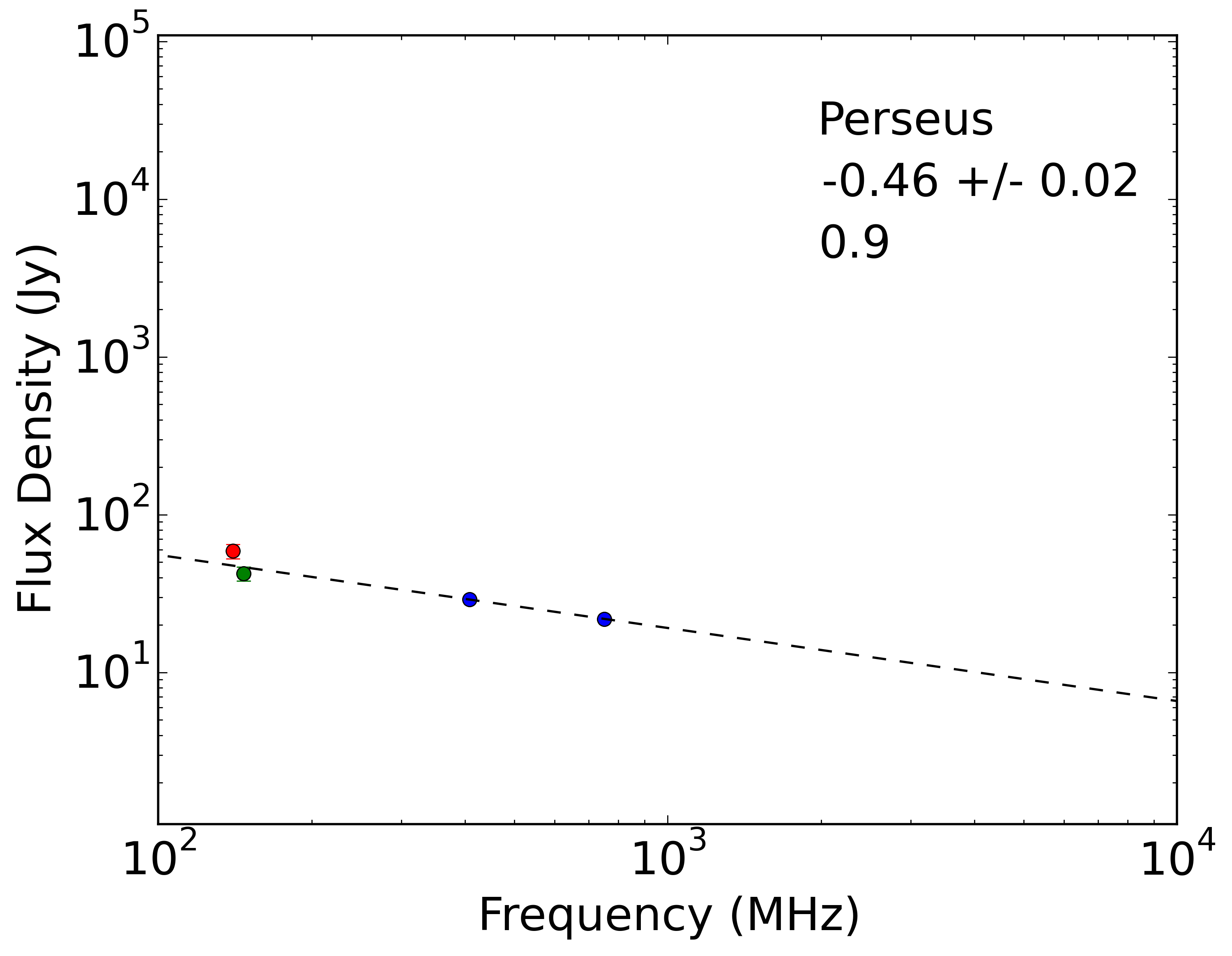}

\includegraphics[width=2.30in]{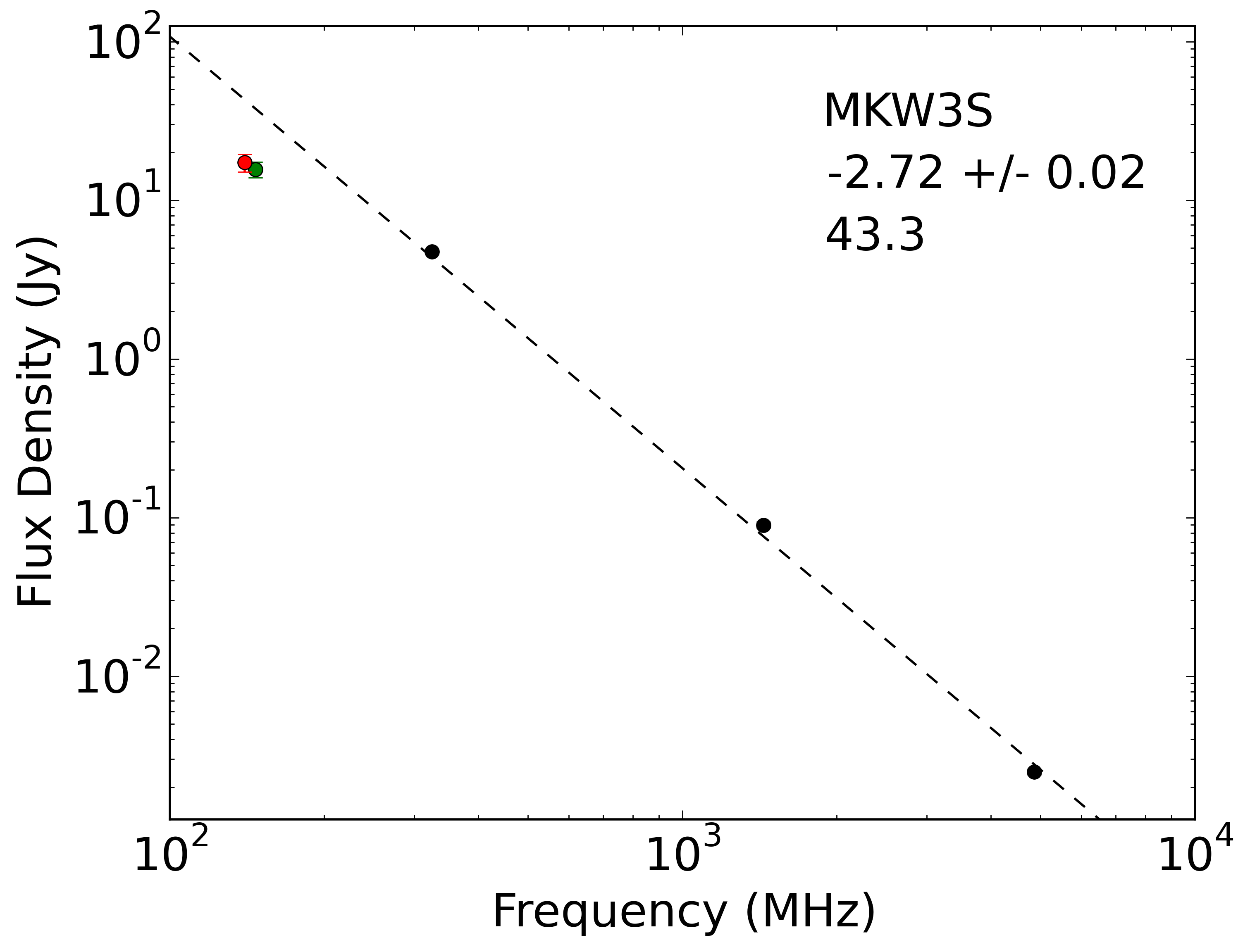}
\includegraphics[width=2.30in]{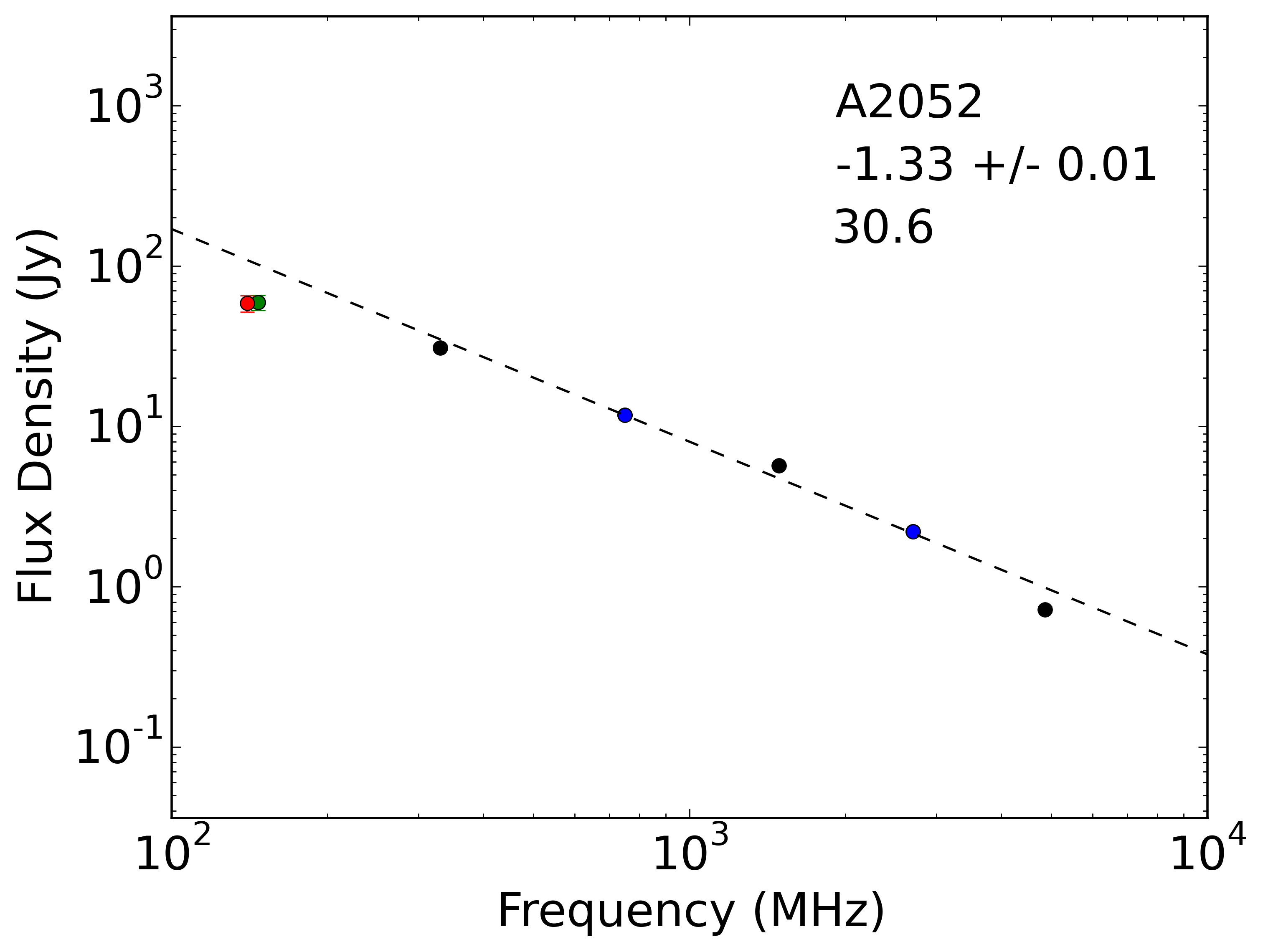}
\includegraphics[width=2.30in]{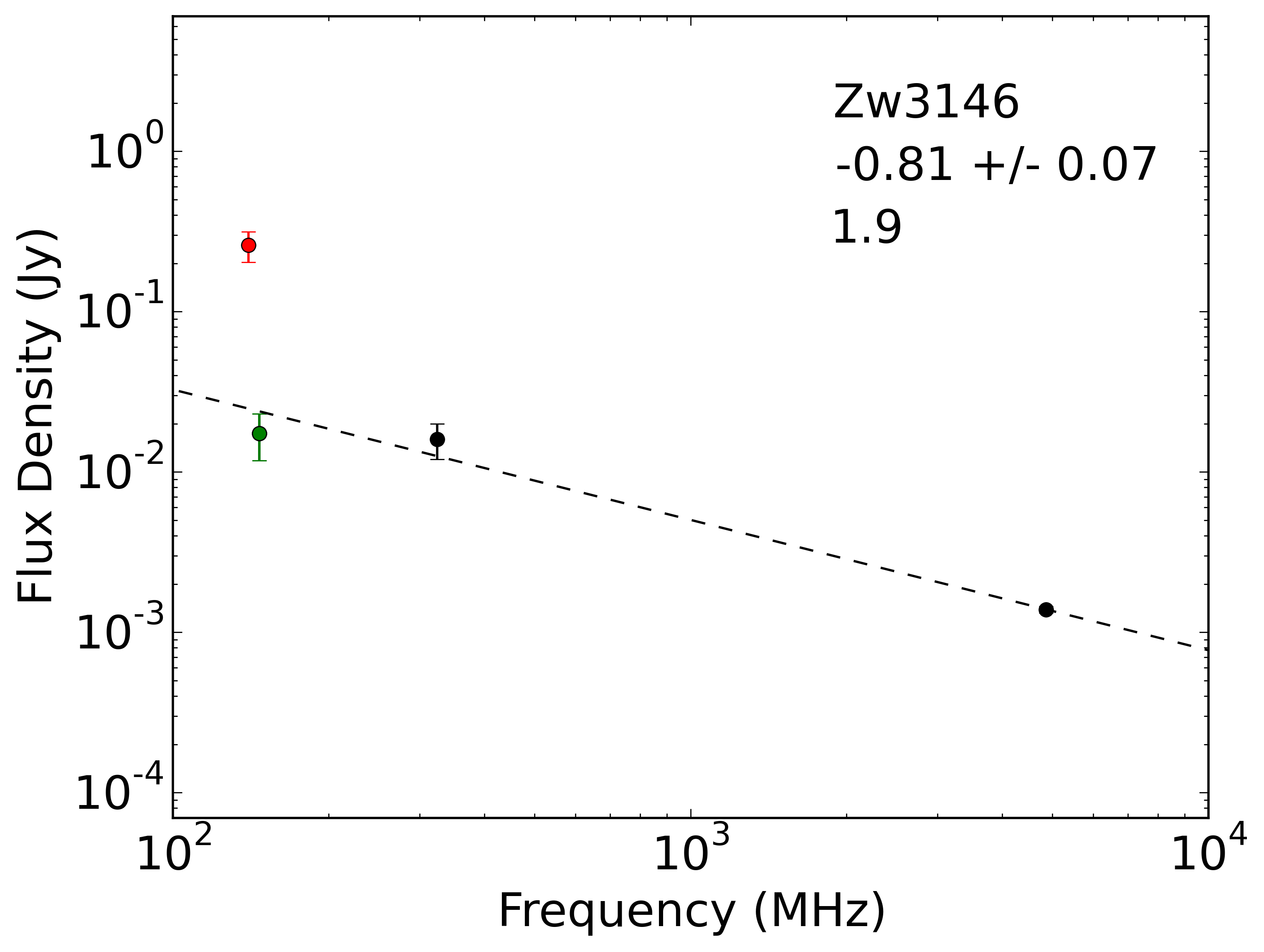}

\includegraphics[width=2.30in]{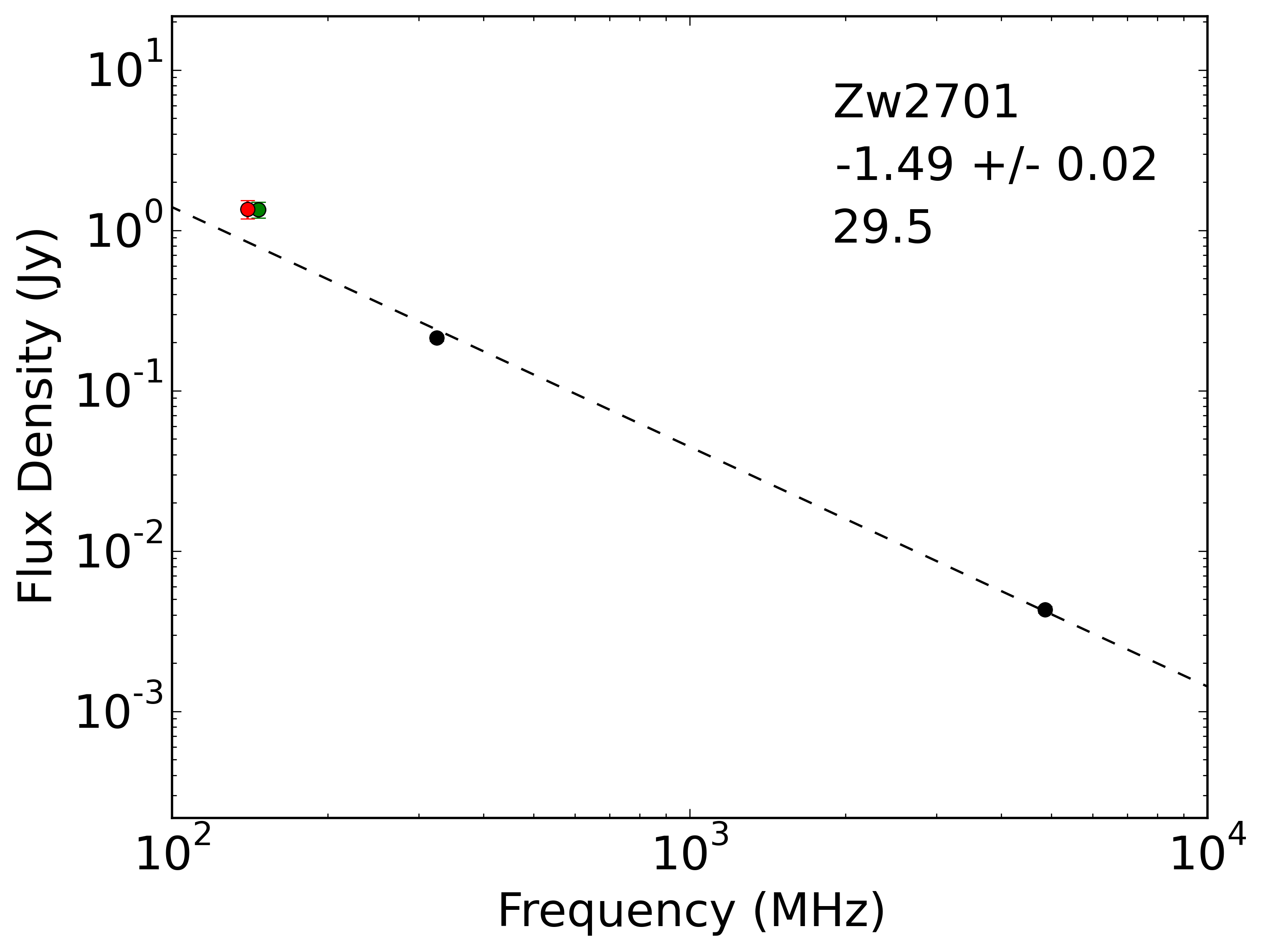}
\includegraphics[width=2.30in]{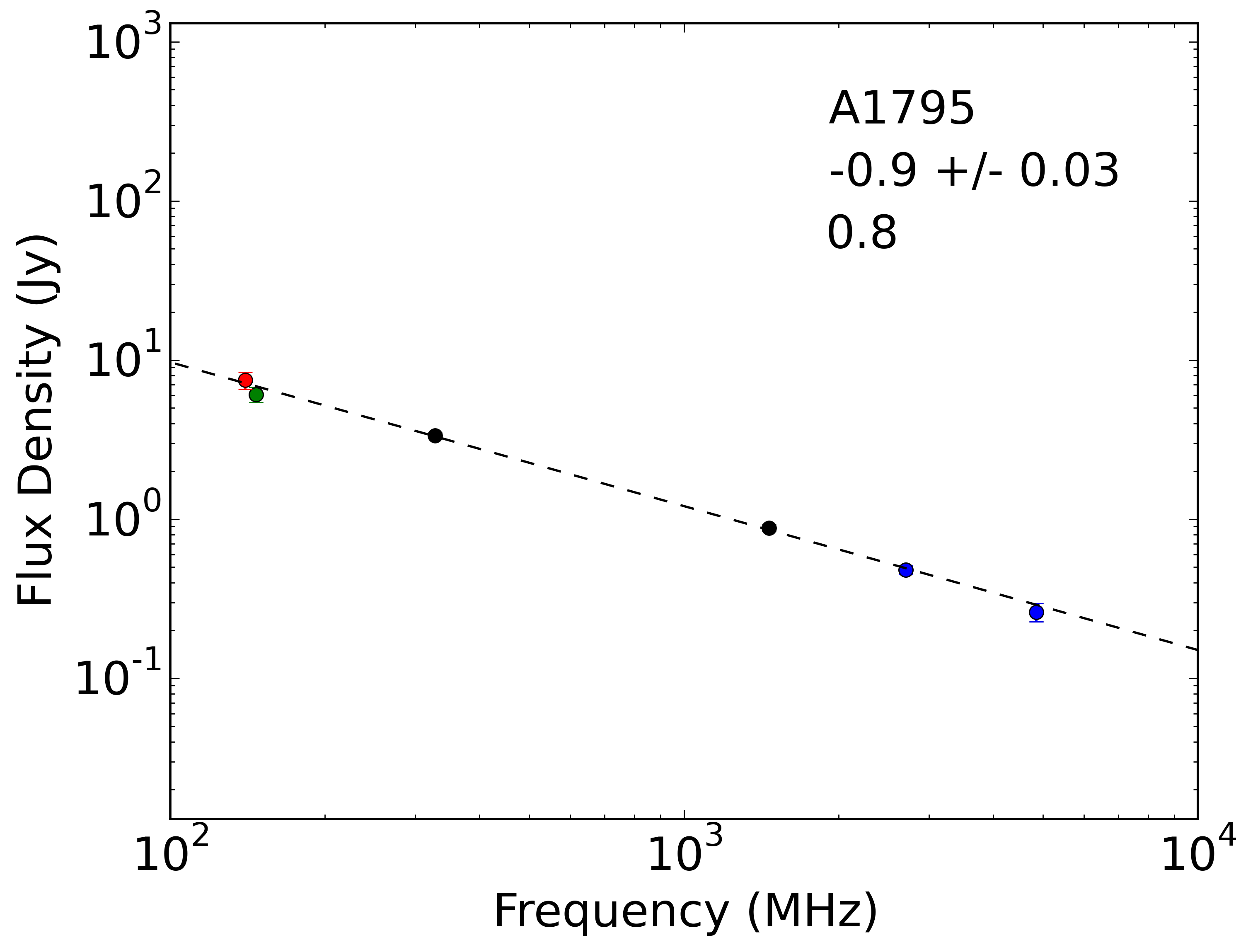}
\includegraphics[width=2.30in]{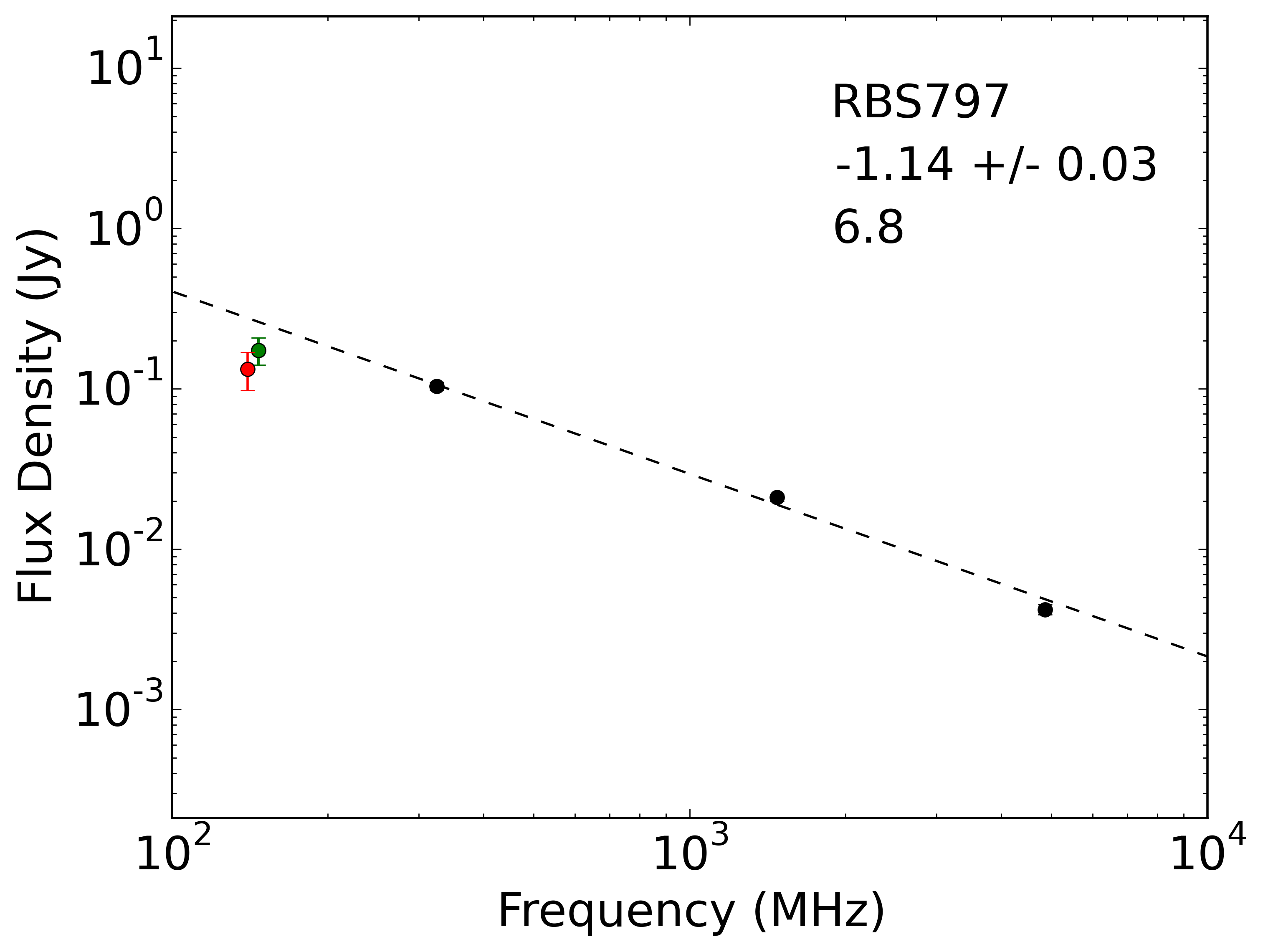}

\includegraphics[width=2.30in]{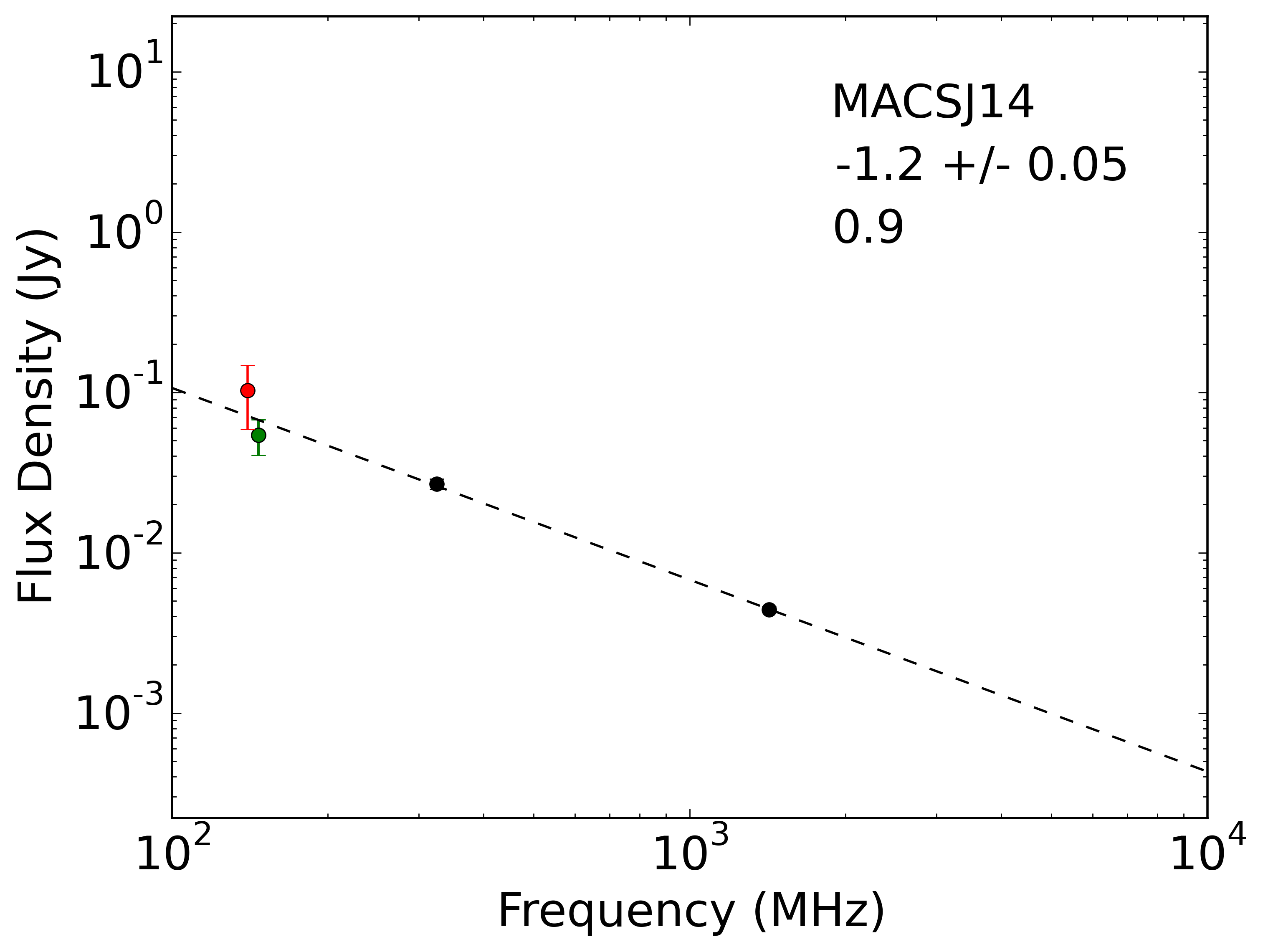}
\includegraphics[width=2.30in]{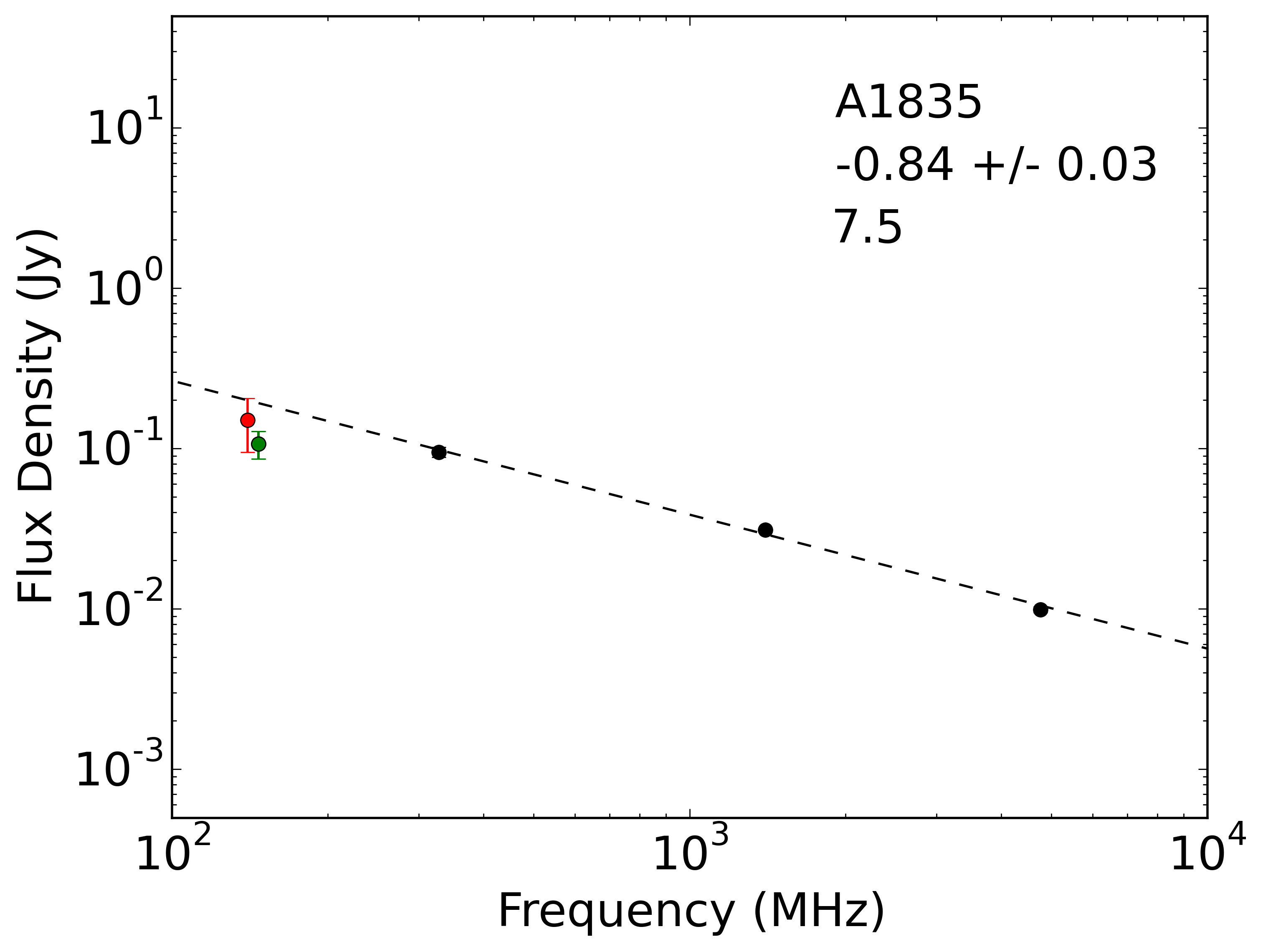}

\end{center}
\caption{\small SED power law fits for the source in the MF-14 sample. The MSSS (red) and TGSS (green) points come from our measurements. The VLA (black) points are from \cite{Birzan2008}. The higher frequency literature points are presented in blue. The dashed line shows the best fitted power law. The MSSS points are not taken into account for the fit. The value at the top right corner of each plot indicates the derived spectral index with its error and the reduced chi-squared of the fit.
\label{fig:sed_msss_tgss_vlss}}
\vspace{0.15in}
\end{figure*}

\newpage

\begin{figure*}[t]
\begin{center}
\includegraphics[width=2.3in]{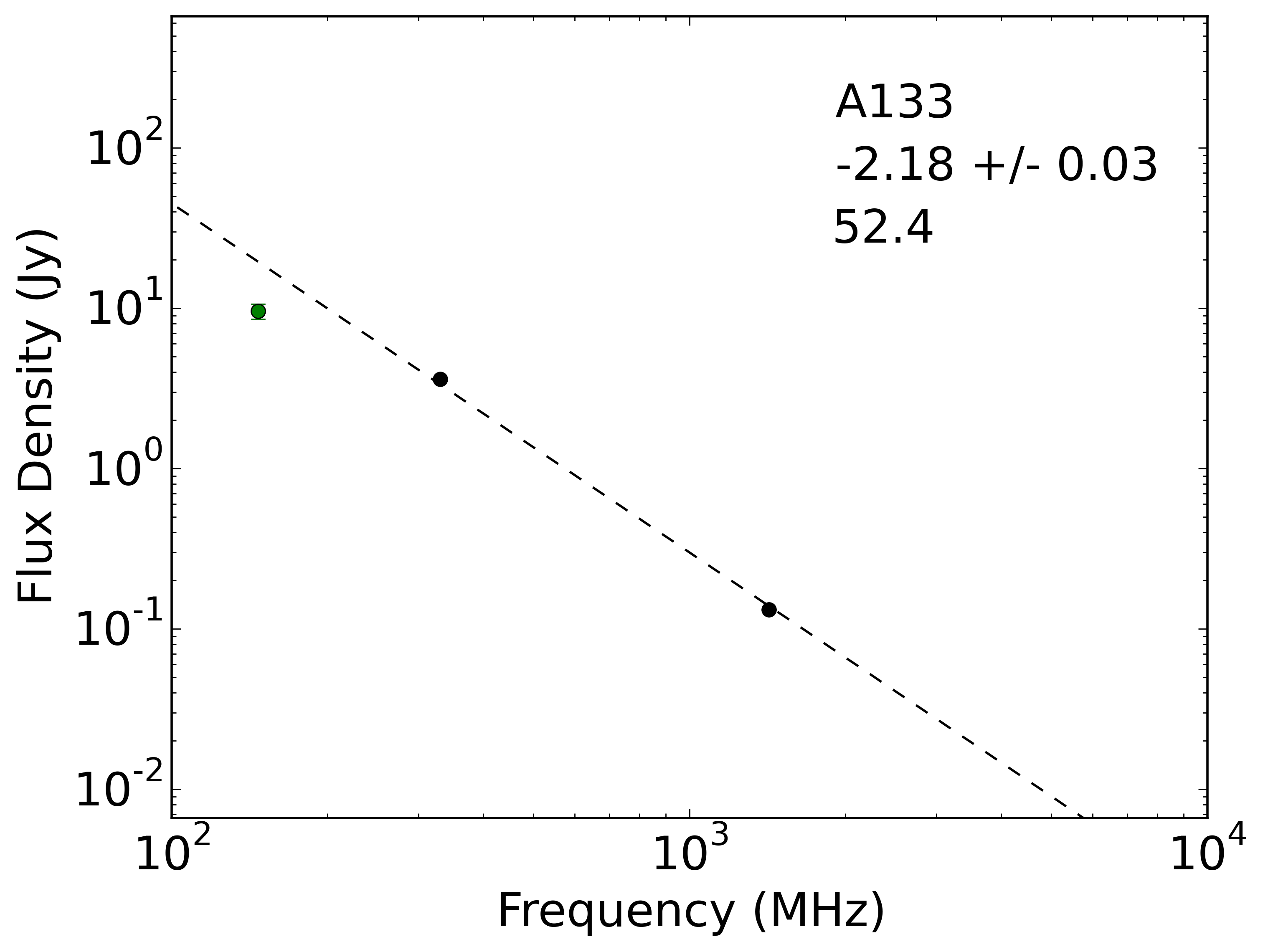}
\includegraphics[width=2.3in]{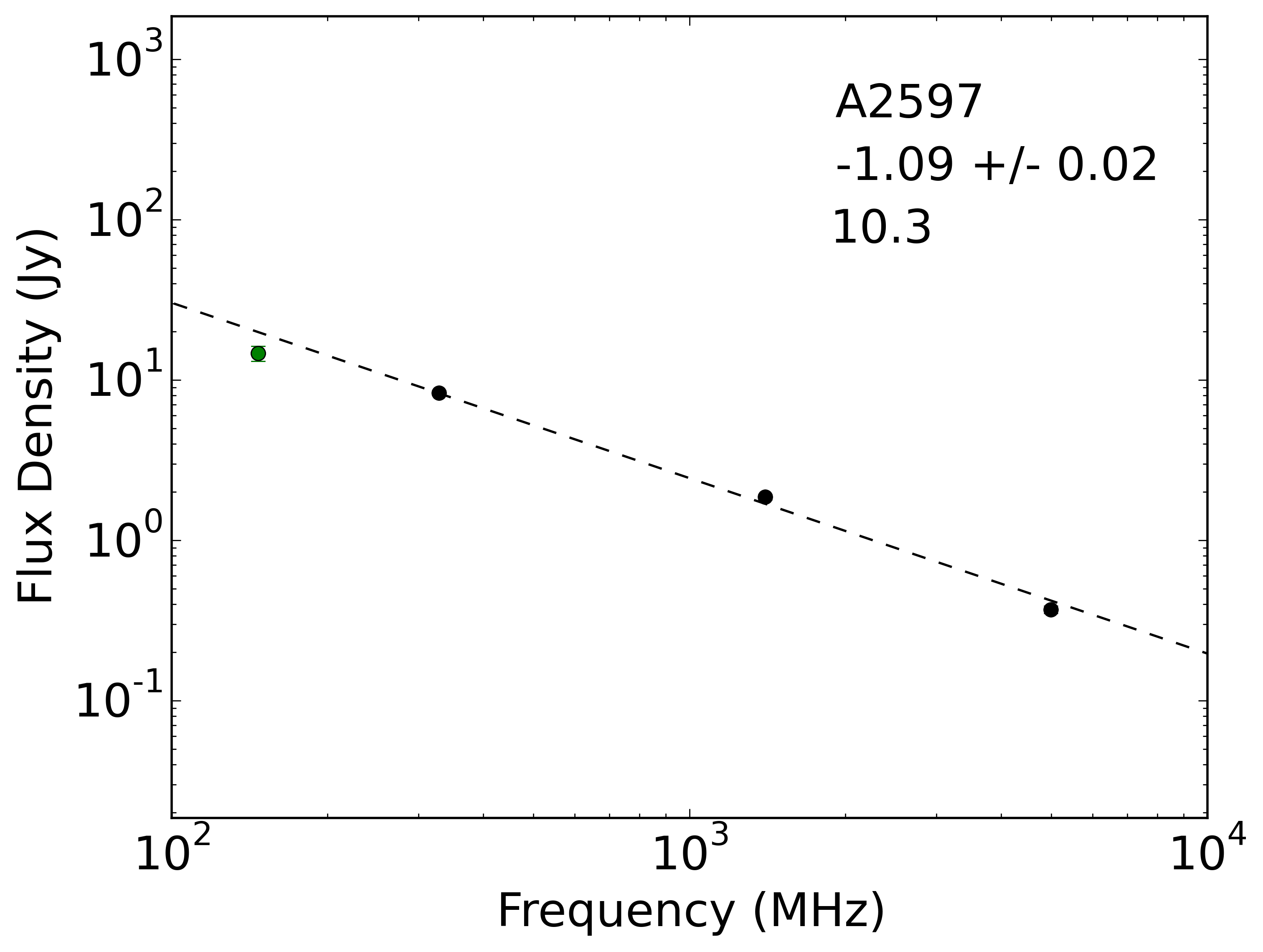}
\includegraphics[width=2.3in]{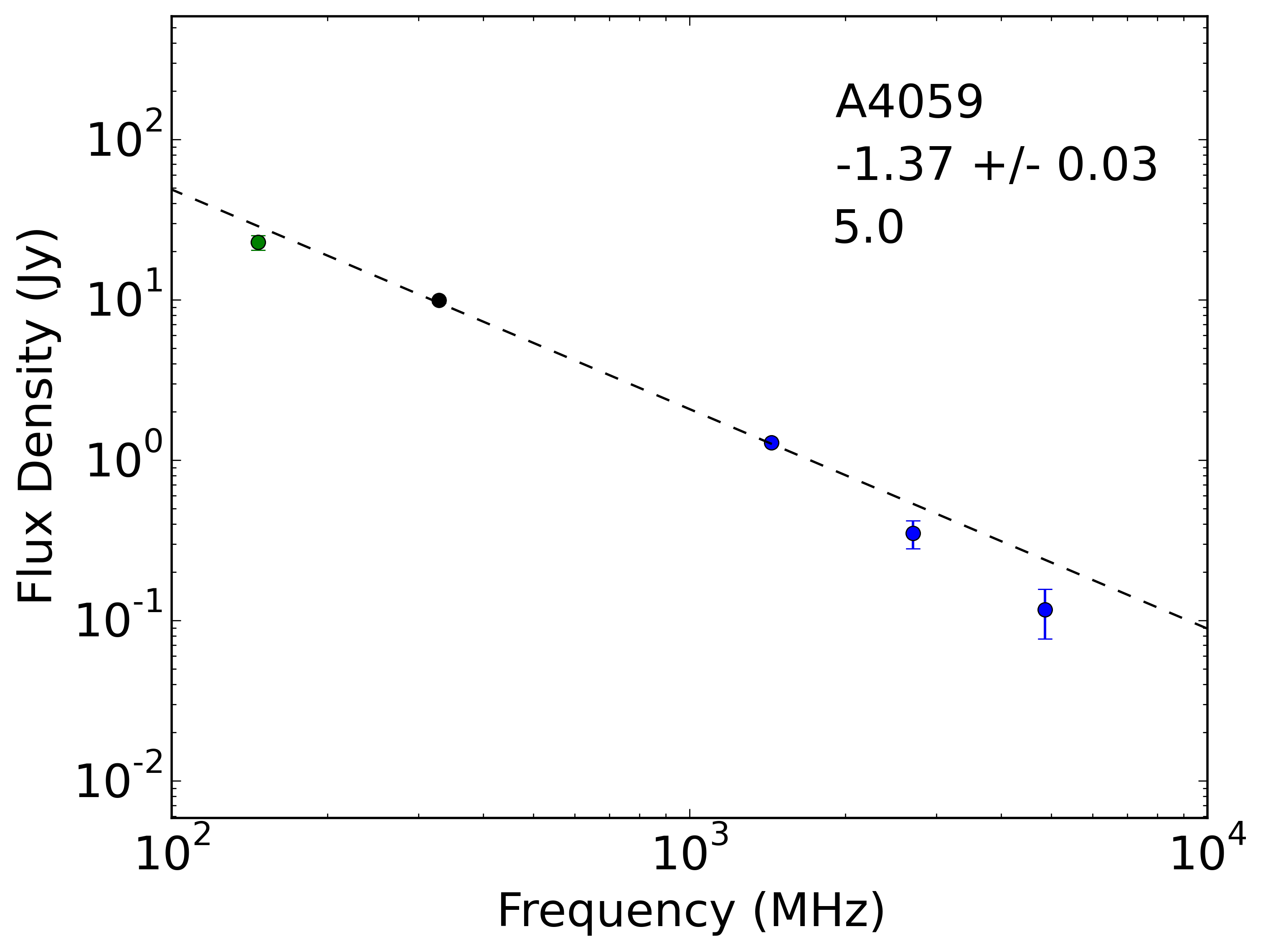}

\includegraphics[width=2.3in]{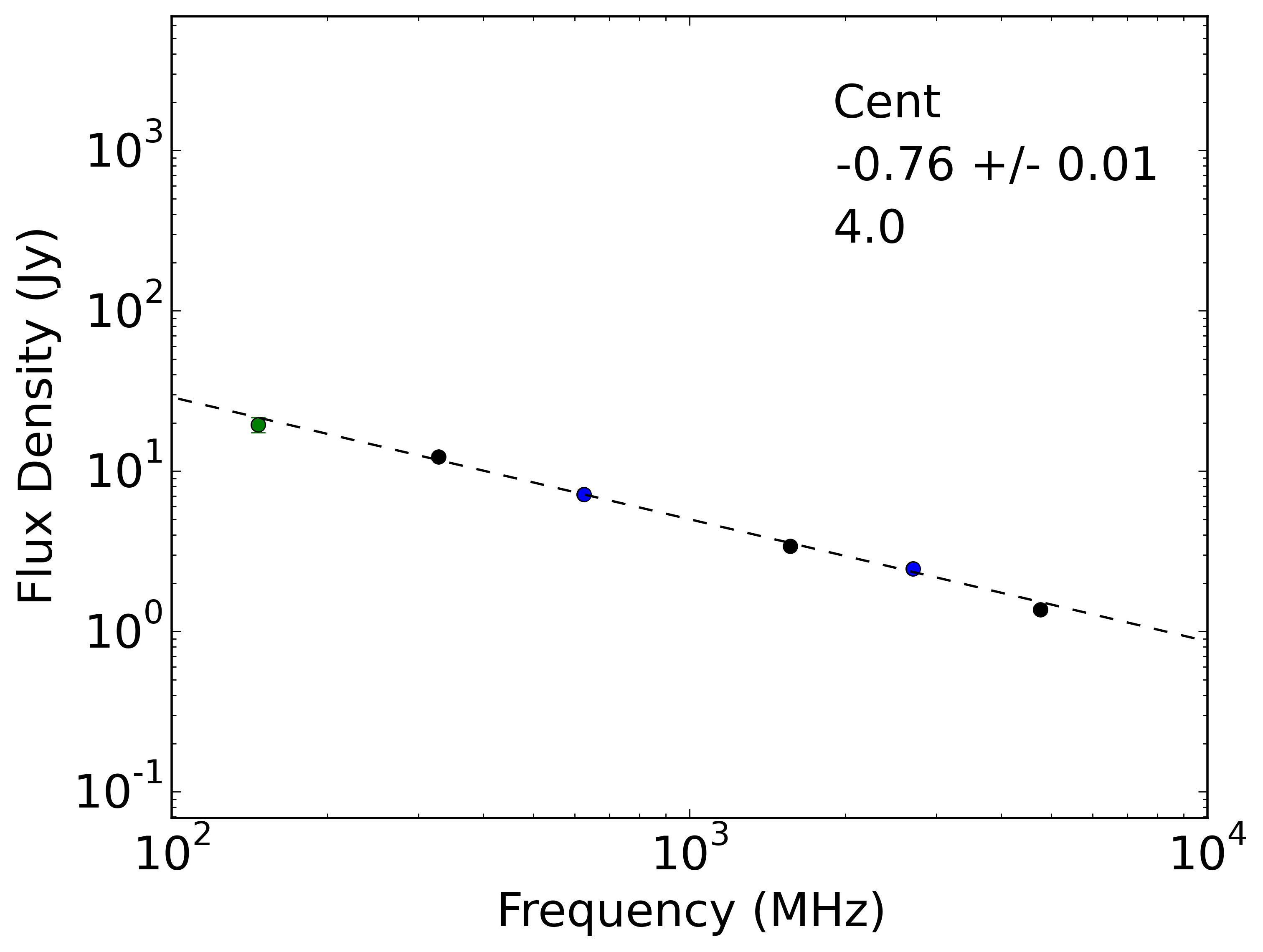}
\includegraphics[width=2.3in]{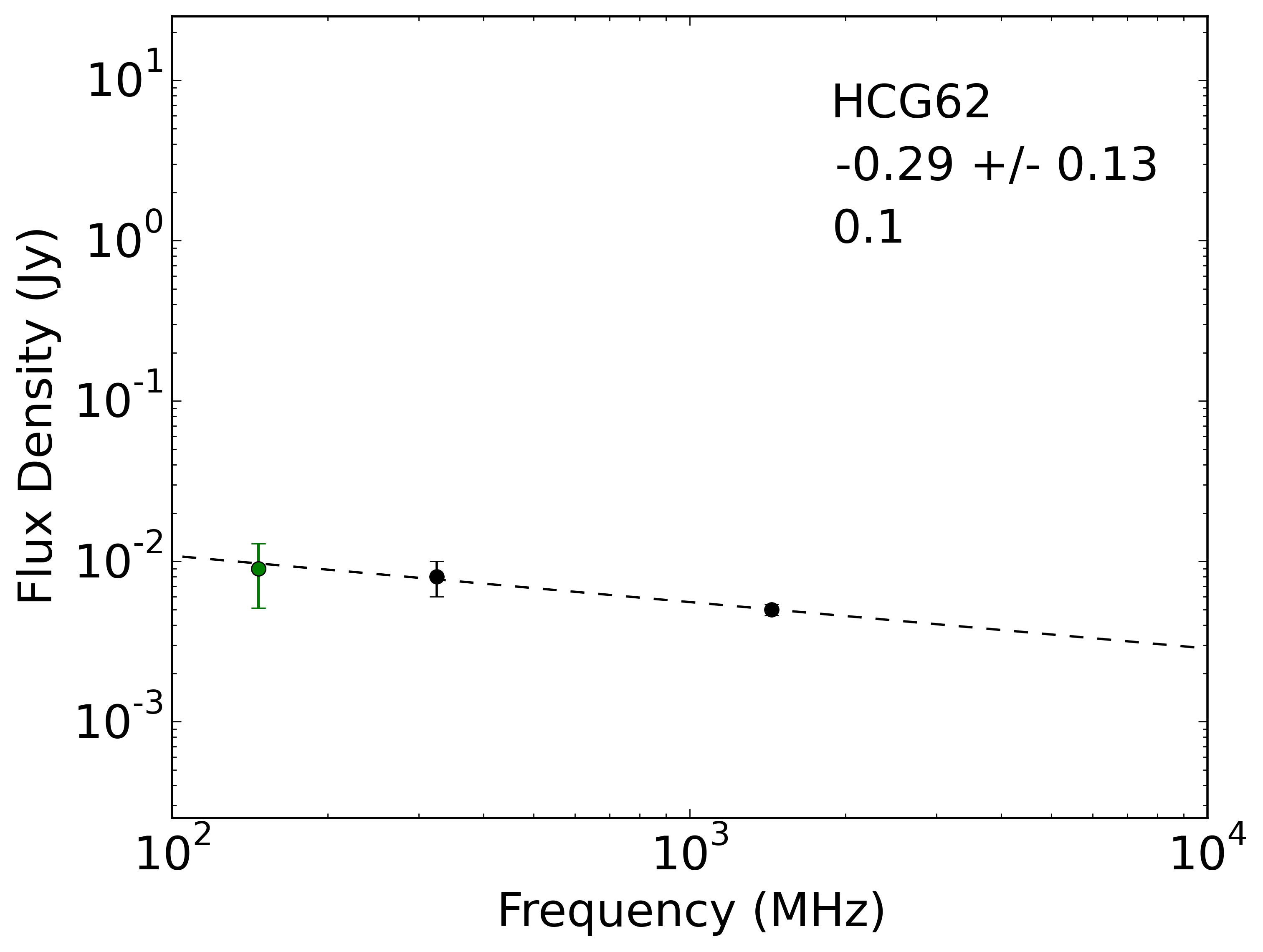}
\includegraphics[width=2.3in]{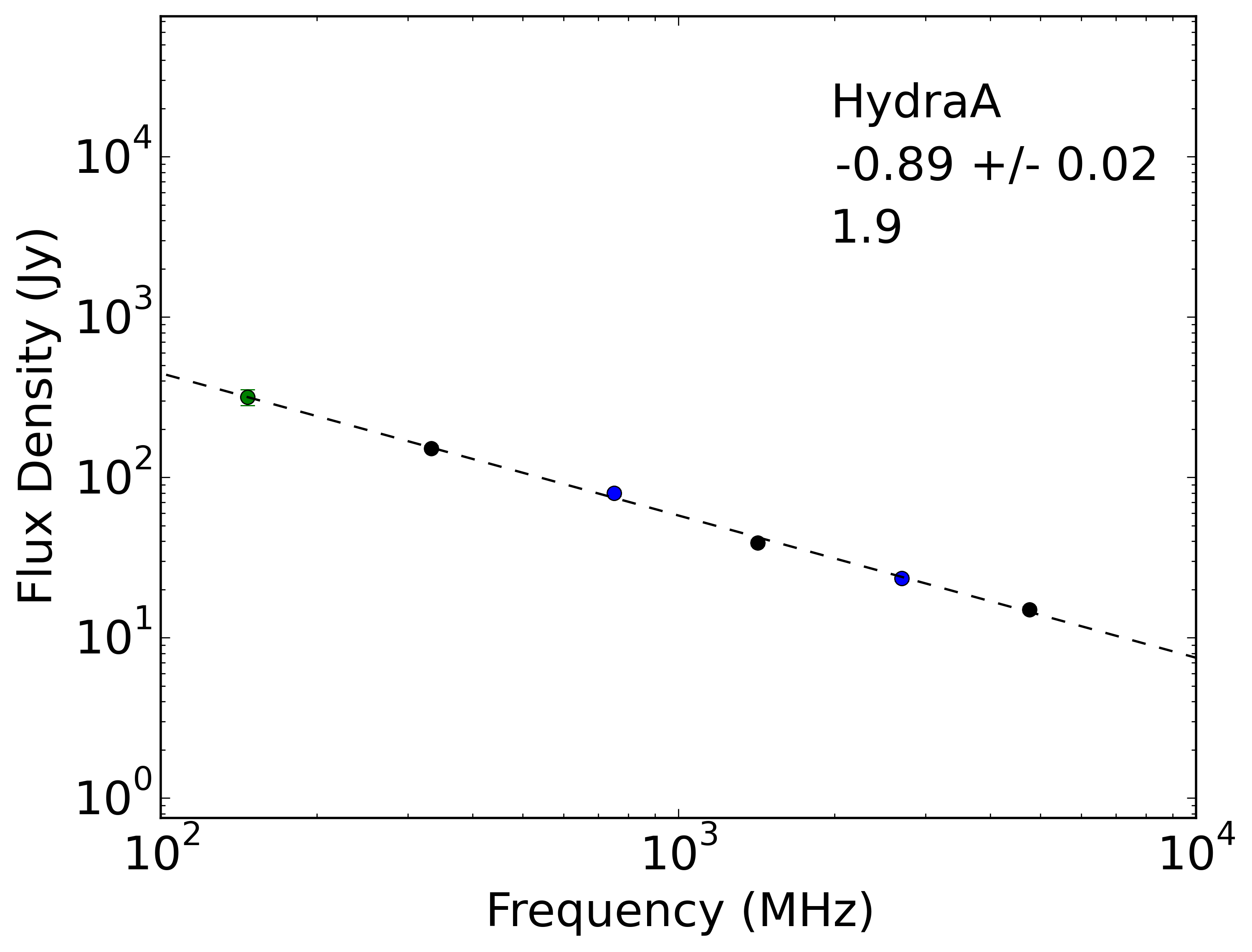}

\includegraphics[width=2.3in]{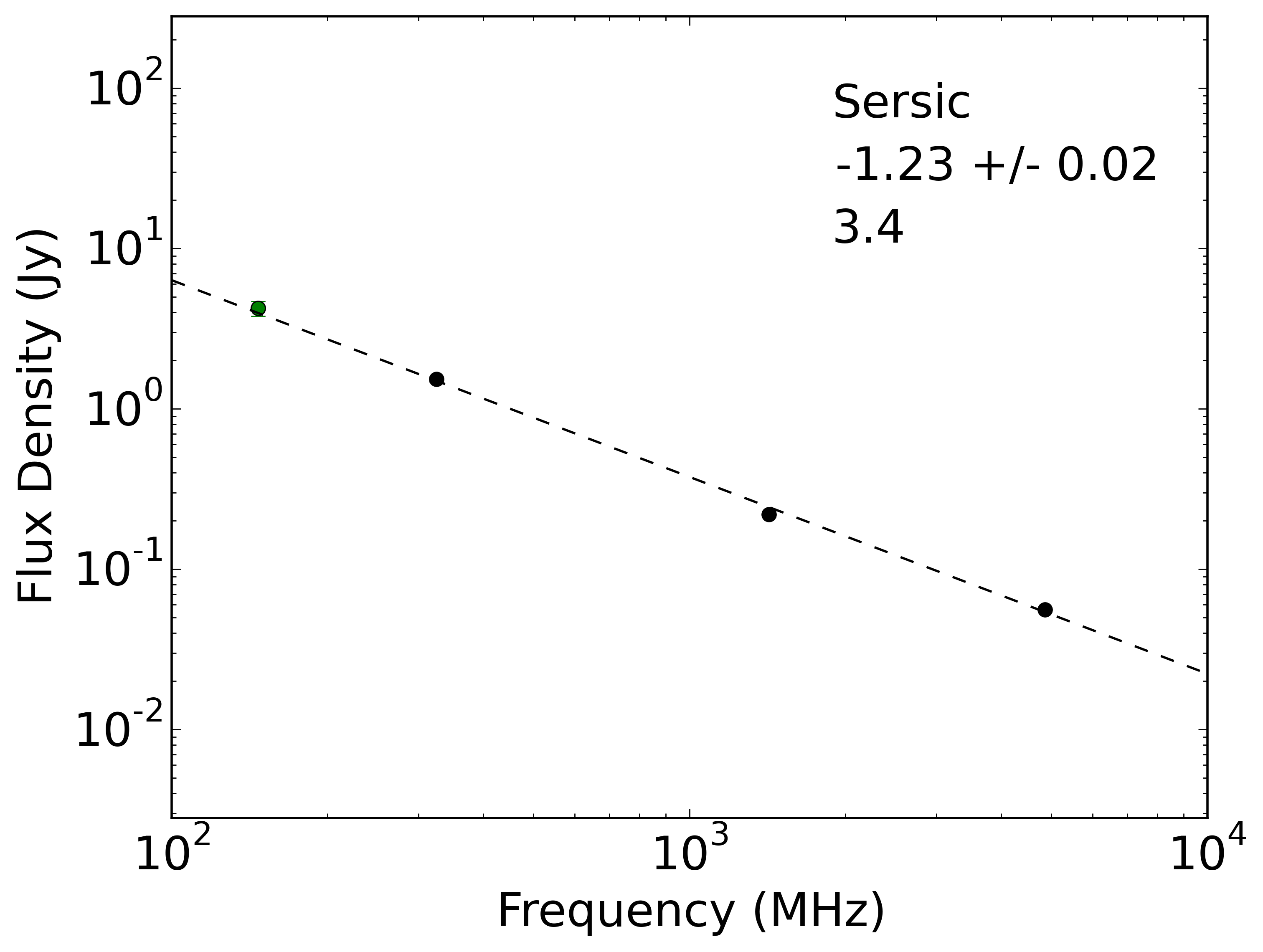}
\includegraphics[width=2.30in]{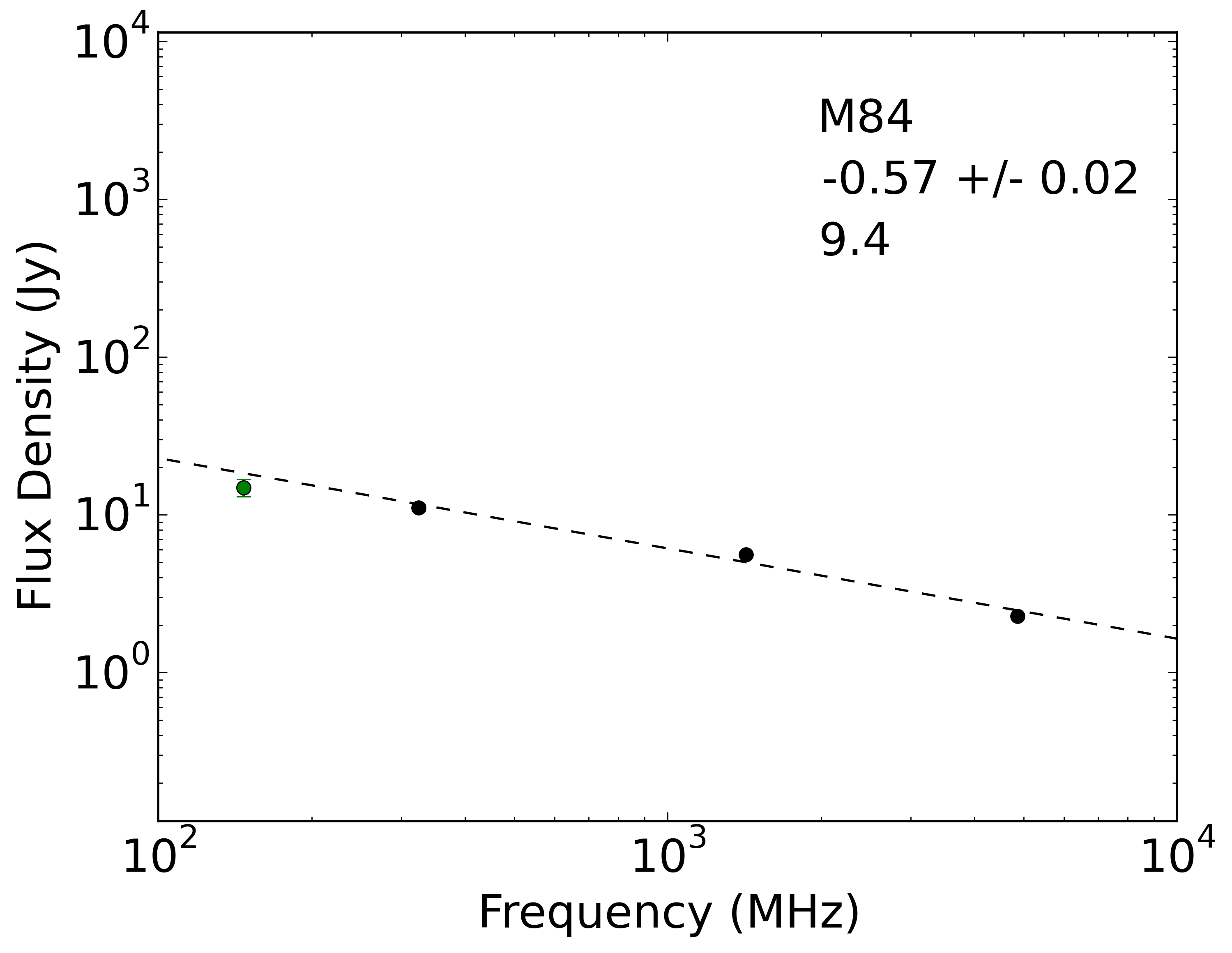}
\includegraphics[width=2.30in]{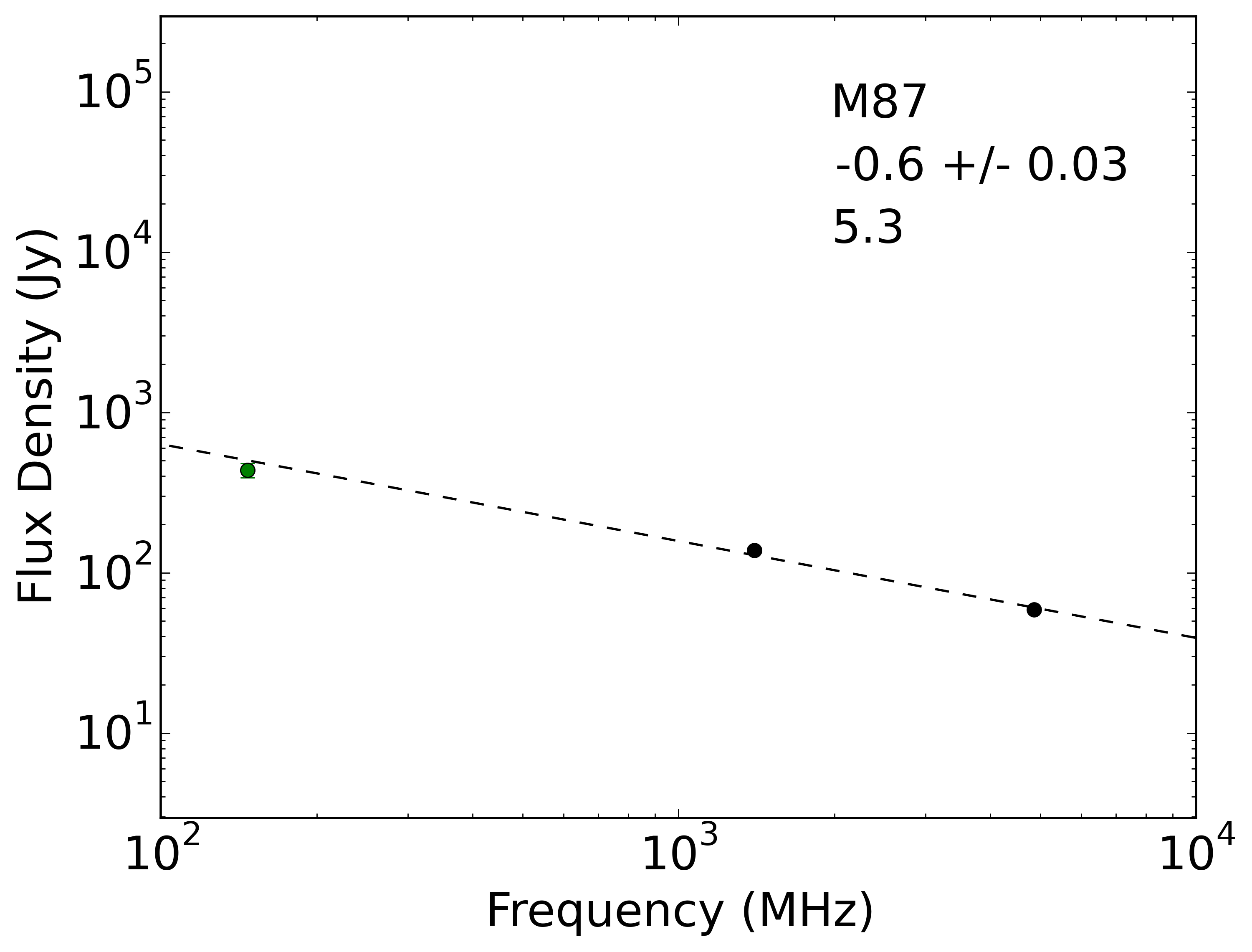}

\end{center}
\caption{\small SED power law fits for the sources in the TF-23 sample but not in the MF-14 sample. TGSS (green) points come from our measurements. The VLA (black) points are from \cite{Birzan2008}. The higher frequency literature points are presented in blue. The dashed line shows the best fitted power law. The value at the top right corner of each plot indicates the derived spectral index with its error and the reduced chi-squared of the fit.
\label{fig:nonMSSS_sed_tgss_vlss}}
\vspace{0.15in}
\end{figure*}

\end{appendices}

\end{document}